\pdfoutput=1

\documentclass{emulateapj}

\usepackage{url}
\usepackage{epstopdf}
\usepackage{multirow}
\usepackage{amsmath}
\usepackage{natbib}
\usepackage{color}
\usepackage{booktabs}
\usepackage{tabularx}
\usepackage[driverfallback=dvipdfmx, breaklinks, colorlinks, citecolor=blue, linkcolor=blue, menucolor=blue, urlcolor=blue]{hyperref}
\usepackage{longtable}
\usepackage{rotating}

\slugcomment{}

\shorttitle{Revealing the Nature of Blazar Radio Cores through Multi-Frequency Polarization Observations}
\shortauthors{Park et al.}

\begin{document}

\title{Revealing the Nature of Blazar Radio Cores through Multi-Frequency Polarization Observations with the Korean VLBI Network}

\author{Jongho Park\altaffilmark{1}, Minchul Kam\altaffilmark{1}, Sascha Trippe\altaffilmark{1,$\dagger$}, Sincheol Kang\altaffilmark{2,3}, Do-Young Byun\altaffilmark{2,3,4}, Dae-Won Kim\altaffilmark{1}, Juan-Carlos Algaba\altaffilmark{1,2}, Sang-Sung Lee\altaffilmark{2,3}, Guang-Yao Zhao\altaffilmark{2}, Motoki Kino\altaffilmark{5,6}, Naeun Shin\altaffilmark{1}, Kazuhiro Hada\altaffilmark{7,8}, Taeseok Lee\altaffilmark{1}, Junghwan Oh\altaffilmark{1}, Jeffrey A. Hodgson\altaffilmark{2}, and Bong Won Sohn\altaffilmark{2,3,4}}
\affil{$^1$Department of Physics and Astronomy, Seoul National University, Gwanak-gu, Seoul 08826, South Korea; jhpark@astro.snu.ac.kr, trippe@astro.snu.ac.kr\\
$^2$Korea Astronomy and Space Science Institute, 776 Daedeok-daero, Yuseong-gu, Daejeon 34055, Korea\\
$^3$Korea University of Science and Technology, 217 Gajeong-ro, Yuseong-gu, Daejeon 34113, Korea\\
$^4$Yonsei University, Yonsei-ro 50, Seodaemun-gu, Seoul 03722, Korea\\
$^5$National Astronomical Observatory of Japan, 2-21-1 Osawa, Mitaka, Tokyo, 181-8588, Japan\\
$^6$Kogakuin University, Academic Support Center, 2665-1 Nakano, Hachioji, Tokyo 192-0015, Japan\\
$^7$Mizusawa VLBI Observatory, National Astronomical Observatory of Japan, 2-21-1 Osawa, Mitaka, Tokyo 181-8588, Japan\\
$^8$Department of Astronomical Science, The Graduate University for Advanced Studies (SOKENDAI), 2-21-1 Osawa, Mitaka, Tokyo 181-8588, Japan
$^\dagger$Corresponding author}
\received{...}
\accepted{...}

\begin{abstract}
  We study the linear polarization of the radio cores of eight blazars simultaneously at 22, 43, and 86 GHz with observations obtained by the Korean VLBI Network (KVN) in three epochs between late 2016 and early 2017 in the frame of the Plasma-physics of Active Galactic Nuclei (PAGaN) project. We investigate the Faraday rotation measure (RM) of the cores; the RM is expected to increase with observing frequency if core positions depend on frequency due to synchrotron self-absorption. We find a systematic increase of RMs at higher observing frequencies in our targets. The RM--$\nu$ relations follow power-laws with indices distributed around 2, indicating conically expanding outflows serving as Faraday rotating media. Comparing our KVN data with contemporaneous optical polarization data from the Steward Observatory for a few sources, we find indication that the increase of RM with frequency saturates at frequencies of a few hundreds GHz. This suggests that blazar cores are physical structures rather than simple $\tau=1$ surfaces. A single region, e.g. a recollimation shock, might dominate the jet emission downstream of the jet launching region. We detect a sign change in the observed RMs of CTA 102 on a time scale of $\approx$1 month, which might be related to new superluminal components emerging from its core undergoing acceleration/deceleration and/or bending. We see indication for quasars having higher core RMs than BL Lac objects, which could be due to denser inflows/outflows in quasars.
\end{abstract}


\section{Introduction \label{sect1}}

Blazars, characterized by violent flux variability across the entire electromagnetic spectrum, are a sub-class of active galactic nuclei (AGNs) which show highly collimated, one-sided relativistic jets (see \citealt{UP1995} for a review). Large-scale magnetic fields which are strongly twisted in the inner part of the accretion disc or the black hole's ergosphere play a crucial role in launching and powering of relativistic jets \citep{BZ1977, BP1982}. Jets appear to be gradually accelerated and collimated magneto-hydrodynamically \citep{VK2004, komissarov2007, komissarov2009, AN2012, TT2013, hada2013, asada2014, mertens2016, hada2017, walker2018} and they are directly linked to accretion process onto supermassive black holes \citep{marscher2002, chatterjee2009, chatterjee2011, ghisellini2014, PT2017}. Their parsec-scale radio morphology is characterized by (a) the `VLBI core', a (radio) bright, optically thick, compact feature, and (b) an extended, optically thin, jet (e.g., \citealt{fromm2013}).

The nature of the core is a matter of ongoing debate. The standard Blandford \& K\"{o}nigl jet model describes the core as the upstream region where the conical jet becomes optically thin, i.e., at unity optical depth (e.g., \citealt{BK1979}). In this scenario, the observed core position shifts closer to the physical location of the jet base at higher observing frequencies -- the well-known `core shift effect' \citep{lobanov1998}. Core shift has been observed in blazars (e.g., \citealt{OG2009b, sokolovsky2011, algaba2012, pushkarev2012, fromm2013, hovatta2014}) as well as in nearby radio galaxies (e.g., \citealt{hada2011, martividal2011}), supporting the idea that the radio core marks the transition between the optically thick and thin jet regimes.

However, the plain `optical depth interpretation' of the radio core ignores the physical structure of AGN jets. Especially, a standing conical shock, located at the end of the jet acceleration and collimation zone (e.g., \citealt{marscher2008}), is expected (see also \citealt{PC2013a, PC2013b} for a discussion of the transition region from a parabolic to a conical jet shape that is dominating jet synchrotron emission of blazars). Such a (quasi-)stationary feature -- a `recollimation shock' -- may appear when there is a mismatch between the gas pressures in the jet and the confining medium (e.g., \citealt{DM1988, gomez1995, gomez1997, agudo2001, mizuno2015, marti2016}). Observations of the nearby radio galaxy M87 indeed reveal a stationary feature (known as HST-1) at the end of the jet collimation region \citep{AN2012}, showing blazar-like activity such as rapid variability and high energy emission \citep{cheung2007}. In addition, recent studies have discovered that most $\gamma$-ray flares in blazars occur when new (apparently) superluminal jet components pass through the core (\citealt{JM2016}, see also \citealt{ramakrishnan2014, casadio2015, rani2015} for the case of individual sources and \citealt{jorstad2001, leontavares2011} for investigation of statistical significance between the two phenomena). This indicates that the core supplies the jet plasma electrons with large amounts of energy, with the possible formation of a shock as the source of high energy emission.

At first glance, these two models and corresponding observational support seem to be in contradiction. This conflict is resolved if the core consists of (a) a standing shock which is optically thin only at (sub)-mm wavelengths, plus (b) extended jet flows downstream of the shock. In this case, there is no core shift expected at millimeter wavelengths where the core becomes transparent. Interestingly, a recent study which used a bona-fide astrometric technique showed that the core shift between 22 and 43 GHz for BL Lacertae is significantly smaller than the expected one from lower frequency data, indicating that the core at these frequencies might be identified with a recollimation shock \citep{dodson2017}. However, a number of previous studies did not find such a trend at the same frequencies (e.g., \citealt{OG2009b, algaba2012, fromm2013}). This might be because the core position accuracy of previous VLBI observations is comparable to the expected amount of core shift at those frequencies.

\begin{deluxetable*}{cccccccc}[t!]

\tablecaption{KVN Observation Log}
\tablehead{
\colhead{Project code} & \colhead{Obs. date} &
\colhead{Frequency} &
\colhead{D Term cal.} & \colhead{$\sigma_{\rm D} [\%]$} & \colhead{EVPA cal.} &
\colhead{$\Delta\chi$ [$^{\circ}$]} & \colhead{$\sigma_{\Delta\chi}$ [$^{\circ}$]} \\
(1) & (2) & (3) & (4) & (5) & (6) & (7) & (8)
}
\startdata
\multirow{2}{0.15\columnwidth}{\centering p16st01i} & \multirow{2}{0.15\columnwidth}{\centering2016 Dec 09} & 21.650 GHz & OJ 287 & 1.22 & 3C 279, OJ 287 & -28.79 & 1.04\\
& & 86.600 GHz & 3C 84 & 2.25 & 3C 279, OJ 287 & 112.76 & 0.90\\
\multirow{2}{0.15\columnwidth}{\centering p16st01j} & \multirow{2}{0.15\columnwidth}{\centering2016 Dec 10} & 43.300 GHz & OJ 287 & 0.70 & 3C 279, OJ 287 & -19.52 & 2.37 \\
& & 86.600 GHz & OJ 287 & 1.86 & 3C 279, OJ 287 & 125.86 & 2.65\\\hline
\multirow{2}{0.15\columnwidth}{\centering p17st01a} & \multirow{2}{0.15\columnwidth}{\centering2017 Jan 16} & 21.650 GHz & 3C 84 & 0.57 & 3C 279, OJ 287 & 11.53 & 0.94\\
& & 86.600 GHz & OJ 287 & 1.25 & 3C 279, OJ 287 & 5.55 & 1.80\\
\multirow{2}{0.15\columnwidth}{\centering p17st01b} & \multirow{2}{0.15\columnwidth}{\centering2017 Jan 17} & 43.100 GHz & 3C 84 & 1.07 & 3C 279, OJ 287 & 54.29 & 0.33 \\
& & 129.300 GHz & $-$ & $-$ & $-$ & $-$ & $-$\\\hline
\multirow{2}{0.15\columnwidth}{\centering p17st01e} & \multirow{2}{0.15\columnwidth}{\centering2017 Mar 22} & 21.650 GHz & 3C 84 & 1.04 & 3C 279, OJ 287 & -28.72 & 0.87\\
& & 86.600 GHz & OJ 287 & 1.59 & 3C 279, OJ 287, 3C 454.3 & 32.42 & 1.04\\
\multirow{2}{0.15\columnwidth}{\centering p17st01f} & \multirow{2}{0.15\columnwidth}{\centering2017 Mar 24} & 43.100 GHz & OJ 287 & 0.90 & 3C 279, OJ 287, BL LAC & 9.07 & 1.37 \\
& & 129.300 GHz & $-$ & $-$ & $-$ & $-$ & $-$
\enddata
\tablecomments{(1) KVN observations project code (2) observation date (3) \emph{starting} observing frequency (4) sources used for instrumental polarization calibration (5) Errors of D-terms estimated by comparing the D-terms obtained from different instrumental polarization calibrators (6) sources used for EVPA calibration by comparing with contemporaneous KVN single dish observations (7) amount of EVPA rotation applied (8) Errors of EVPA calibration estimated from $\Delta\chi$ values of different sources in (6). We note that we did not include the 129 GHz observations in our analysis due to the complicated polarization calibration at this frequency. A dedeciated study for investigating the capability of polarimetry at this frequency will be presented elsewhere.}\label{Information}

\end{deluxetable*}

An alternative route is provided by multi-frequency polarimetric observations of the core that provide RMs, defined as $\rm EVPA_{obs} = EVPA_{int} + RM\lambda^2$, where $\rm EVPA_{obs}$ and $\rm EVPA_{int}$ are observed and intrinsic electric vector position angles (EVPAs) of linearly polarized emission and $\lambda$ is observing wavelength. If the core is the $\tau = 1$ surface of a continuous conical jet and the jet is in a state of energy equipartition, then the core RM obeys the relation $|{\rm RM_{\rm core, \nu}}| \propto \nu^a$, where $a$ is the power-law index of the electron density distribution given by $N_e \propto d^{-a}$, with $d$ being the distance from the jet base \citep{jorstad2007}. In this scenario, we observe polarized emission from regions closer to the jet base at higher frequencies due to the core shift effect, where one may expect higher particle densities and stronger magnetic fields. Looking at this argument the other way around, we would expect \emph{no increase} in RM as a function of frequency at millimeter wavelengths \emph{if} the core is indeed a standing recollimation shock. This provides the opportunity to uncover the nature of blazar VLBI cores, and thus the intrinsic structure of blazar jets, through multi-frequency polarimetric observations at millimeter wavelengths.

At centimeter wavelengths, many studies showed that the power law index $a$ is usually distributed around $a = 2$ (e.g., \citealt{OG2009a, algaba2013, kravchenko2017}), corresponding to a spherical or conical outflow \citep{jorstad2007}. A conical outflow is more likely than a spherical one because \citet{pushkarev2017} showed that conical jet geometries are common in blazars. $a\approx2$ found in many blazars is in agreement with the fact that many blazars show core shift at these wavelengths. However, to the best of our knowledge, there are only a few studies of the core RM of blazars at (sub-)mm frequencies. \cite{jorstad2007} analyzed 7, 3, and 1 mm polarization data and obtained an average $\langle a \rangle = 1.8 \pm 0.5$ by comparing with other studies done at cm wavelengths. This result indicates that the dependence of RM on observing frequency might continue up to mm wavelengths. Some of their sources are not fitted well by $\lambda^2$ laws even at the highest frequencies, indicating that a frequency dependence of RM exists even at around 1 mm. Another study using the IRAM 30-m telescope at 3 and 1~mm found RMs (a few times $10^4 \rm\ rad/m^2$) that are much larger than those at cm wavelengths (a few hundred $\rm rad/m^2$, \citealt{hovatta2012}), albeit within large errors (\citealt{agudo2014}, see also \citealt{agudo2018a, agudo2018b, thum2018}).

A recent observation with the Atacama Large Millimeter Array (ALMA) at 1~mm has revealed a very high rotation measure of $(3.6\pm0.3)\times10^5\rm \ rad/m^2$ in 3C 273 with the core RM scaling with frequency like $\rm|RM|\propto\nu^{1.9\pm0.2}$ from cm to mm wavelengths. \cite{martividal2015} observed even larger RMs ($\approx 10^8 \rm\ rad/m^2$ in the rest frame) in the gravitationally lensed quasar PKS 1830--211 through ALMA observations at up to 300 GHz (about 1 THz in the rest frame). These results may suggest that (i) blazar core RMs rapidly increase as a function of frequency, as predicted by \cite{jorstad2007}; (ii) polarized (sub-)mm radiation might originate near the jet base, not from a recollimation shock (which presumably is located quite far from the jet base). However, it is uncertain whether this is a common behaviour of blazars or if these quasars are special. Therefore, a systematic study of blazar core RMs with multi-frequency polarimetric observations at (sub)mm wavelengths is necessary.

The Korean VLBI Network (KVN) has the unique capability of observing simultaneously at four frequencies, 22, 43, 86, and 129 GHz, or at two of these frequencies in dual polarization mode \citep{lee2011,lee2014}. Thanks to the simultaneous observation at multiple frequencies, one can overcome rapid phase variations at high frequencies caused by tropospheric delay that reduce the coherence time by applying the fringe solutions obtained at lower frequencies to higher ones, i.e., frequency phase transfer (FPT) \citep{rioja2011, rioja2014, algaba2015, zhao2018}. This technique increases the fringe detection rate to values larger than 80\% even at 129 GHz for sources brighter than $\approx0.5$ Jy, making the KVN a powerful instrument for multi-frequency mm polarimetry of AGNs.

In early 2017, we launched a KVN large program, the Plasma-physics of Active Galactic Nuclei (PAGaN) project (see \citealt{kim2015, oh2015} for related studies), for monitoring about 14 AGNs at the four KVN frequencies in dual polarization mode almost every month.\footnote{\url{https://radio.kasi.re.kr/kvn/ksp.php\#ksp003}} One of the main scientific goals of the project is a systematic study of RMs of blazars at mm wavelengths and their evolution in time. In this paper, we present the results from three observation epochs located between late 2016 and early 2017, which were performed as test observations for the initiation of the large program. We describe observations and data calibration in Section~\ref{sect2}. Results are shown and discussed in Section~\ref{sect3} and~\ref{sect4}, respectively. In Section~\ref{sect5}, we summarize our findings.

\begin{sidewaystable*}[p]
  \vskip-100mm
  \caption{Our targets and the polarization properties of their cores}\label{table:source}%
  \newcolumntype{R}{>{\raggedright \arraybackslash} X}
  \newcolumntype{S}{>{\centering \arraybackslash} X}
  \newcolumntype{T}{>{\raggedleft \arraybackslash} X}
  \begin{tabularx}{\linewidth} {>{\setlength\hsize{.20\hsize}}R >{\setlength\hsize{.13\hsize}}R >{\setlength\hsize{.10\hsize}}R >{\setlength\hsize{.10\hsize}}R >{\setlength\hsize{.40\hsize}}S>{\setlength\hsize{.29\hsize}}S >{\setlength\hsize{.29\hsize}}S >{\setlength\hsize{.20\hsize}}S >{\setlength\hsize{.22\hsize}}S >{\setlength\hsize{.20\hsize}}S >{\setlength\hsize{.21\hsize}}S >{\setlength\hsize{.20\hsize}}S>{\setlength\hsize{.25\hsize}}S>{\setlength\hsize{.20\hsize}}S>{\setlength\hsize{.47\hsize}}S>{\setlength\hsize{.47\hsize}}S
  >{\setlength\hsize{.15\hsize}}S }  
  \toprule
    \multirow{2}{.20\hsize}{Source} & \multirow{2}{.13\hsize}{Type} & \multirow{2}{.10\hsize}{z} & \multirow{2}{.10\hsize}{$l$} & \multirow{2}{.35\hsize}{\centering Epoch} & \multirow{2}{.25\hsize}{\centering $\alpha_{22/43}$} & \multirow{2}{.25\hsize}{\centering $\alpha_{43/86}$}& \multicolumn{2}{c}{22 GHz} & \multicolumn{2}{c}{43 GHz} & \multicolumn{2}{c}{86 GHz} & \multirow{2}{.20\hsize}{\centering $\beta$} & \multirow{2}{.40\hsize}{\centering $\rm RM_{22/43}[\rm rad/m^2]$} & \multirow{2}{.40\hsize}{\centering $\rm RM_{43/86}$} & \multirow{2}{.15\hsize}{\centering $a$} \\
 & & & & & & & $m [\%]$ & $\chi$ [$^{\circ}$] & $m$ & $\chi$ & $m$ & $\chi$ & &\tabularnewline
 & \centering(1) & \centering(2) & \centering(3) & & \centering(4)  & \centering(5)  & & & & & &  & \centering(6) & \centering(7) & \centering(8) & \centering(9) \tabularnewline
  \midrule
  \midrule

\multirow{3}{.20\hsize}{3C279}&\multirow{3}{.15\hsize}{FSRQ}&\multirow{3}{.10\hsize}{0.54}&\multirow{3}{.10\hsize}{20}&12/09-10/16&$-0.4\pm0.2$&$-0.3\pm0.2$&9.9$\pm$0.1&26$\pm$1&10.3$\pm$0.5&33$\pm$3&10.5$\pm$2.2&37$\pm$2&-0.05&(-2.0$\pm$1.0)$\times10^3$&(-4.0$\pm$4.5)$\times10^3$&1.0$\pm$1.8\tabularnewline
&&&&01/16-17/17&$-0.3\pm0.2$&$-0.7\pm0.2$&8.8$\pm$0.1&28$\pm$1&8.8$\pm$0.4&31$\pm$2&8.2$\pm$0.9&38$\pm$3&0.03&(-6.5$\pm$6.5)$\times10^2$&(-8.4$\pm$4.1)$\times10^3$&3.7$\pm$1.6\tabularnewline
&&&&03/22-24/17&$-0.1\pm0.2$&$-0.4\pm0.2$&8.6$\pm$0.1&30$\pm$1&11.2$\pm$0.2&36$\pm$2&12.4$\pm$0.4&49$\pm$3&-0.31&(-1.7$\pm$0.6)$\times10^3$&(-1.4$\pm$0.4)$\times10^4$&3.1$\pm$0.6\tabularnewline\midrule
\multirow{3}{.20\hsize}{OJ287}&\multirow{3}{.15\hsize}{BLO}&\multirow{3}{.15\hsize}{0.31}&\multirow{3}{.10\hsize}{14}&12/09-10/16&$-0.2\pm0.2$&$-0.4\pm0.2$&4.6$\pm$0.1&-2$\pm$1&6.1$\pm$0.1&-1$\pm$2&7.4$\pm$0.5&-2$\pm$2&-0.38&(-3.8$\pm$5.6)$\times10^2$&(1.2$\pm$2.5)$\times10^3$&1.7$\pm$3.7\tabularnewline
&&&&01/16-17/17&$0.0\pm0.2$&$-0.7\pm0.2$&6.6$\pm$0.1&-26$\pm$1&8.3$\pm$0.1&-33$\pm$1&10.9$\pm$0.1&-36$\pm$2&-0.73&(1.5$\pm$0.2)$\times10^3$&(2.8$\pm$1.6)$\times10^3$&1.0$\pm$0.7\footnote{This power-law index is obtained by including the RM at 15/22 GHz; see Figure~\ref{result} and Section~\ref{oj287} for details.}\tabularnewline
&&&&03/22-24/17&$-0.2\pm0.2$&$-0.3\pm0.2$&5.0$\pm$0.1&-63$\pm$1&7.6$\pm$0.1&-63$\pm$1&8.3$\pm$0.2&-57$\pm$1&-0.43&(-0.3$\pm$3.6)$\times10^2$&(-4.8$\pm$1.5)$\times10^3$&7.1$\pm$14.6\tabularnewline\midrule
\multirow{2}{.20\hsize}{CTA102}&\multirow{2}{.15\hsize}{FSRQ}&\multirow{2}{.15\hsize}{1.04}&\multirow{2}{.10\hsize}{26}&12/09-10/16&$0.3\pm0.2$&$0.3\pm0.2$&1.6$\pm$0.1&-83$\pm$3&1.5$\pm$0.2&-91$\pm$5&1.8$\pm$0.4&-117$\pm$11&0.02&(4.3$\pm$2.6)$\times10^3$&(5.2$\pm$2.4)$\times10^4$&3.6$\pm$1.1\tabularnewline
&&&&01/16-17/17&$0.4\pm0.2$&$0.1\pm0.2$&1.1$\pm$0.1&-63$\pm$2&1.4$\pm$0.2&0$\pm$3&1.4$\pm$0.1&36$\pm$8&-0.13&(-3.2$\pm$0.2)$\times10^4$&(-7.2$\pm$1.6)$\times10^4$&1.2$\pm$0.3\tabularnewline\midrule
\multirow{2}{.20\hsize}{3C345}&\multirow{2}{.20\hsize}{FSRQ}&\multirow{2}{.20\hsize}{0.59}&\multirow{2}{.10\hsize}{21}&12/09-10/16&$-0.3\pm0.2$&$-0.6\pm0.2$&2.6$\pm$0.2&-111$\pm$2&3.9$\pm$0.2&-74$\pm$3&6.6$\pm$0.8&-62$\pm$3&-0.65&(-1.2$\pm$0.1)$\times10^4$&(-1.4$\pm$0.5)$\times10^4$&0.3$\pm$0.5\tabularnewline
&&&&01/16-17/17&$-0.2\pm0.2$&$-0.7\pm0.2$&3.7$\pm$0.1&-88$\pm$1&5.8$\pm$0.2&-74$\pm$1&9.0$\pm$0.7&-61$\pm$2&-0.65&(-4.4$\pm$0.5)$\times10^3$&(-1.6$\pm$0.3)$\times10^4$&1.9$\pm$0.3\tabularnewline\midrule
\multirow{2}{.20\hsize}{1749+096}&\multirow{2}{.20\hsize}{BLO}&\multirow{2}{.20\hsize}{0.32}&\multirow{2}{.10\hsize}{15}&12/09-10/16&$0.0\pm0.2$&$-0.2\pm0.2$&4.2$\pm$0.1&2$\pm$1&3.2$\pm$0.1&-1$\pm$3&2.2$\pm$0.5&-11$\pm$3&0.42&(4.5$\pm$6.2)$\times10^2$&(8.8$\pm$3.6)$\times10^3$&4.3$\pm$2.1\tabularnewline
&&&&03/22-24/17&$-0.1\pm0.2$&$-0.4\pm0.2$&4.6$\pm$0.2&-10$\pm$2&2.9$\pm$0.1&-22$\pm$2&2.5$\pm$0.8&-30$\pm$6&0.63&(2.6$\pm$0.5)$\times10^3$&(6.5$\pm$5.1)$\times10^3$&1.3$\pm$1.2\tabularnewline\midrule
0235+164&FSRQ&0.94&25&03/22-24/17&$0.1\pm0.2$&$0.0\pm0.2$&2.8$\pm$0.1&48$\pm$1&3.3$\pm$0.1&51$\pm$2&2.4$\pm$0.3&59$\pm$3&-0.09&(-1.4$\pm$1.0)$\times10^3$&(-1.5$\pm$0.6)$\times10^4$&3.5$\pm$1.2\tabularnewline\midrule
BLLAC&BLO&0.07&4&03/22-24/17&$0.0\pm0.2$&$-0.9\pm0.2$&3.5$\pm$0.1&38$\pm$1&4.1$\pm$0.3&31$\pm$2&3.8$\pm$0.4&19$\pm$3&-0.11&(1.0$\pm$0.3)$\times10^3$&(6.4$\pm$1.9)$\times10^3$&2.7$\pm$0.6\tabularnewline\midrule
1633+38&FSRQ&1.81&27&03/22-24/17&$-0.2\pm0.2$&$-0.2\pm0.2$&1.5$\pm$0.2&31$\pm$3&1.4$\pm$0.2&24$\pm$4&2.2$\pm$0.7&18$\pm$6&-0.09&(6.3$\pm$4.8)$\times10^3$&(2.4$\pm$2.7)$\times10^4$&1.9$\pm$2.0\tabularnewline
  \bottomrule    
  \end{tabularx}
   \tablecomments{(1) Source types: FSRQ: flat spectrum radio quasars; BLO: BL Lac objects; taken from the MOJAVE web site. (2) Redshifts are taken from the NASA/IPAC Extragalactic Database (NED). (3) Linear size probed for each source in units of parsec corresponding to the typical size of the minor axis of the KVN beam at 22 GHz, using the scaling factors provided by the MOJAVE web site. (4) Total intensity spectral index, defined as $S_{\nu} \propto \nu^{\alpha}$ between 22 and 43 GHz. (5) Total intensity spectral index between 43 and 86 GHz. (6) Polarization spectral index, defined as $m\propto\lambda^{\beta}$. (7) RM between 22 and 43 GHz in source rest frame. (8) RM between 43 and 86 GHz in source rest frame. (9) Power-law index in ${\rm |RM|}\propto\nu^a$ (see Figure~\ref{result}).}
  \end{sidewaystable*}

\section{Observations and Data Reduction}
\label{sect2}

We observed a total of 11 sources in the 22, 43, and 86 GHz bands with the KVN on 2016 December 9--10 and in the four bands including the 129 GHz band on 2017 January 16--17 and 2017 March 22--24 with observation time of $\approx48$ hours for each epoch. Since KVN can observe at two frequencies simultaneously in dual polarization mode, we allocated the first half of the observing time to 22/86 GHz observations and the other half to 43/129 GHz. Although we obtained the data at 129 GHz in the two epochs observations, we had a difficulty in polarization calibration of the data and thus we did not include them in this paper. More sophisticated investigation of the 129 GHz data will be presented in a forthcoming paper (Kam et al. in preparation). All sources were observed in 6--15 scans of 5--20 minutes in length, depending on source declination and brightness. We performed cross-scan observations at least twice per hour to correct antenna pointing offsets that might lead to inaccurate correlated amplitudes. The received signals were 2-bit quantized and divided into 4 sub-bands (IFs) of 16 MHz bandwidth for each polarization and each frequency. Mark 5B recorders were used at recording rates of 1024 Mbps. The data were correlated with the DiFX software correlator in the Korea-Japan Correlator Center \citep{lee2015a}. Table~\ref{Information} summarizes our observations.

A standard data post-correlation process was performed with the NRAO Astronomical Image Processing System (AIPS). Potential effects of digital sampling on the amplitudes of cross-correlation spectra were estimated by the AIPS task ACCOR. Amplitude calibration was done by using the antennas gain curves and opacity corrected system temperatures provided by the observatory. The fringe amplitudes were re-normalized by taking into account potential amplitude distortion due to quantization, and the quantization and re-quantization losses \citep{lee2015b}. 

The instrumental delay residuals were removed by using the data in a short time range of bright calibrators, either 3C 279 or 3C 454.3. To apply the FPT technique, a global fringe fitting was performed with a solution interval of 10 seconds for the lower frequency first (22 or 43 GHz), which led us to very high fringe detection rates $\gtrsim 95\%$ in most cases. Then, we transferred the obtained fringe solutions to the simultaneously observed higher frequency (86 GHz). This process corrects rapidly varying tropospheric errors in the visibility phases at high frequencies (though not the ionospheric errors that vary more slowly). Then, the residual phases have much longer coherence times, typically larger than a few minutes. Thus, we performed a global fringe fitting with a much longer solution interval of $\approx3$ minutes for the high frequency data, which resulted in quite high fringe detection rates -- usually larger than 95\% at 86 GHz for our sources. Bandpass calibration was performed by using scans on bright sources such as 3C 279.

The cross-hand R-L phase and delay offsets were calibrated by using the data for bright sources, such as OJ 287, 3C 84, 3C 279 and 3C 454.3, located within short time ranges, with the task RLDLY. We used the Caltech Difmap package for imaging and phase self-calibration \citep{shepherd1997}. Typical beam sizes are $5.6\times3.2$, $2.8\times1.6$, and $1.4\times0.8$ mas at 22, 43, and 86 GHz, respectively. We determined the feed polarization leakage (D-terms) for each antenna by using the task LPCAL \citep{leppanen1995} with a total intensity model of the D-Term calibrators. 3C 84 usually serves as a good D-Term calibrator thanks to its high flux density and very low degree of linear polarization ($\lesssim$0.5\%) at lower frequencies but less so at high frequency (86 GHz) where its linear polarization becomes non-negligible (Kam et al., in preparation). Thus, we also used a compact, bright, and polarized source OJ 287 at 86 GHz. We chose the D-Term calibrator for each epoch and for each frequency by comparing the behaviour of observed visibility ratios on the complex plane with the D-Term models of different calibrators (see Appendix~\ref{appendixa}). The EVPA calibration was performed by comparing the integrated EVPAs of the VLBI maps of the EVPA calibrators after the instrumental polarization calibration with contemporaneous KVN single dish polarization observations. We performed KVN single dish observations within two days of each VLBI observations as described in \cite{kang2015}. For the 2016 data, we have two 86 GHz data separated by 1 day. We note that the maps for all sources after the calibration are almost identical to each other and we used the average of Stokes I, Q, and U maps of the two data for our further analysis.

Estimating errors for degree of linear polarization ($m$) and EVPA is important but not straightforward. Errors for each polarization quantity can be derived from the following relations:
\begin{equation}
\sigma_p = \frac{\sigma_Q + \sigma_U}{2}
\end{equation}
\begin{equation}
\sigma_{\rm EVPA} = \frac{\sigma_p}{2p}
\end{equation}
\begin{equation}
\sigma_{m} = \frac{\sigma_p}{I}
\end{equation}
where $\sigma_Q$ and $\sigma_U$ denote rms noise in the Stokes Q and U images, respectively, $p = \sqrt{Q^2 + U^2}$, and $m = p/I$ \citep{hovatta2012}. In most cases, random errors are quite small and systematic errors are much more dominant in the above quantities. Imperfect D-term calibration is usually the most dominant source of errors in $m$. For EVPAs, both the D-term uncertainty and EVPA correction error are important. Following \cite{roberts1994}, errors of $m$ and EVPA caused by residual D-terms can be expressed as:
\begin{equation}
\label{eq4}
\sigma_{m, {\rm D}} = \sigma_D(N_aN_{\rm IF}N_s)^{-1/2}
\end{equation}
\begin{equation}
\label{eq5}
\sigma_{\rm EVPA, {\rm D}} \approx \frac{\sigma_{m, {\rm D}}}{2m}
\end{equation}
where $\sigma_D$ is the D-term error, $N_a$ and $N_{\rm IF}$ are the number of antennas and IFs, respectively, and $N_s$ is the number of scans having independent parallactic angles which depend on the source declination. We estimated the D-term errors by comparing the D-terms obtained from different D-Term calibrators (see (5) in Table~\ref{Information} and Appendix~\ref{appendixa}) and estimated $\sigma_{m, {\rm D}}$ and $\sigma_{\rm EVPA, {\rm D}}$ using Equation~\ref{eq4} and~\ref{eq5}. Thanks to the number of IFs being four and the large parallactic angle coverage of our sources, we could achieve errors in $m$ (typically $0.1-0.3\%$) much smaller than the D-Term errors (typically $1-2\%$). We also assessed the EVPA correction error, $\sigma_{\Delta\chi}$, by comparing the amount of EVPA rotation calculated from different EVPA calibrators (see (6) and (8) in Table~\ref{Information}). Then, we added $\sigma_m$ and $\sigma_{m, {\rm D}}$ quadratically for $m$ and $\sigma_{\rm EVPA}$, $\sigma_{\rm EVPA, D}$, and $\sigma_{\Delta\chi}$ quadratically for EVPA. 

In the Appendix, we show the results of D-term calibration and the temporal evolution of the D-terms. The overall D-terms are usually less than $\approx10\%$, except for Ulsan station at 86 GHz which showed D-terms as large as $\approx20\%$. The D-terms obtained from different calibrators are quite consistent with each other, showing standard deviations of $\lesssim2\%$ (see (5) in Table~\ref{Information}). The D-terms are more or less stable over $\approx$ 3 months, showing standard deviations of $\lesssim2\%$. We also compare our KVN 22/43 and 86 GHz data of 3C 273 observed in 2016 December with contemporaneous Very Long Baseline Array (VLBA) 15/43 GHz data, respectively. Both fractional polarization and EVPAs at a few different locations in the jet are in good agreement within errors between the data of the different instruments, considering non-negligible time gaps between the observations and the expected RM of a few hundred $\rm rad/m^2$ in the jet (e.g., \citealt{hovatta2012}).

\begin{figure*}[p]
\centering
\includegraphics[trim=7mm 0mm 6mm 0mm, clip, width = 75mm]{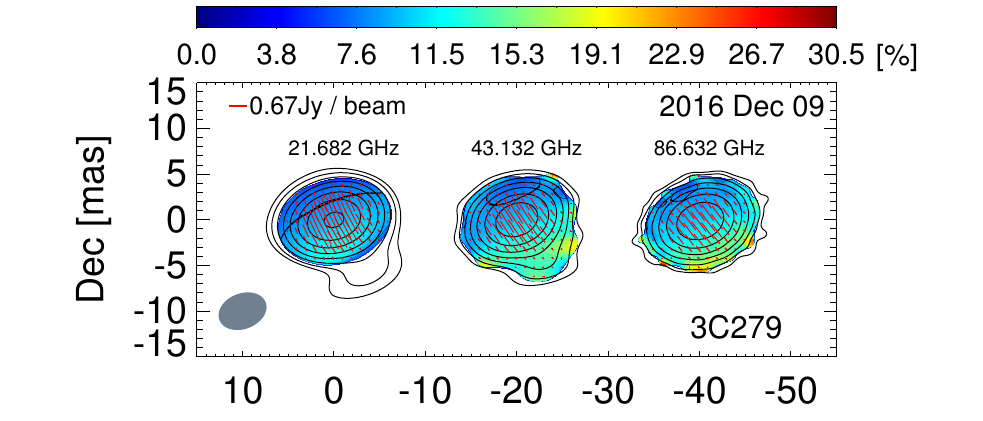}
\includegraphics[trim=7mm 4mm 5mm 13mm, clip, width = 47mm]{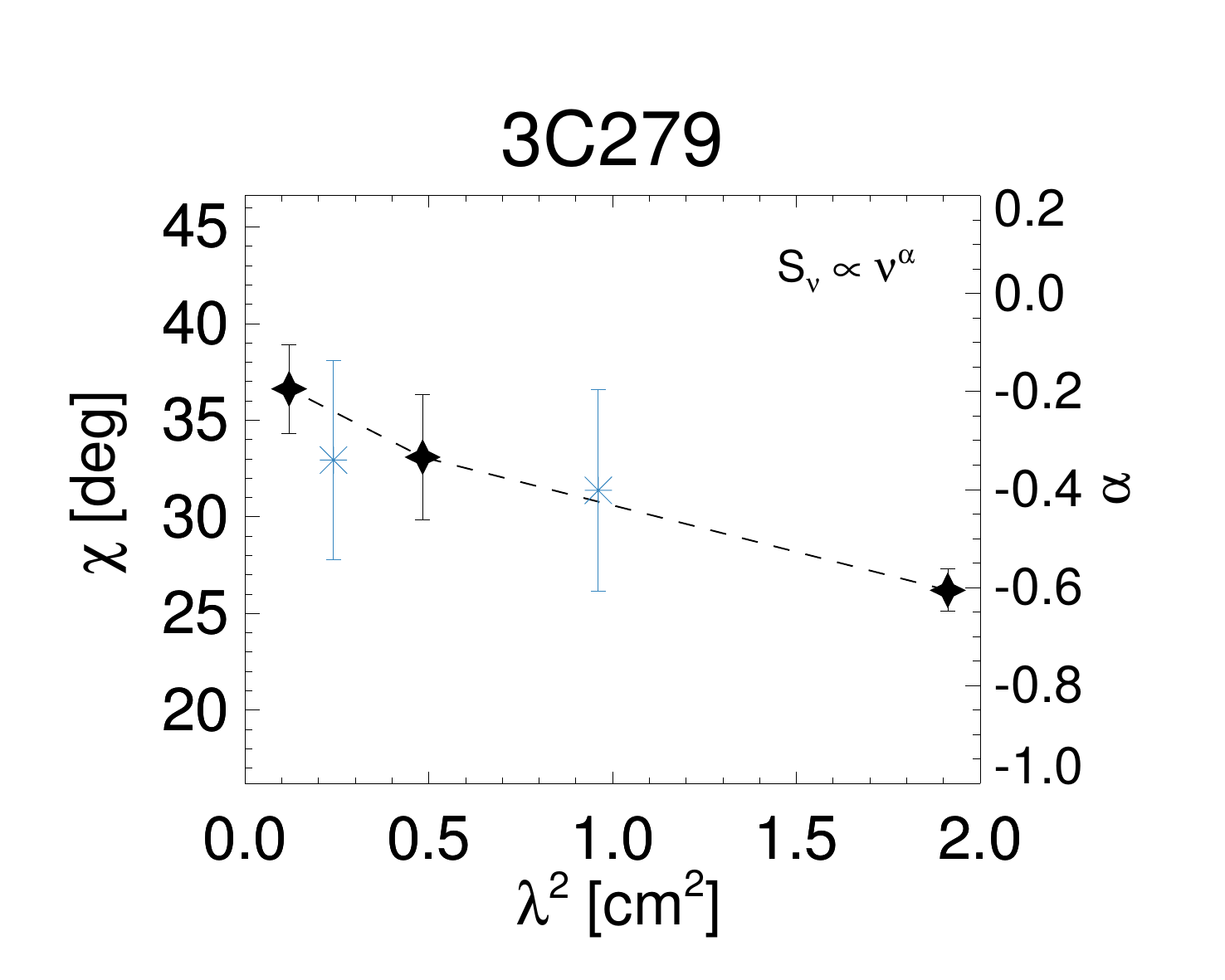}
\includegraphics[trim=5mm 4mm 24mm 13mm, clip, width = 43mm]{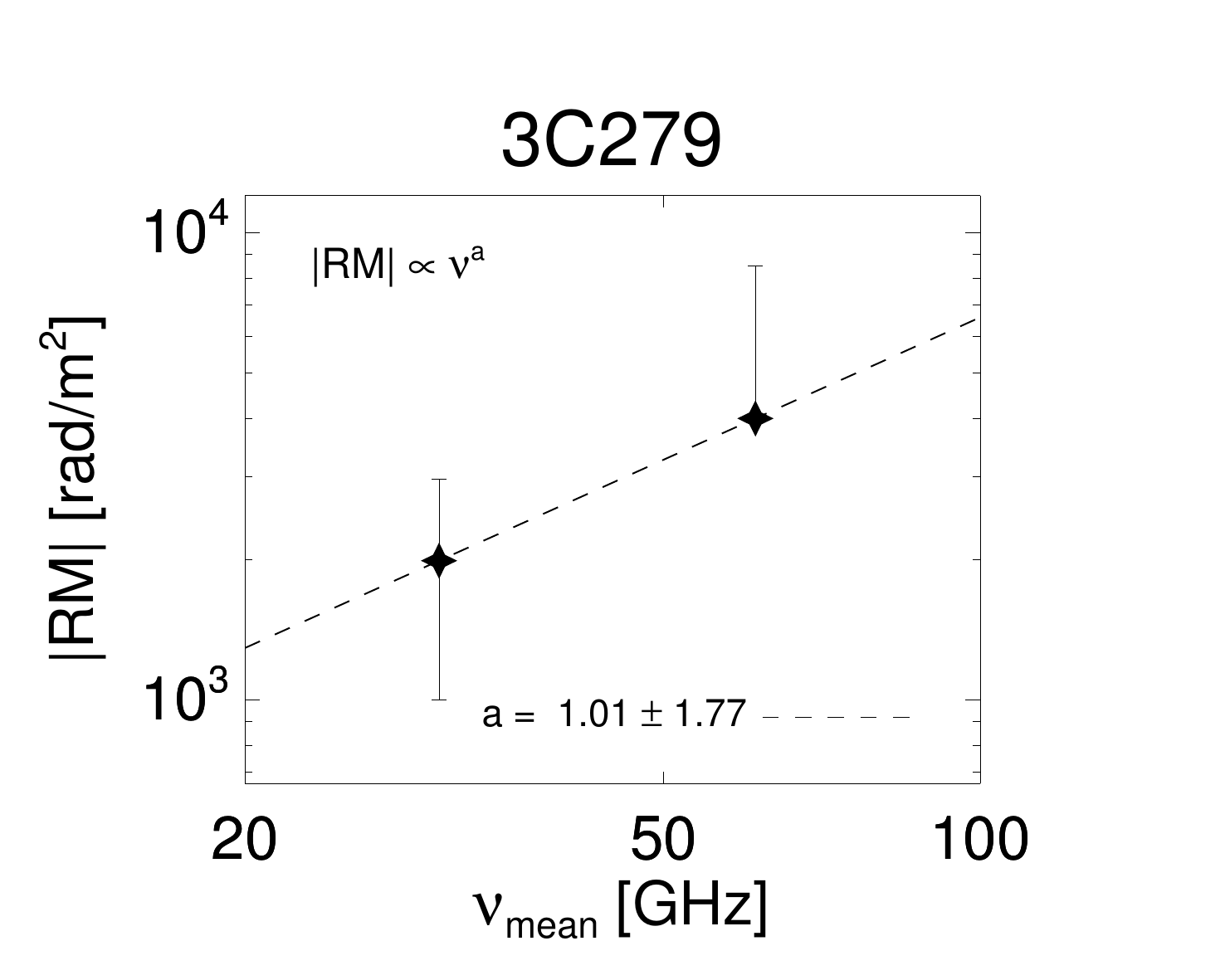}\\[-4mm] 
\includegraphics[trim=7mm 0mm 6mm 0mm, clip, width = 75mm]{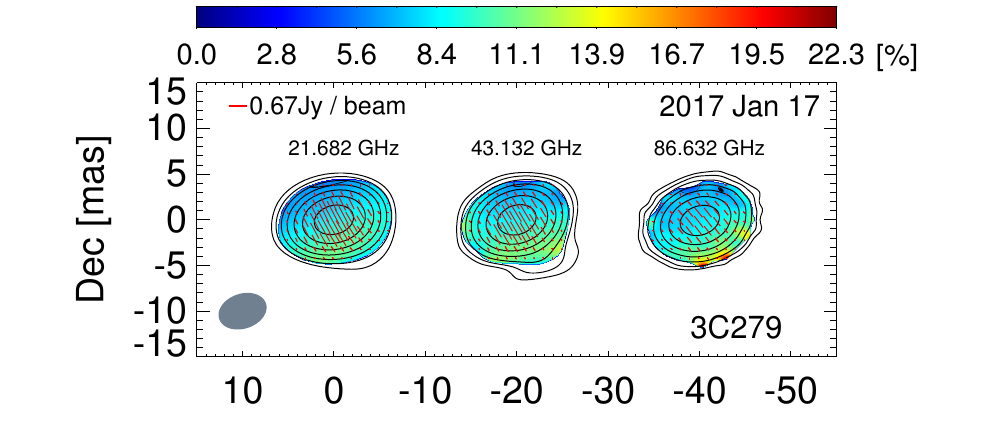}
\includegraphics[trim=7mm 4mm 5mm 1mm, clip, width = 47mm]{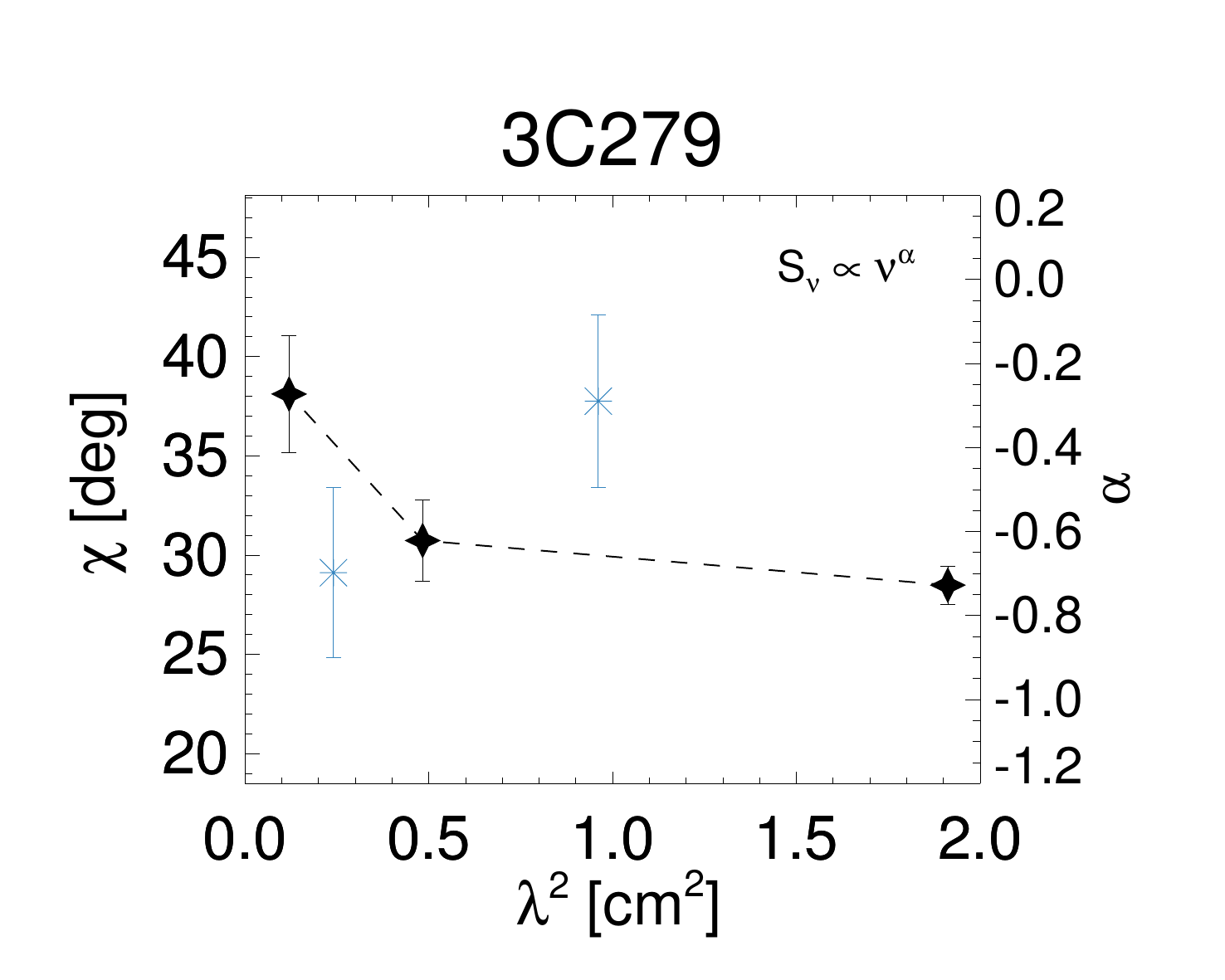}
\includegraphics[trim=5mm 4mm 24mm 1mm, clip, width = 43mm]{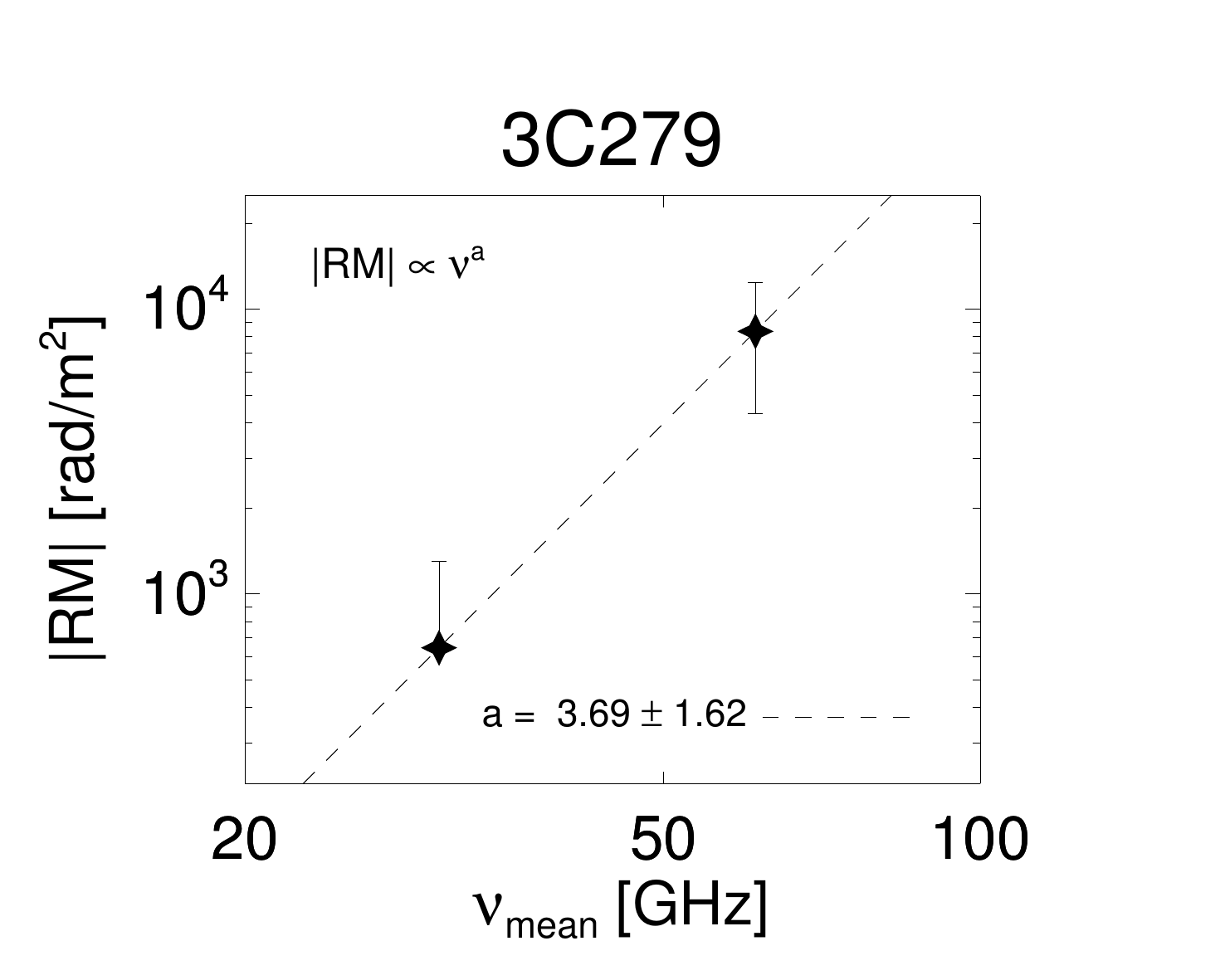}\\[-4mm] 
\includegraphics[trim=7mm 0mm 6mm 0mm, clip, width = 75mm]{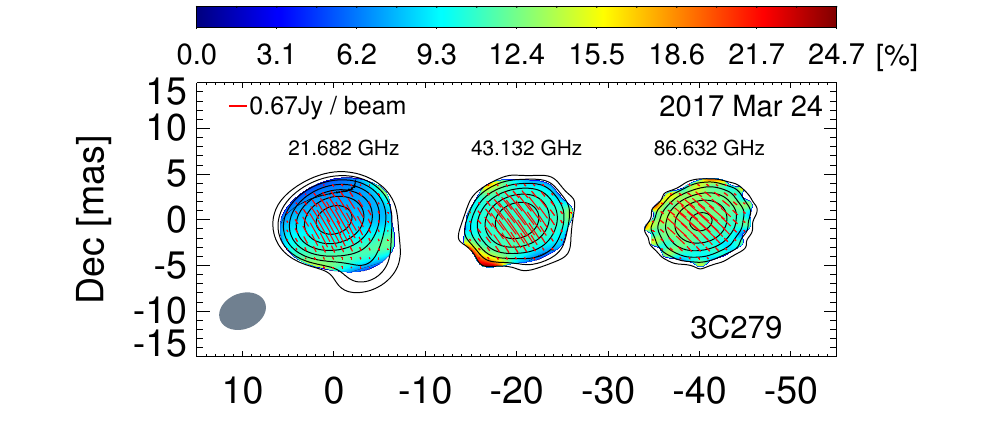}
\includegraphics[trim=7mm 4mm 5mm 1mm, clip, width = 47mm]{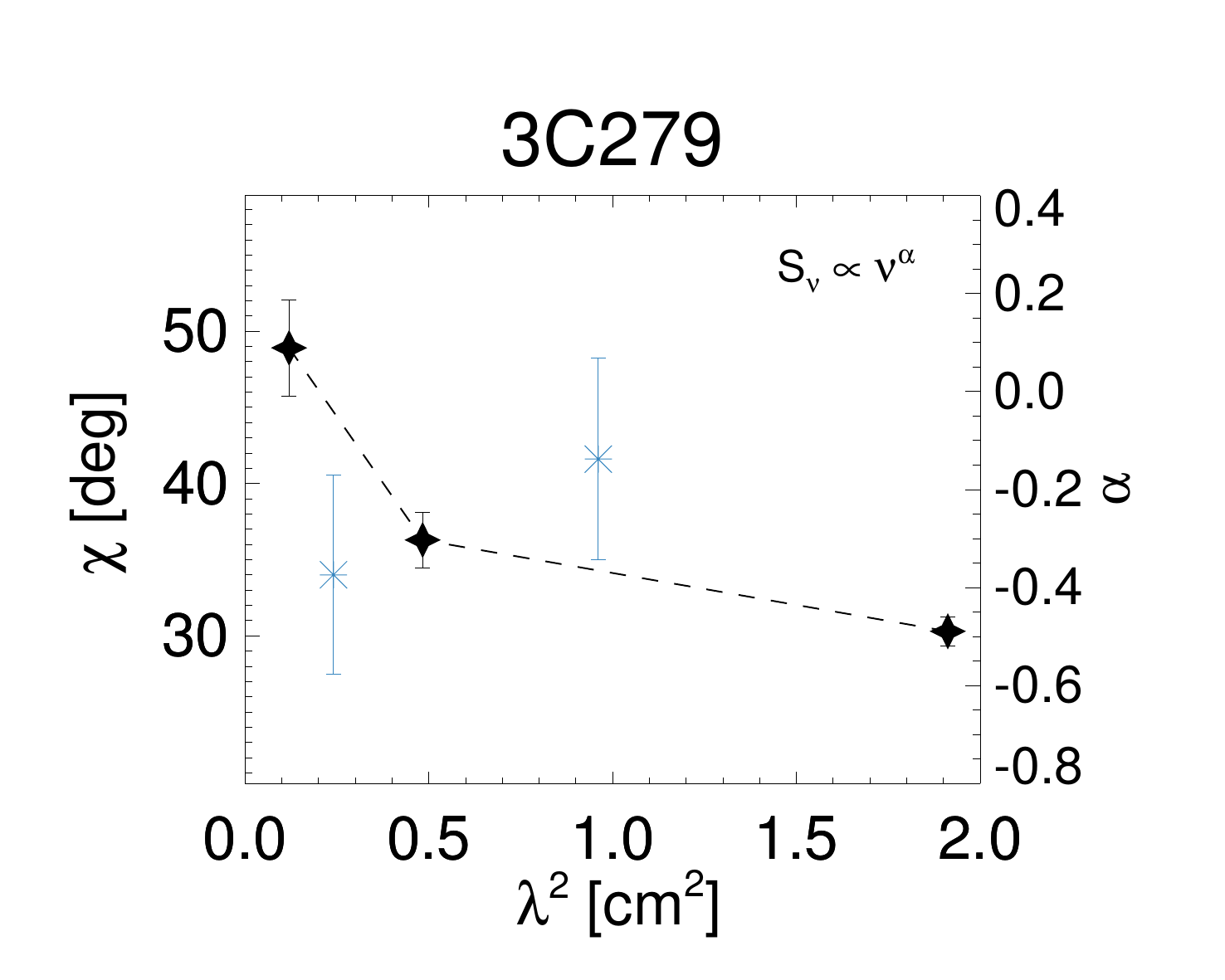}
\includegraphics[trim=5mm 4mm 24mm 1mm, clip, width = 43mm]{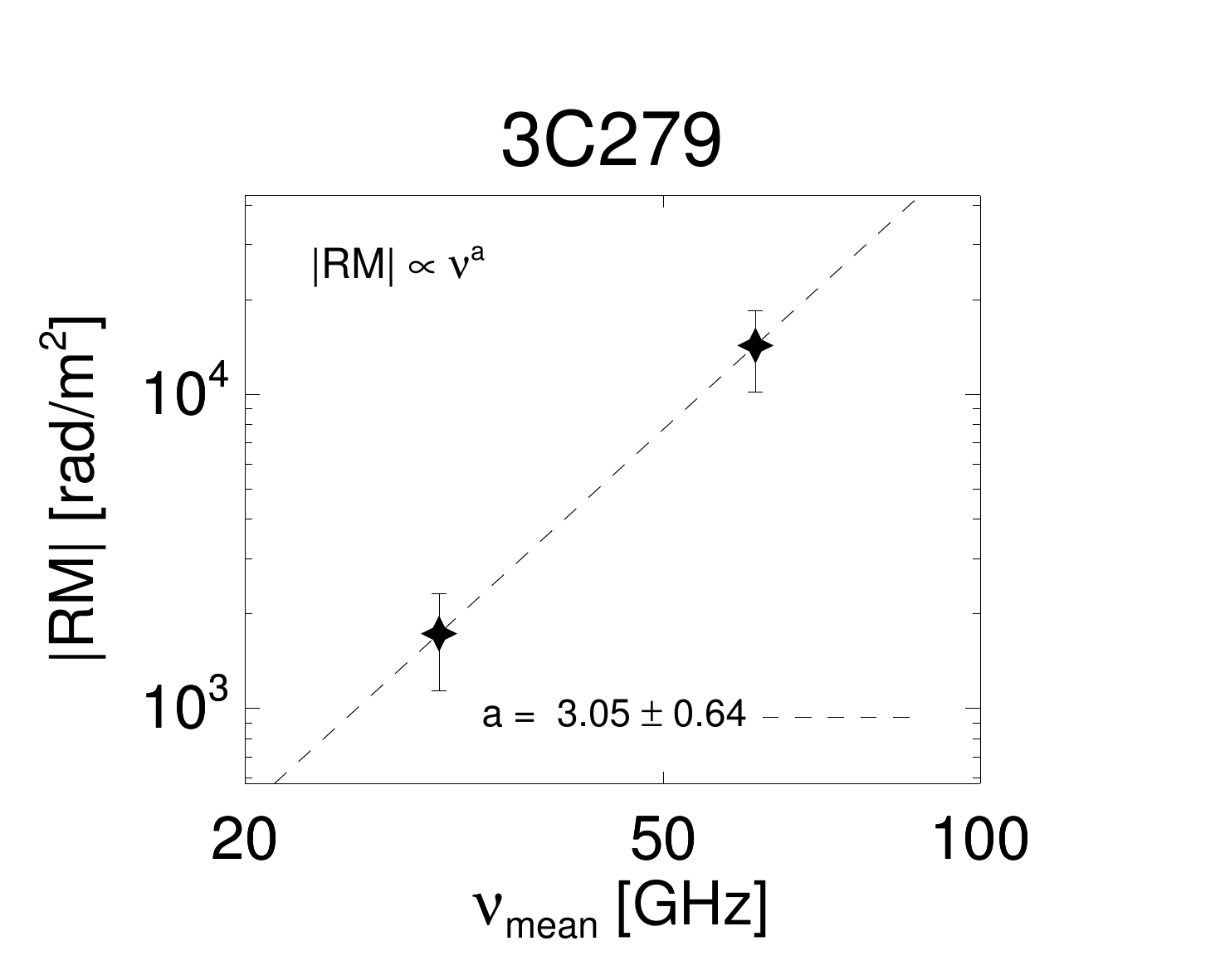}\\[-4mm] 
\includegraphics[trim=7mm 0mm 6mm 0mm, clip, width = 75mm]{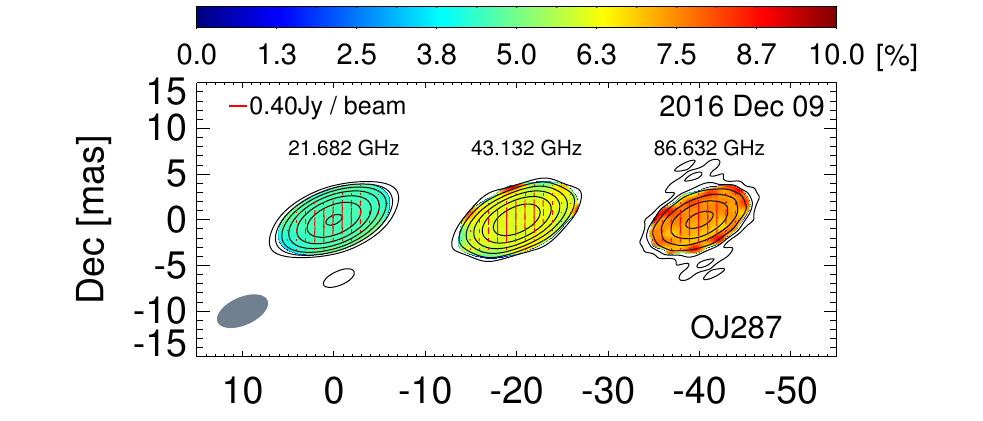}
\includegraphics[trim=7mm 4mm 5mm 1mm, clip, width = 47mm]{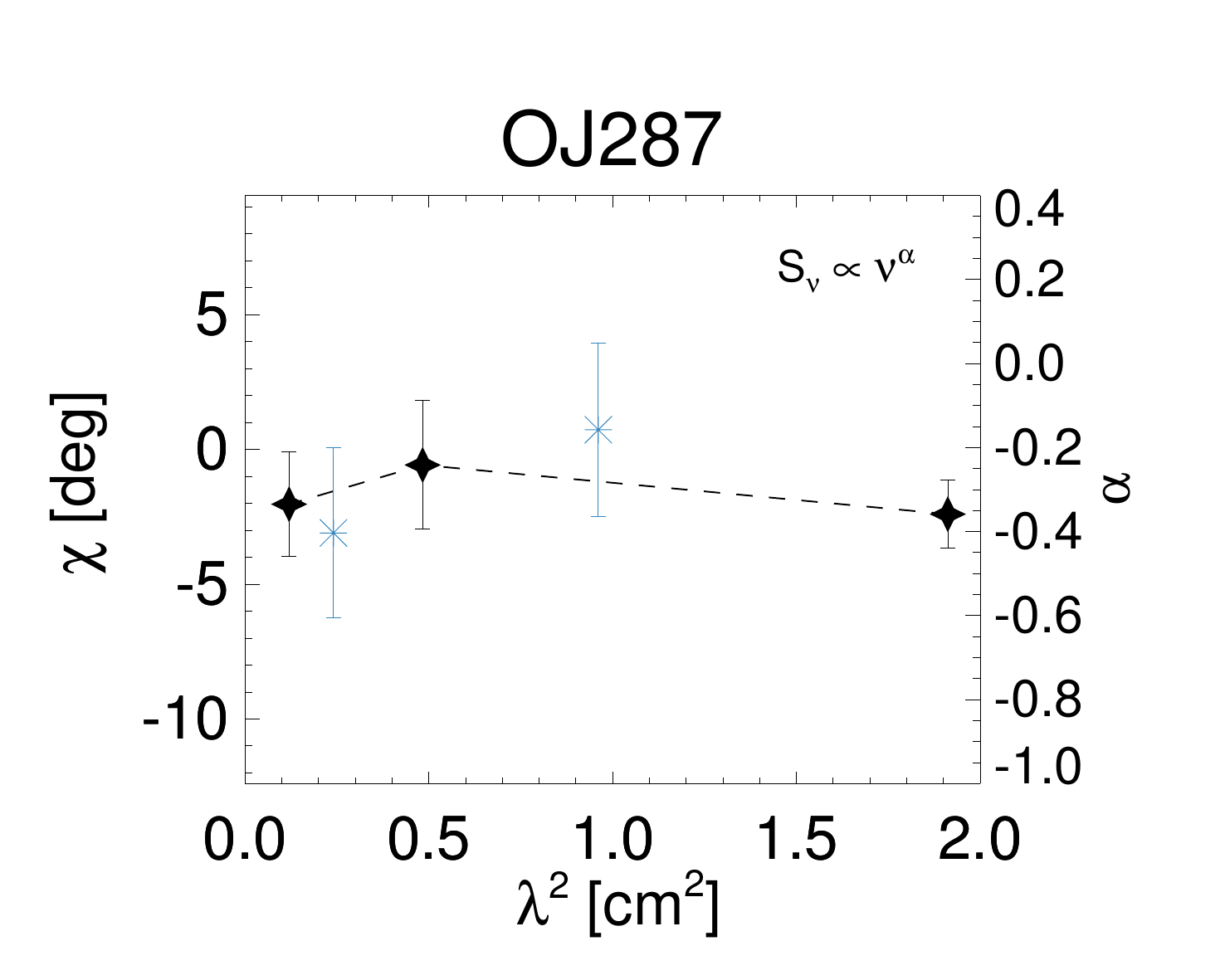}
\includegraphics[trim=5mm 4mm 24mm 1mm, clip, width = 43mm]{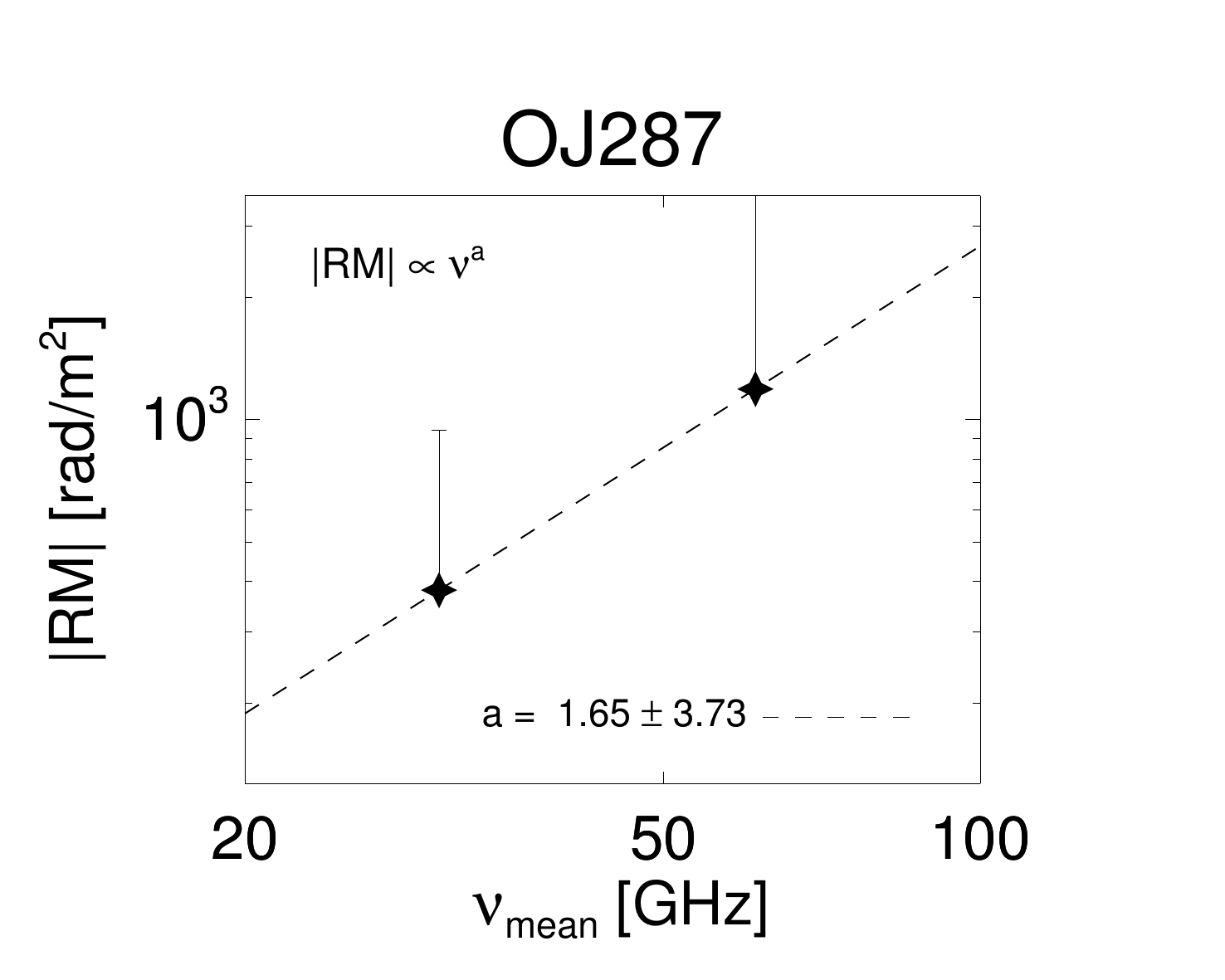}\\[-4mm] 
\includegraphics[trim=7mm 0mm 6mm 0mm, clip, width = 75mm]{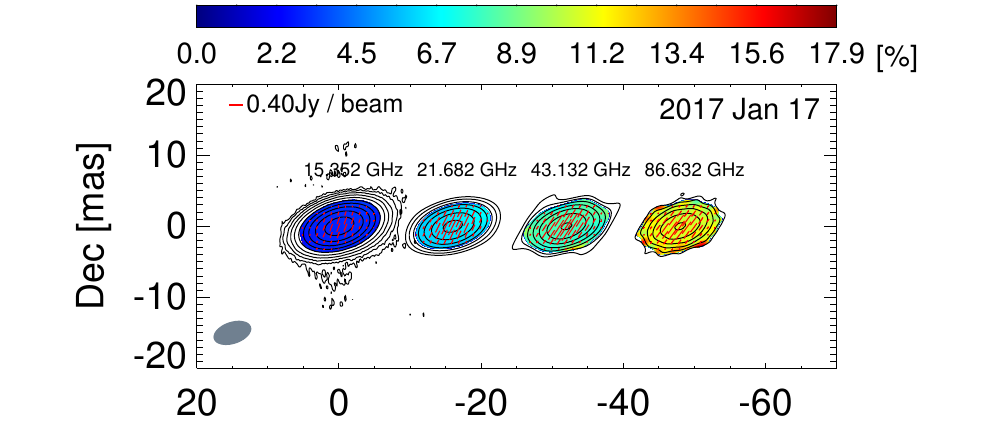}
\includegraphics[trim=7mm 4mm 5mm 1mm, clip, width = 47mm]{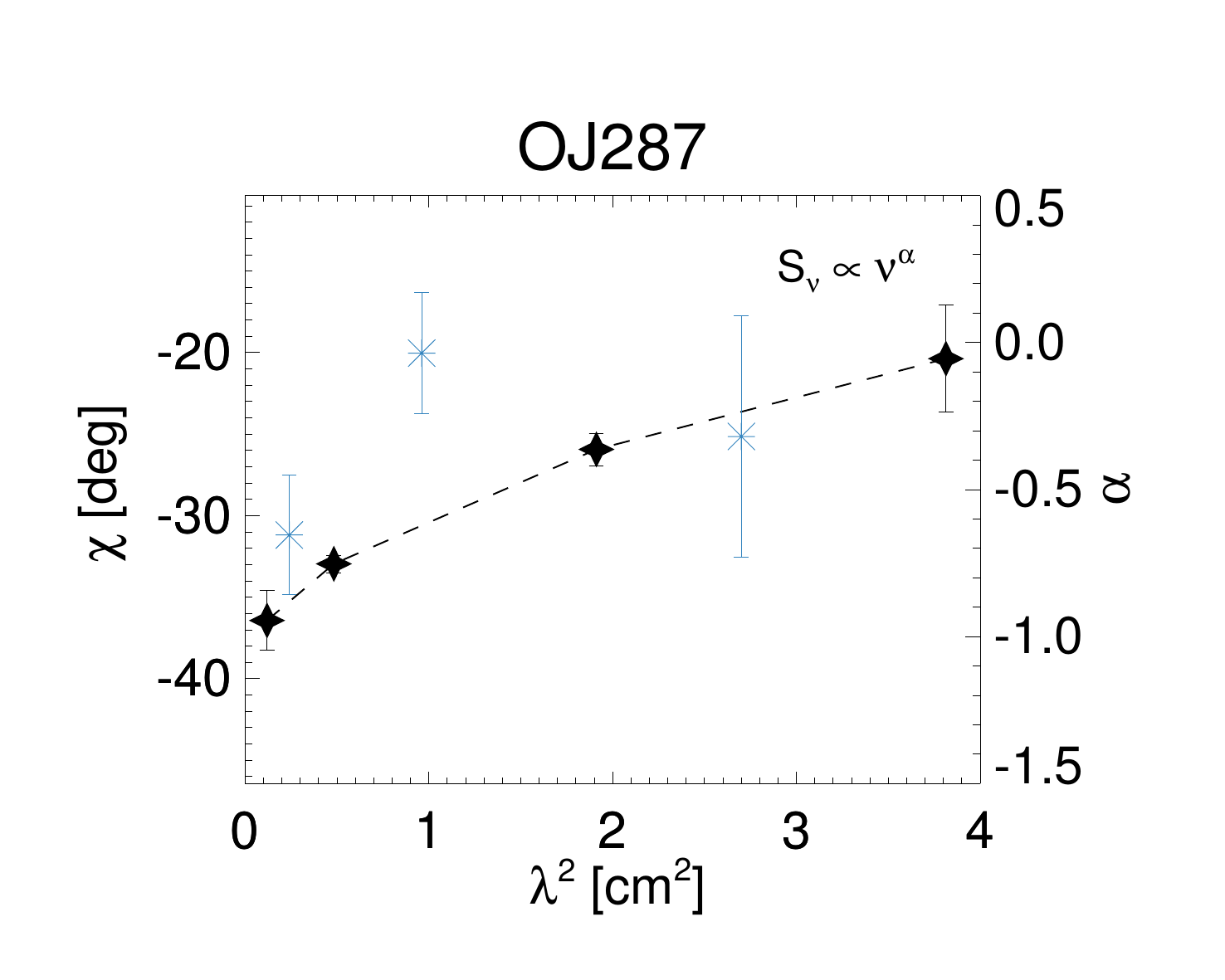}
\includegraphics[trim=5mm 4mm 24mm 1mm, clip, width = 43mm]{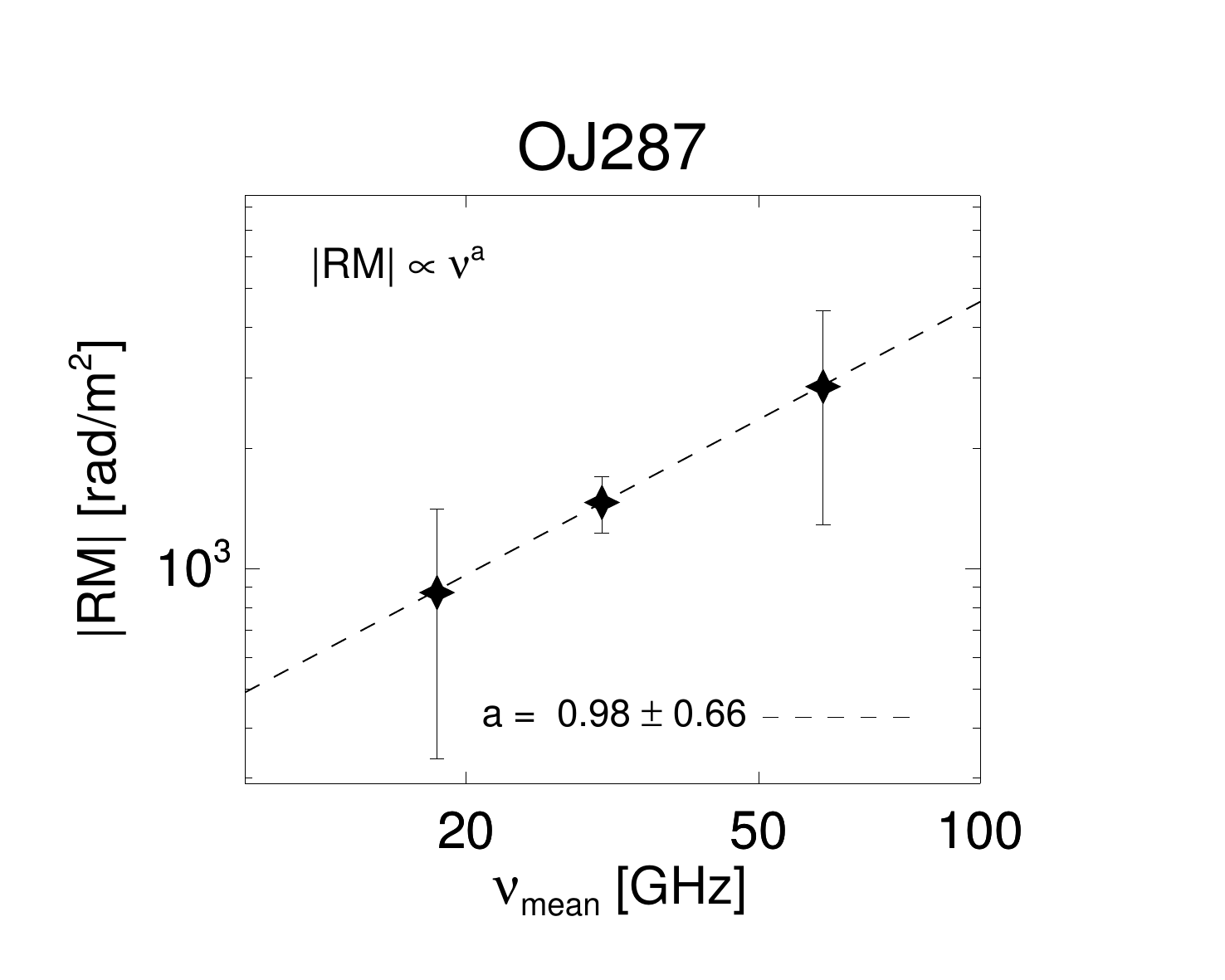}\\[-4mm] 
\includegraphics[trim=7mm 0mm 6mm 0mm, clip, width = 75mm]{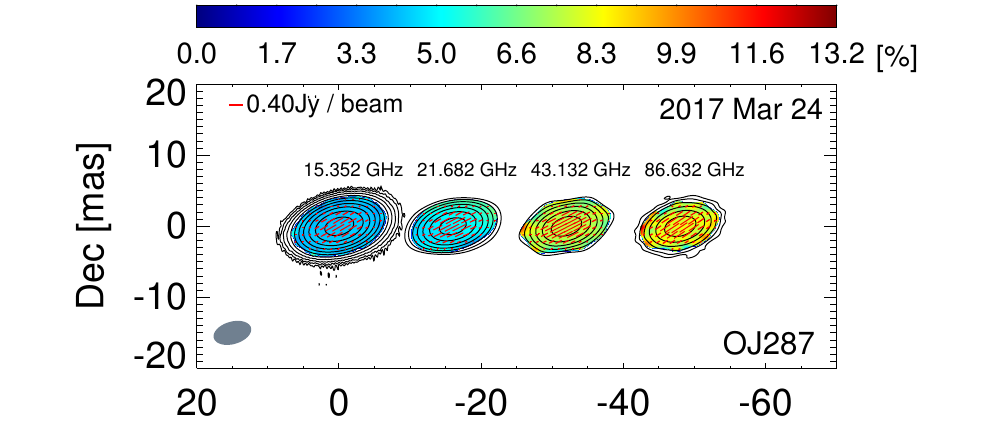}
\includegraphics[trim=7mm 4mm 5mm 1mm, clip, width = 47mm]{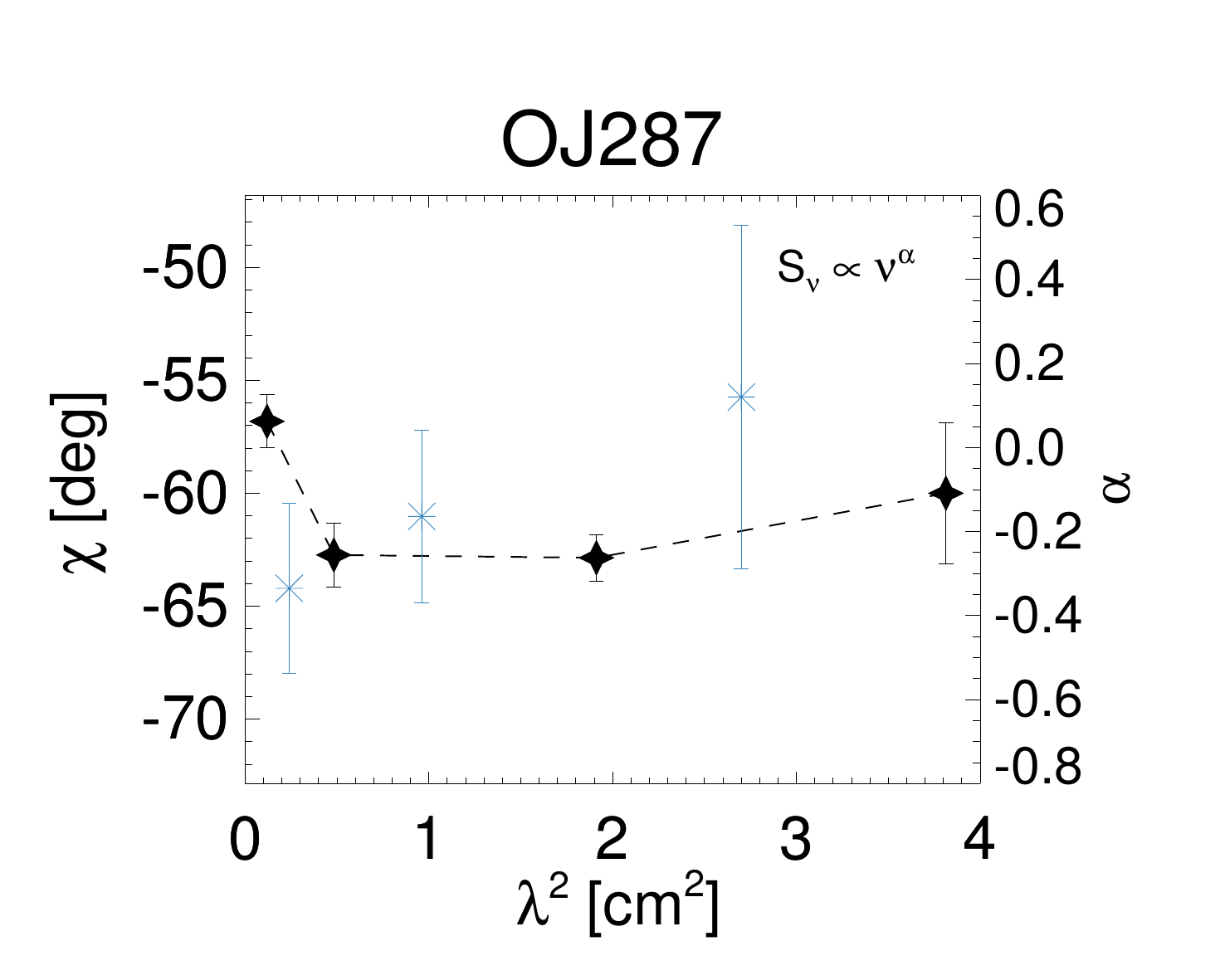}
\includegraphics[trim=5mm 4mm 24mm 1mm, clip, width = 43mm]{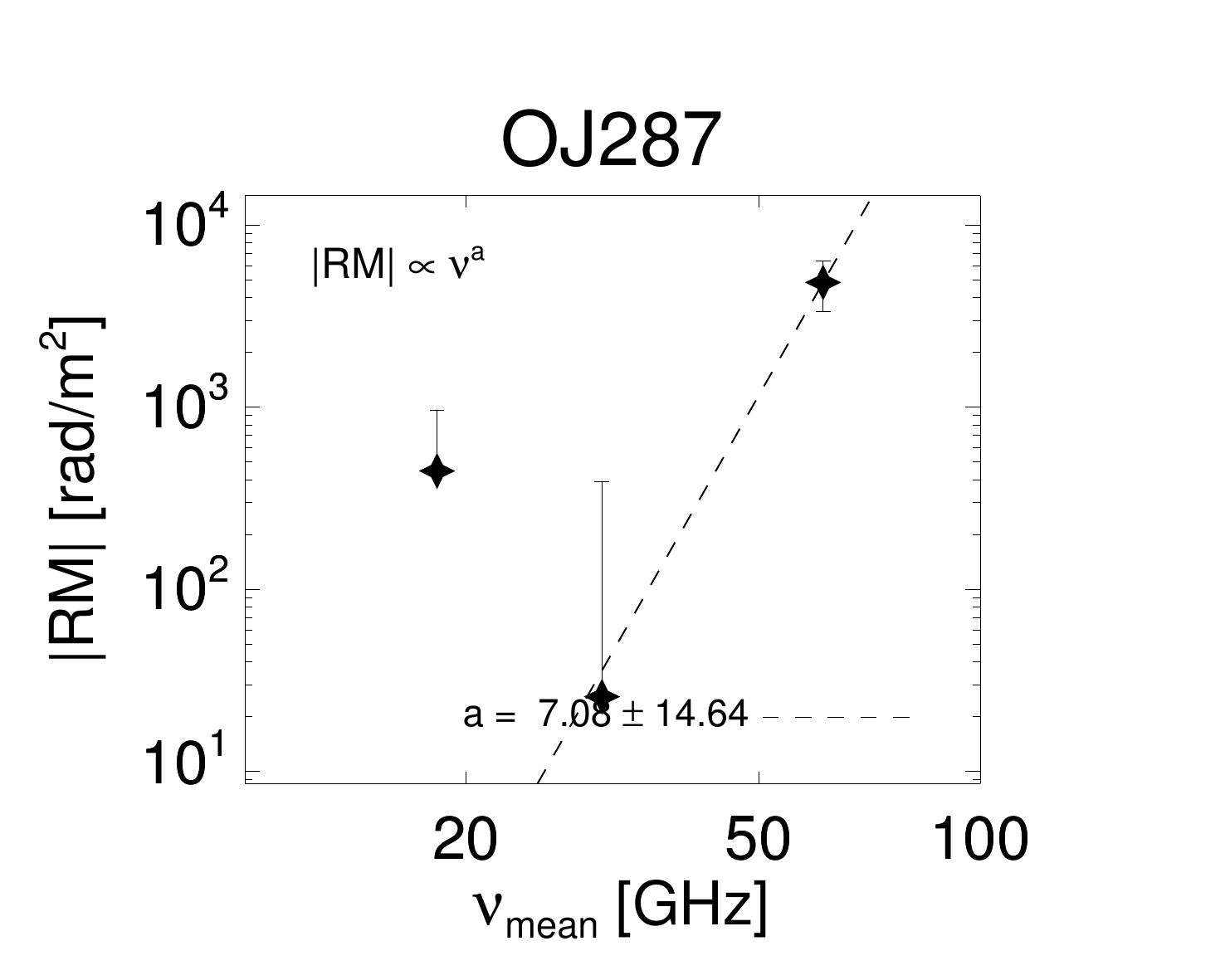}
\caption{\emph{Left}: Contours show naturally weighted CLEAN maps. The lowest contour level starts at 3--5 times the rms noise in the maps. Colors show the degree of linear polarization in units of \%. Red ticks show distribution of EVPAs with their lengths proportional to polarized intensity {\color{red}(not degree of polarization)}. The maps at different observing frequencies are shifted by 20 mas along the x-axis. The gray shaded ellipses show the beam size {\color{red}of the KVN at 22 GHz}. All maps from different frequencies are convolved with this beam. {\color{red}For OJ 287, we include contemporaneous MOJAVE data at 15 GHz in our analysis (see Section~\ref{oj287}).} Observation dates are noted in the upper right corner of each panel. \emph{Center}: {\color{red}EVPA (black diamonds, values on the left axis) and spectral index (blue asterisks, values on the right axis) at the core as function of $\lambda^2$ (geometrical mean $\lambda^2$ for spectral index)}. \emph{Right}: RM as function of geometric mean observing frequency. Each data point is obtained from two adjacent data points {\color{red}for EVPAs} in the corresponding central panel. The dashed lines are power-law functions fitted to the data points, the best-fit power-law indices $a$ are noted. All RM values are rest frame values. \label{result}}
\end{figure*}

\addtocounter{figure}{-1}

\begin{figure*}[p]
\begin{center}
\includegraphics[trim=7mm 0mm 6mm 0mm, clip, width = 80mm]{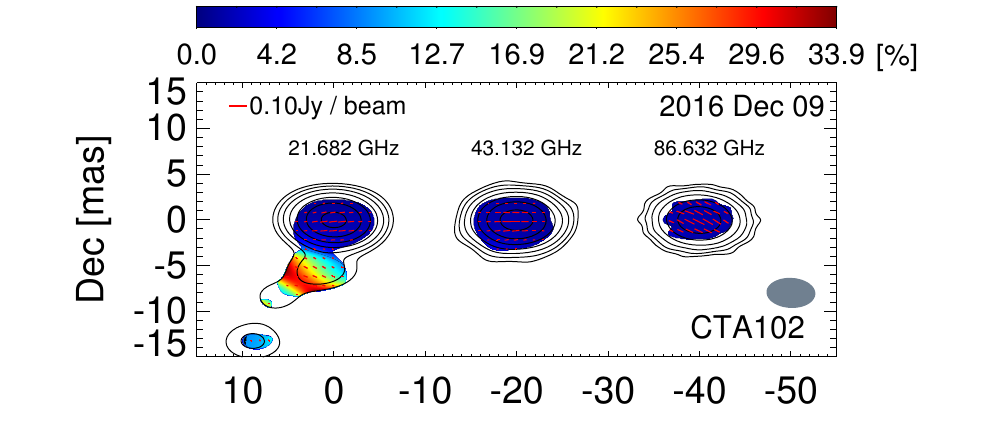}
\includegraphics[trim=3mm 4mm 10mm 13mm, clip, width = 52mm]{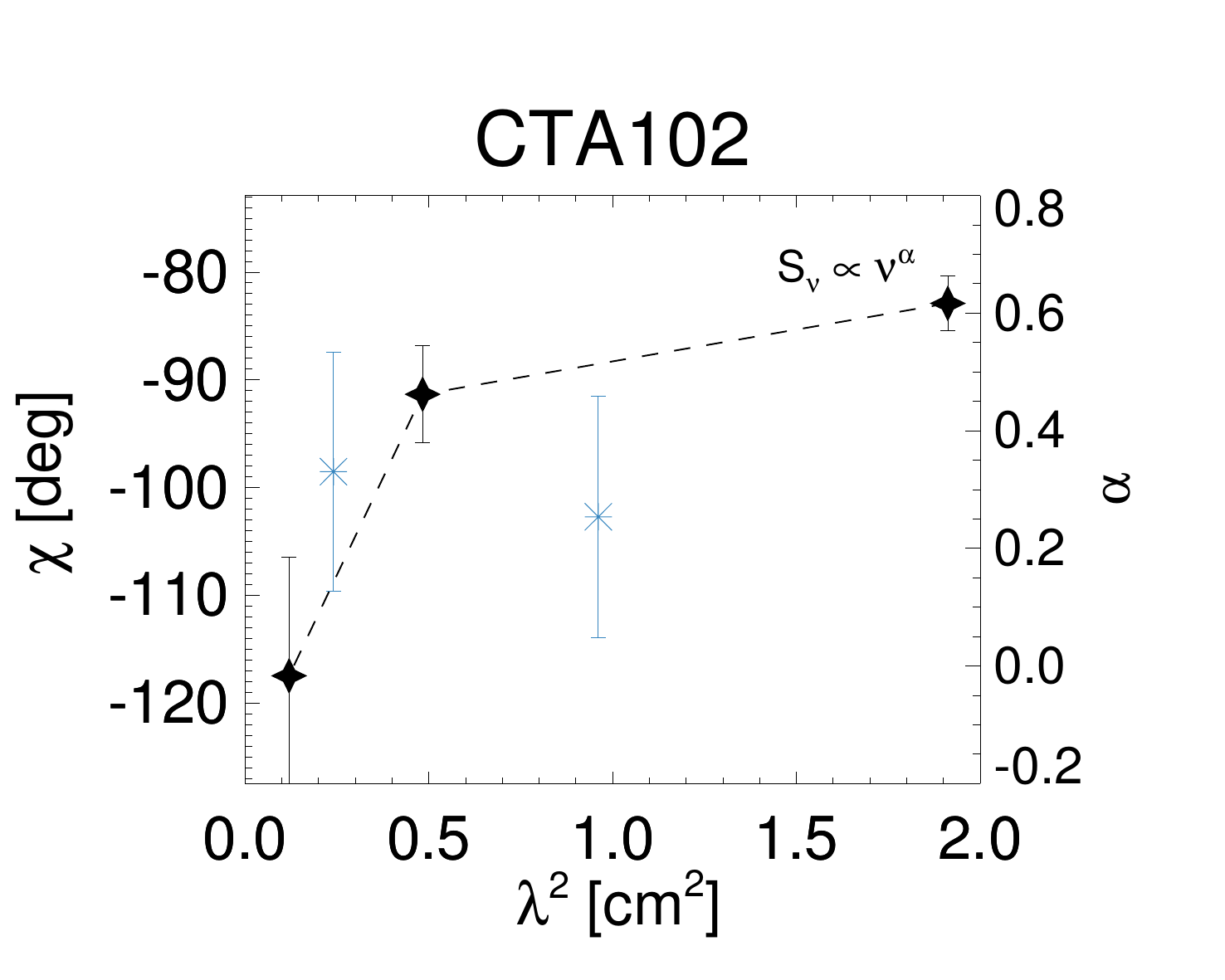}
\includegraphics[trim=5mm 4mm 24mm 13mm, clip, width = 46mm]{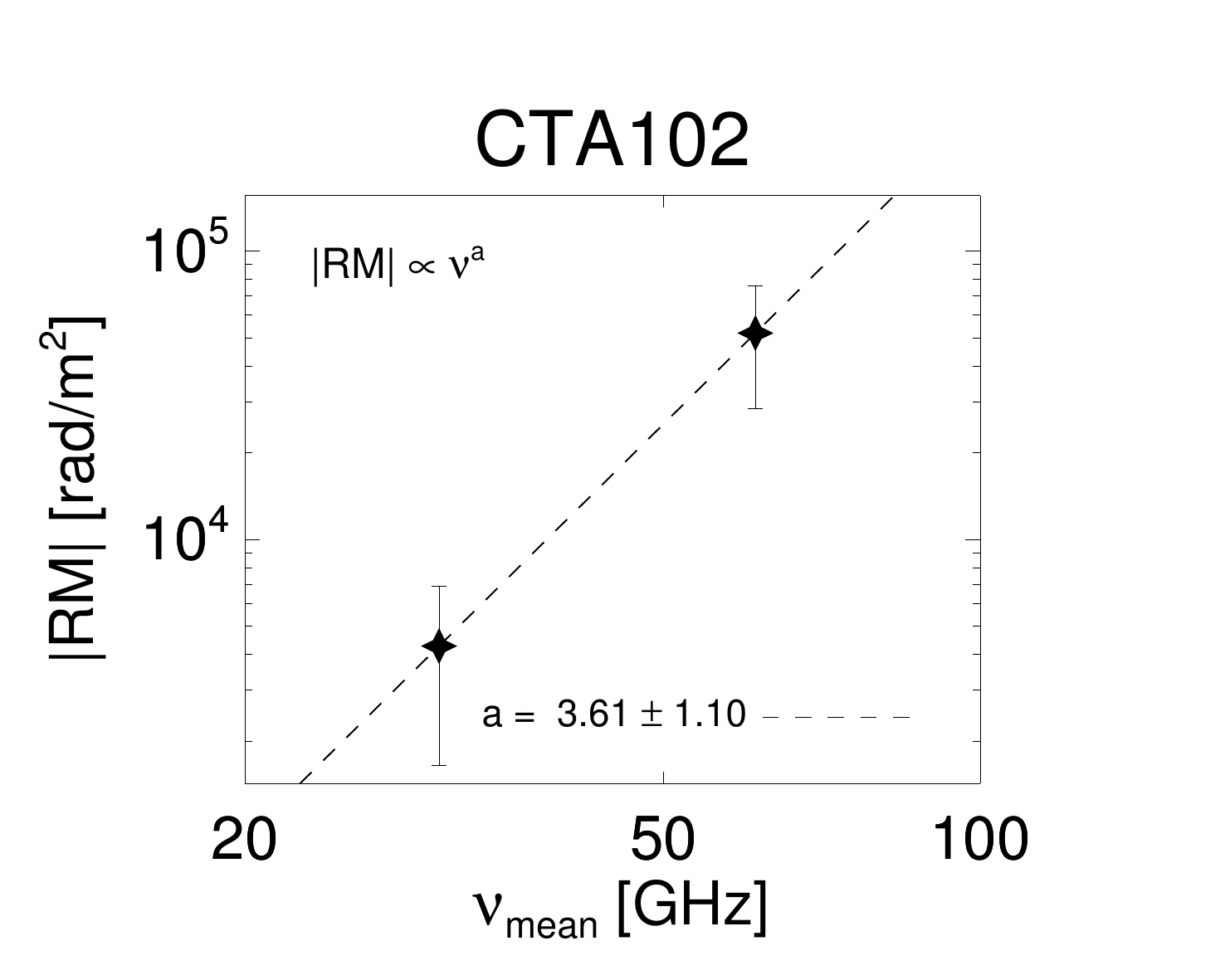}
\includegraphics[trim=7mm 0mm 6mm 0mm, clip, width = 80mm]{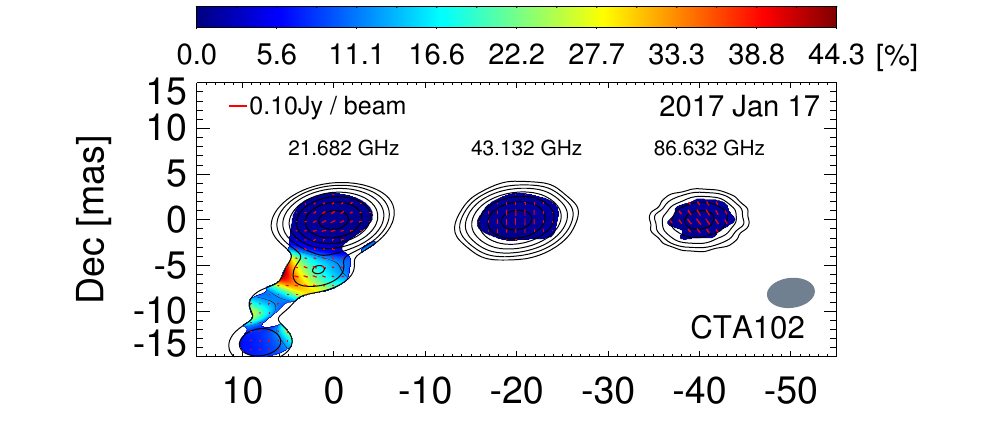}
\includegraphics[trim=7mm 4mm 10mm 13mm, clip, width = 51mm]{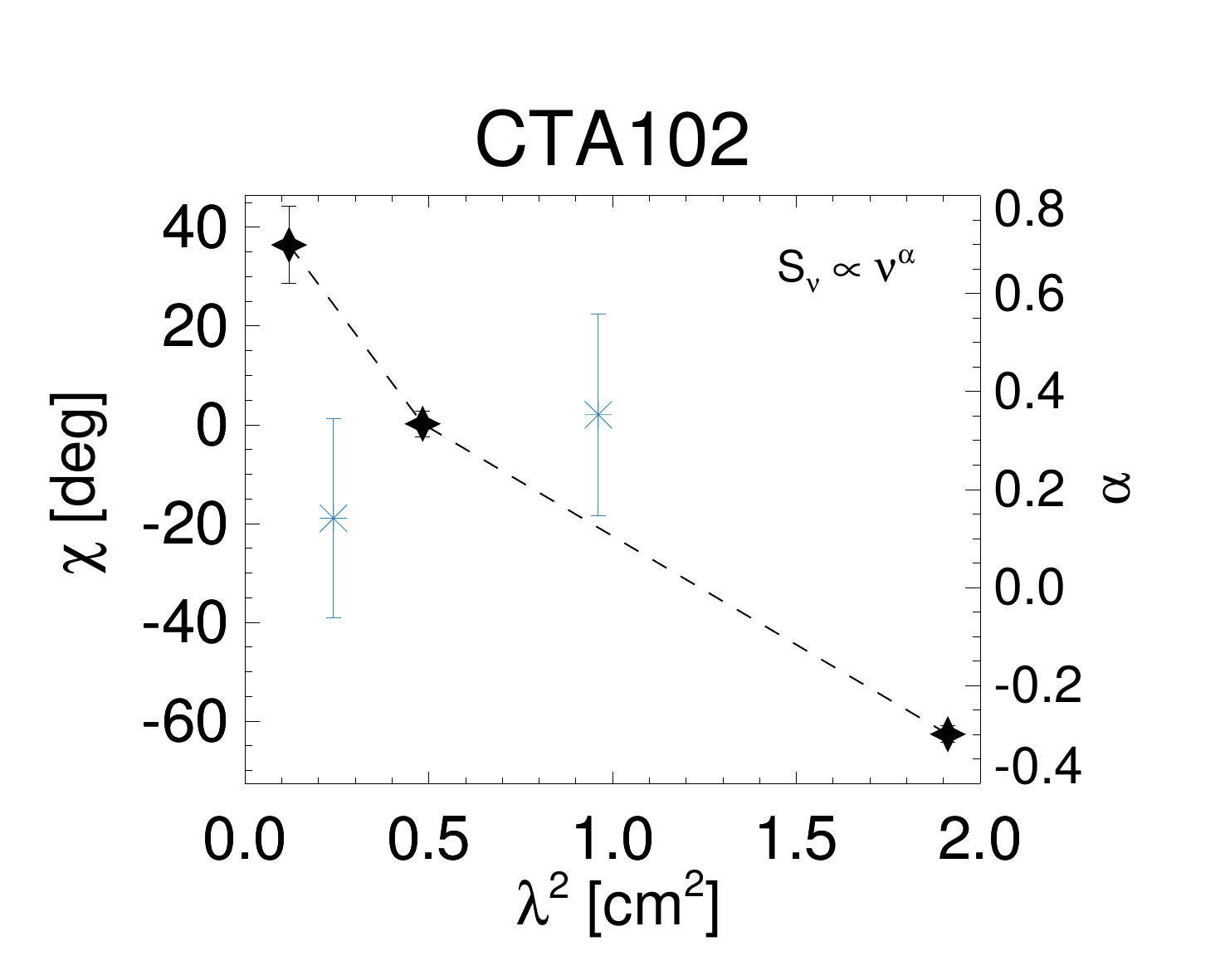}
\includegraphics[trim=5mm 4mm 24mm 13mm, clip, width = 47mm]{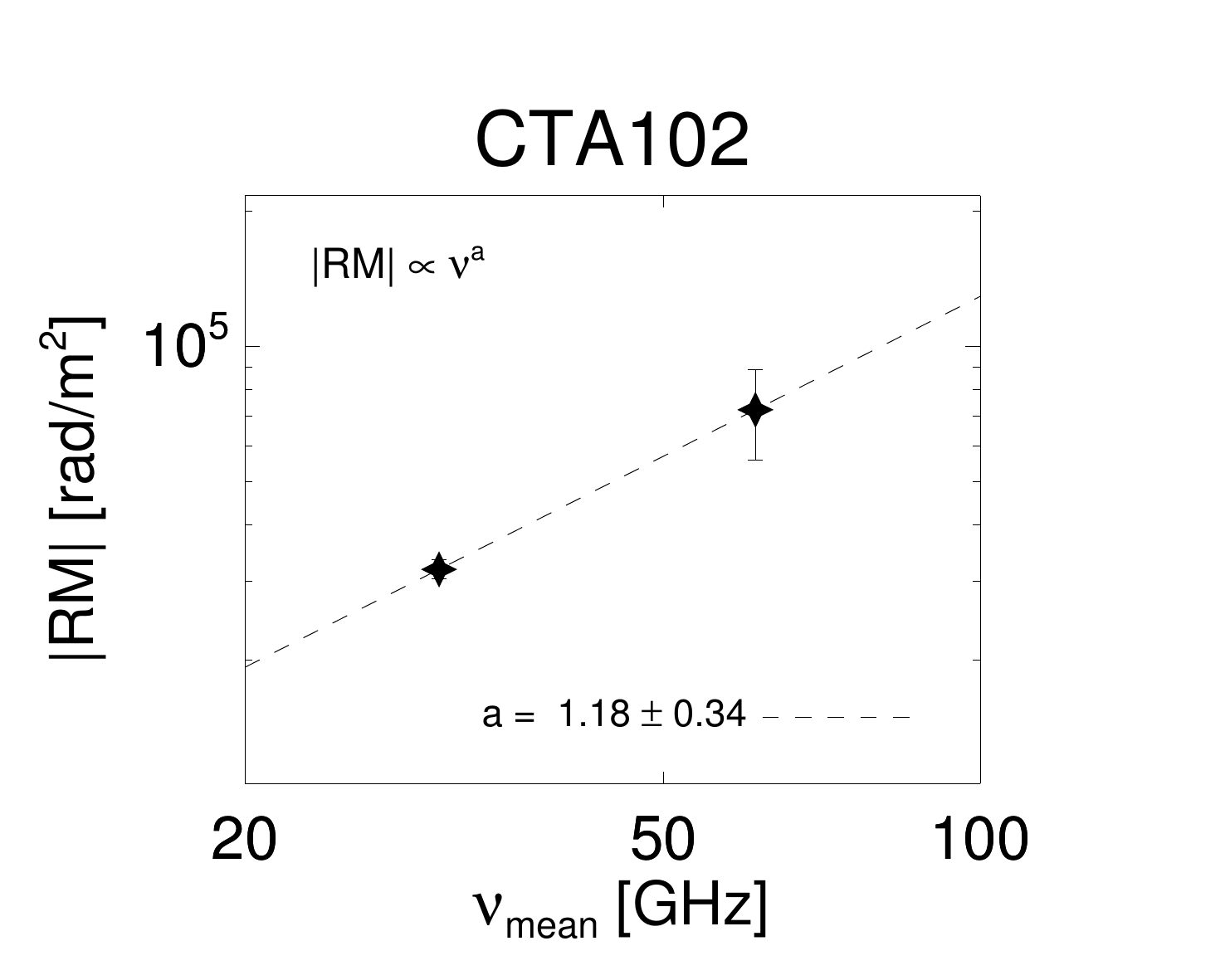}
\includegraphics[trim=7mm 0mm 6mm 0mm, clip, width = 80mm]{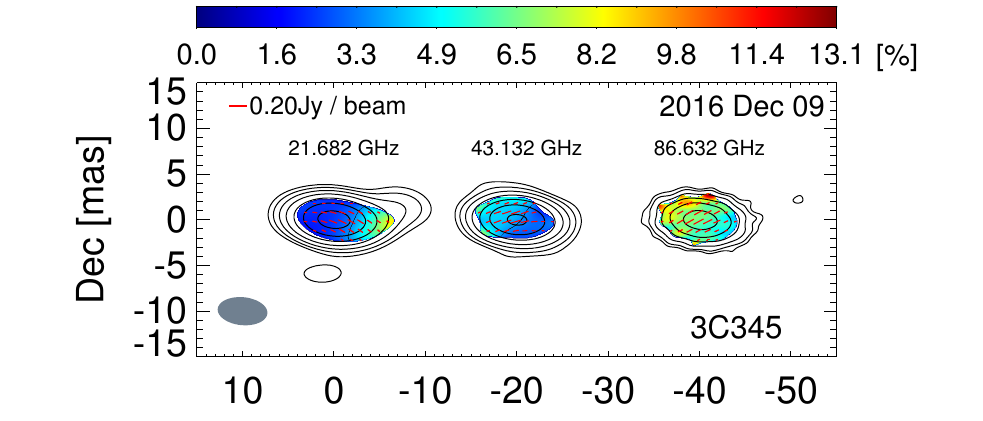}
\includegraphics[trim=3mm 4mm 10mm 13mm, clip, width = 52mm]{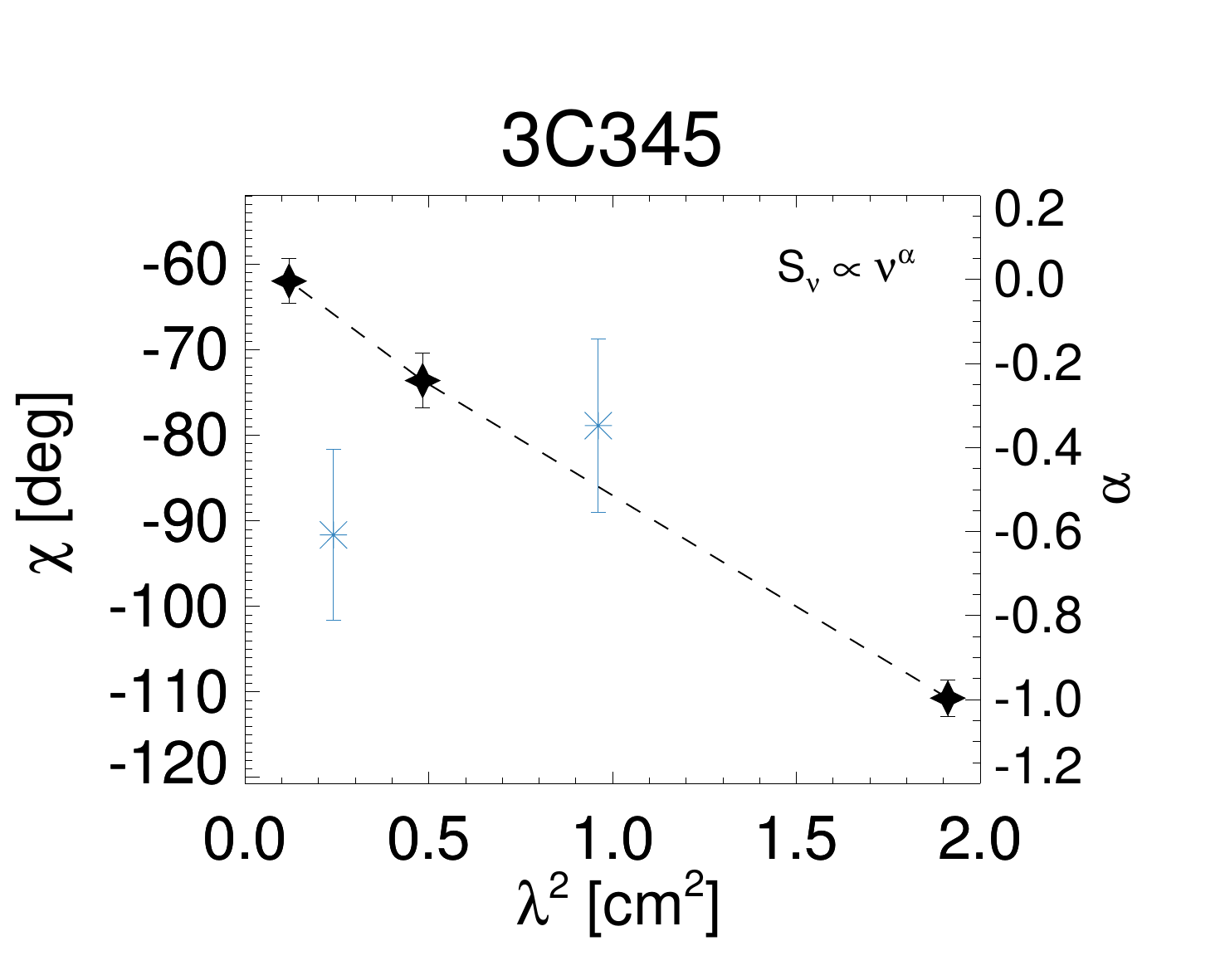}
\includegraphics[trim=5mm 4mm 24mm 13mm, clip, width = 46mm]{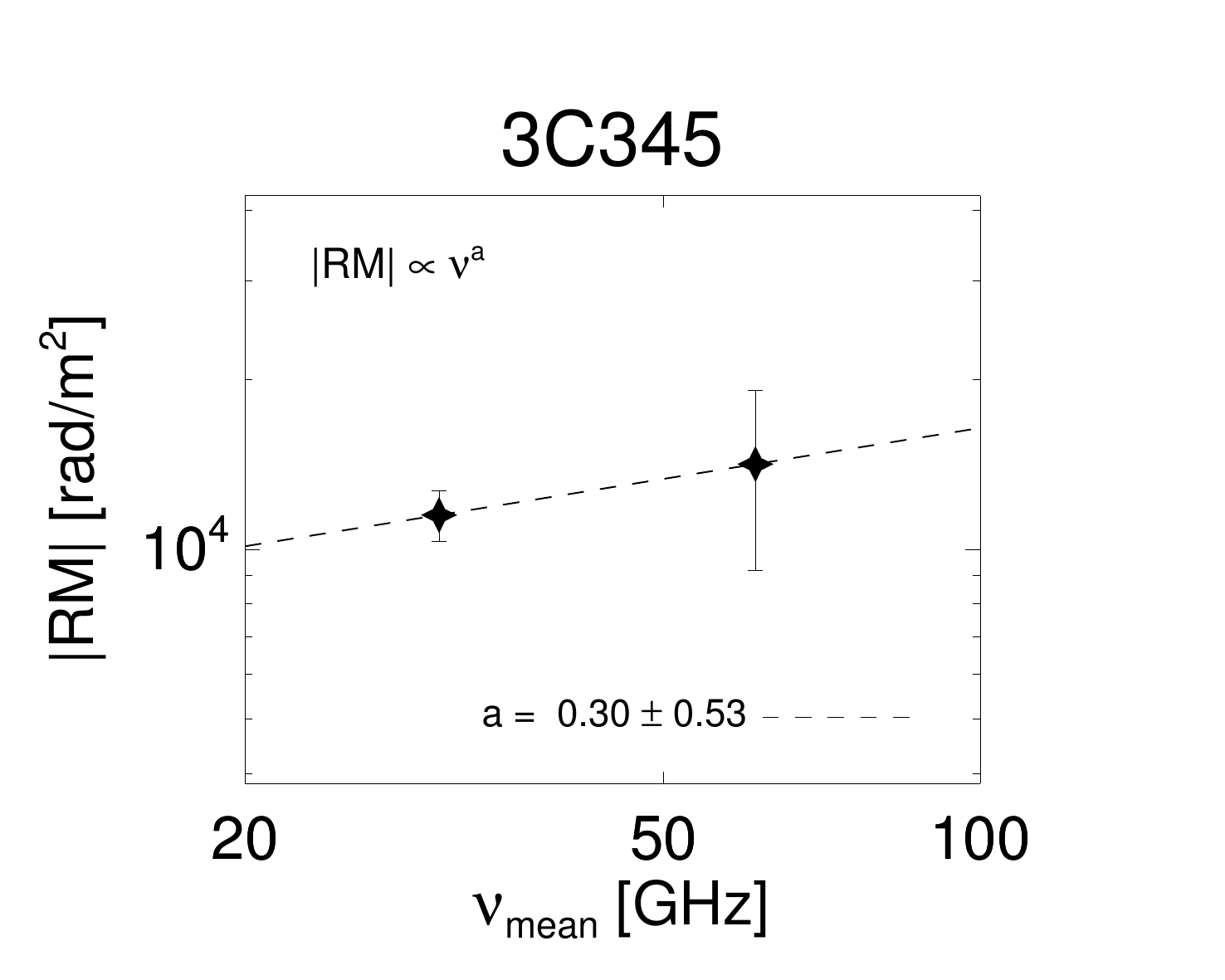}
\includegraphics[trim=7mm 0mm 6mm 0mm, clip, width = 80mm]{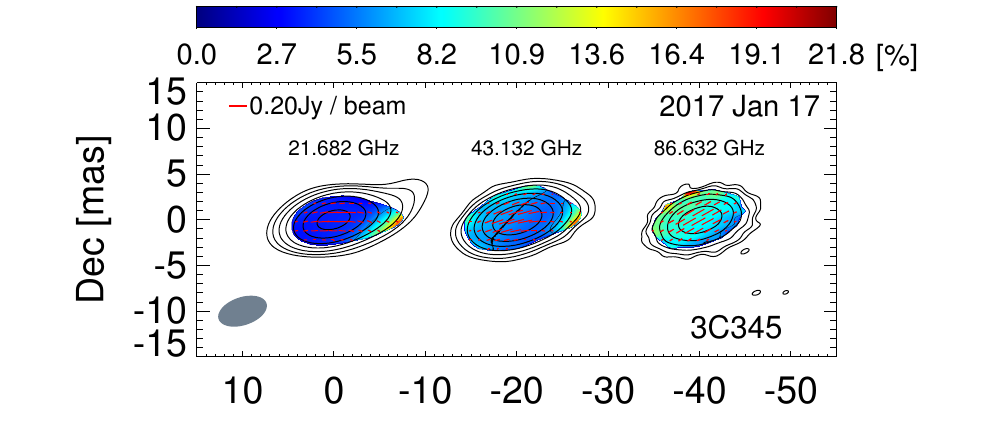}
\includegraphics[trim=7mm 4mm 10mm 13mm, clip, width = 51mm]{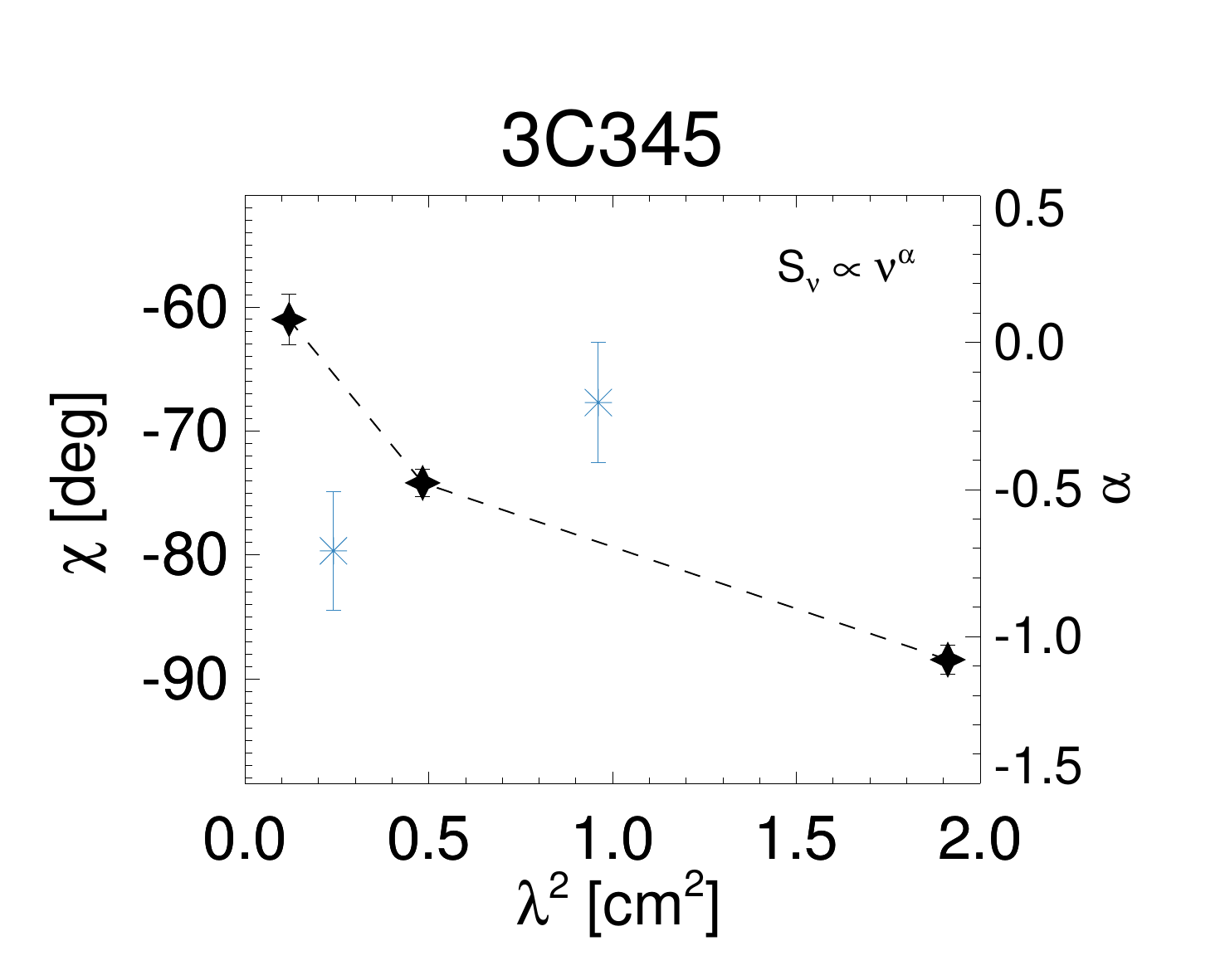}
\includegraphics[trim=5mm 4mm 24mm 13mm, clip, width = 47mm]{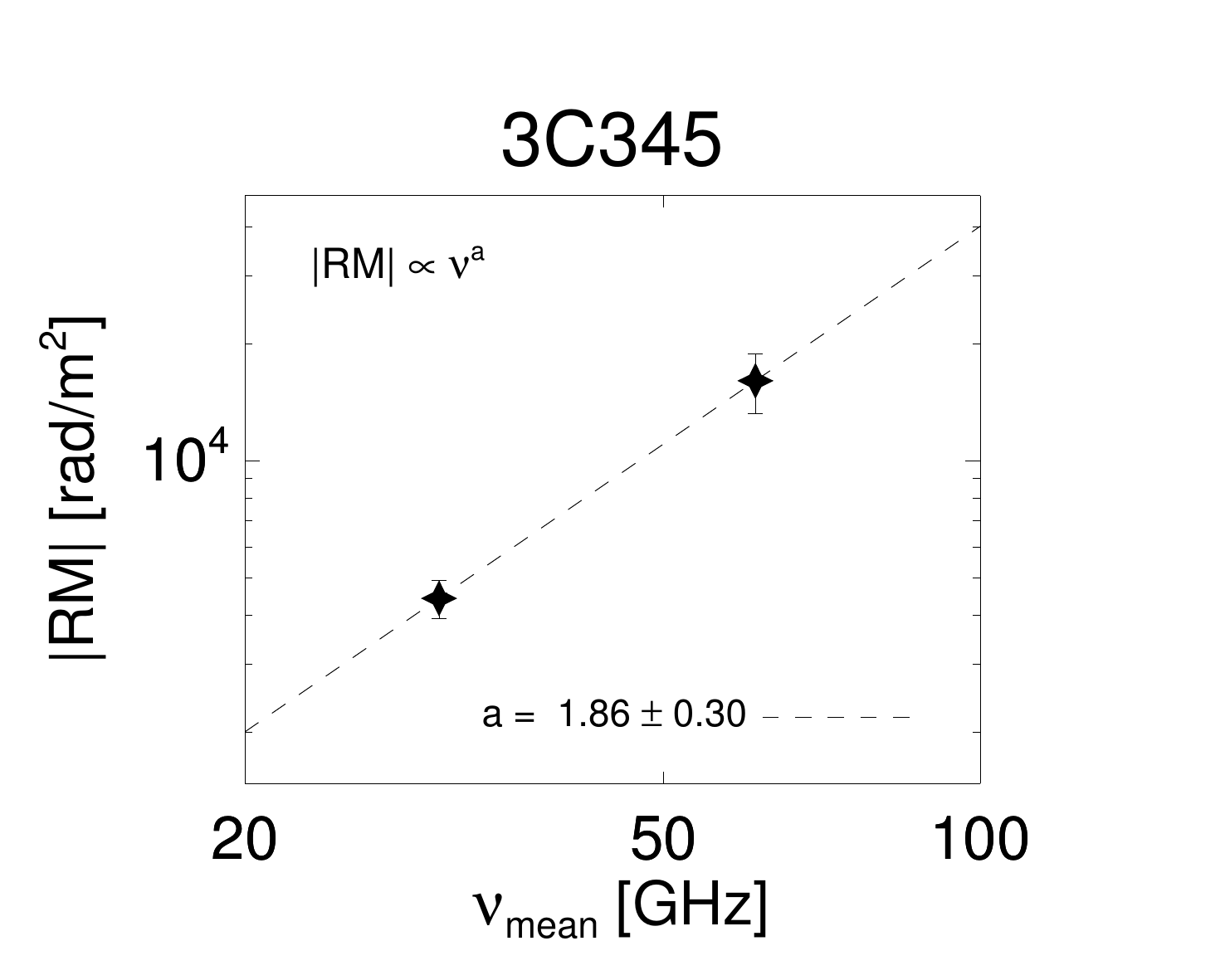}
\includegraphics[trim=7mm 0mm 6mm 0mm, clip, width = 80mm]{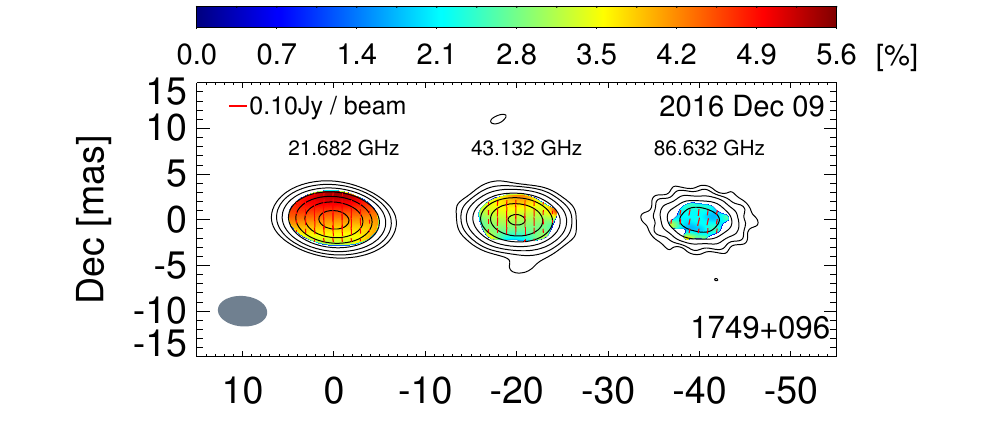}
\includegraphics[trim=7mm 4mm 10mm 13mm, clip, width = 51mm]{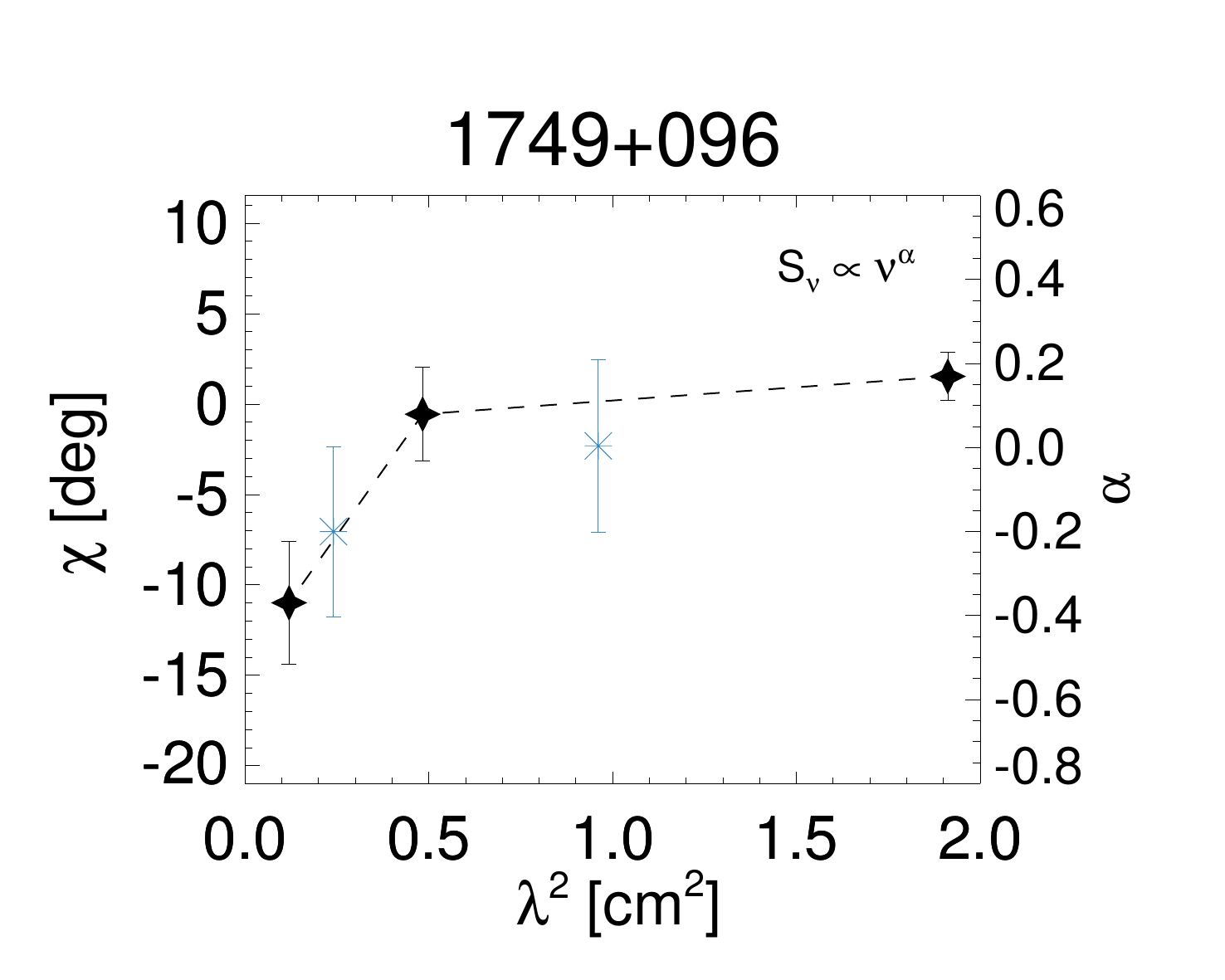}
\includegraphics[trim=5mm 4mm 24mm 13mm, clip, width = 47mm]{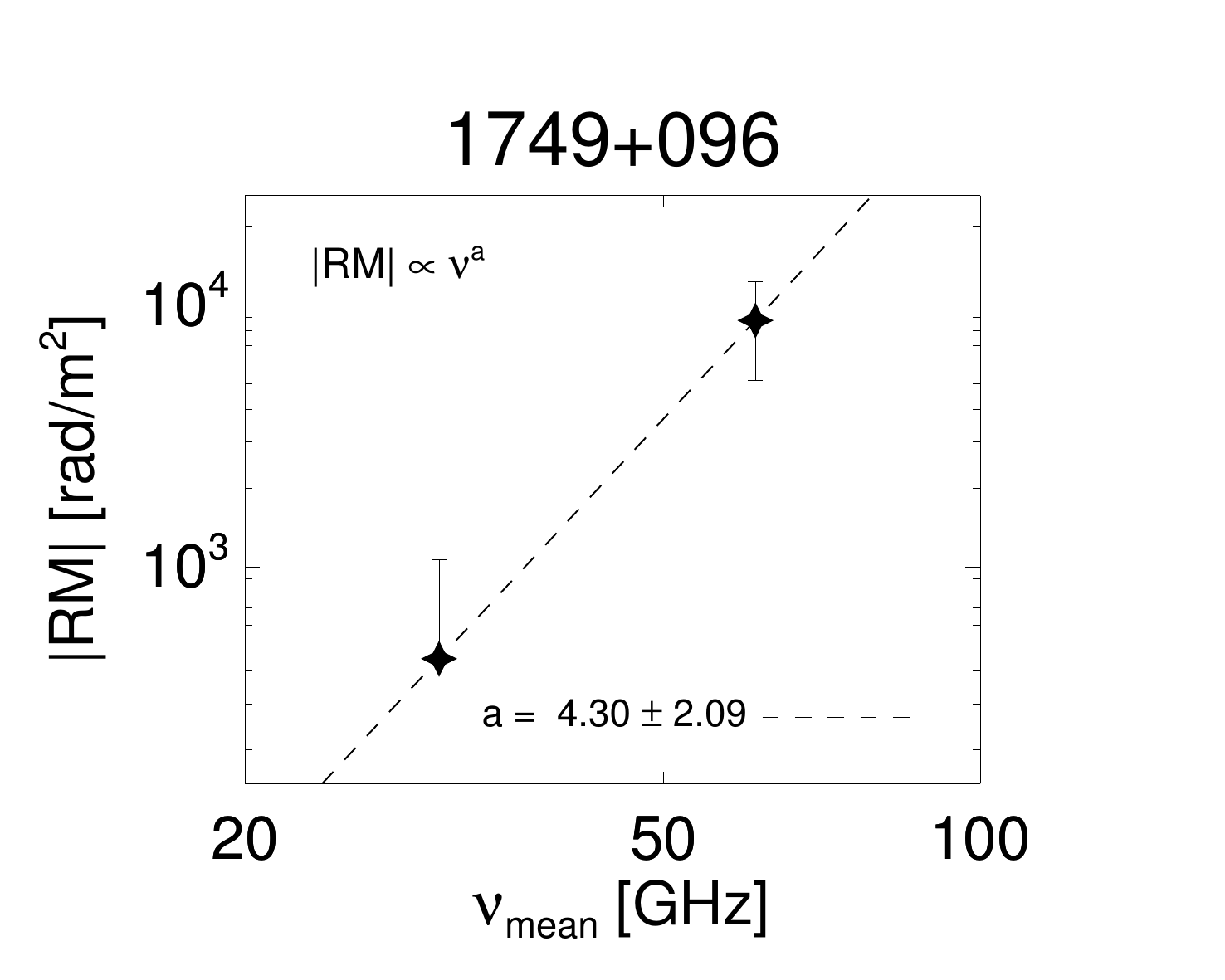}
\includegraphics[trim=7mm 0mm 6mm 0mm, clip, width = 80mm]{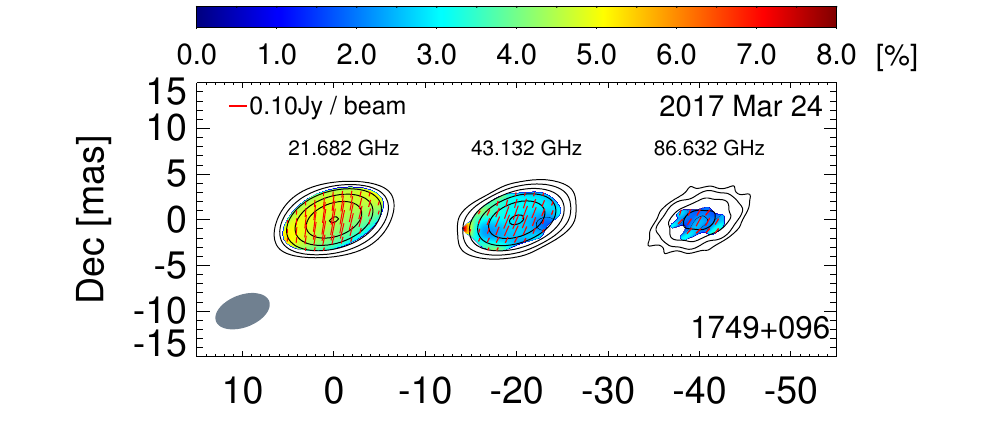}
\includegraphics[trim=7mm 4mm 10mm 13mm, clip, width = 51mm]{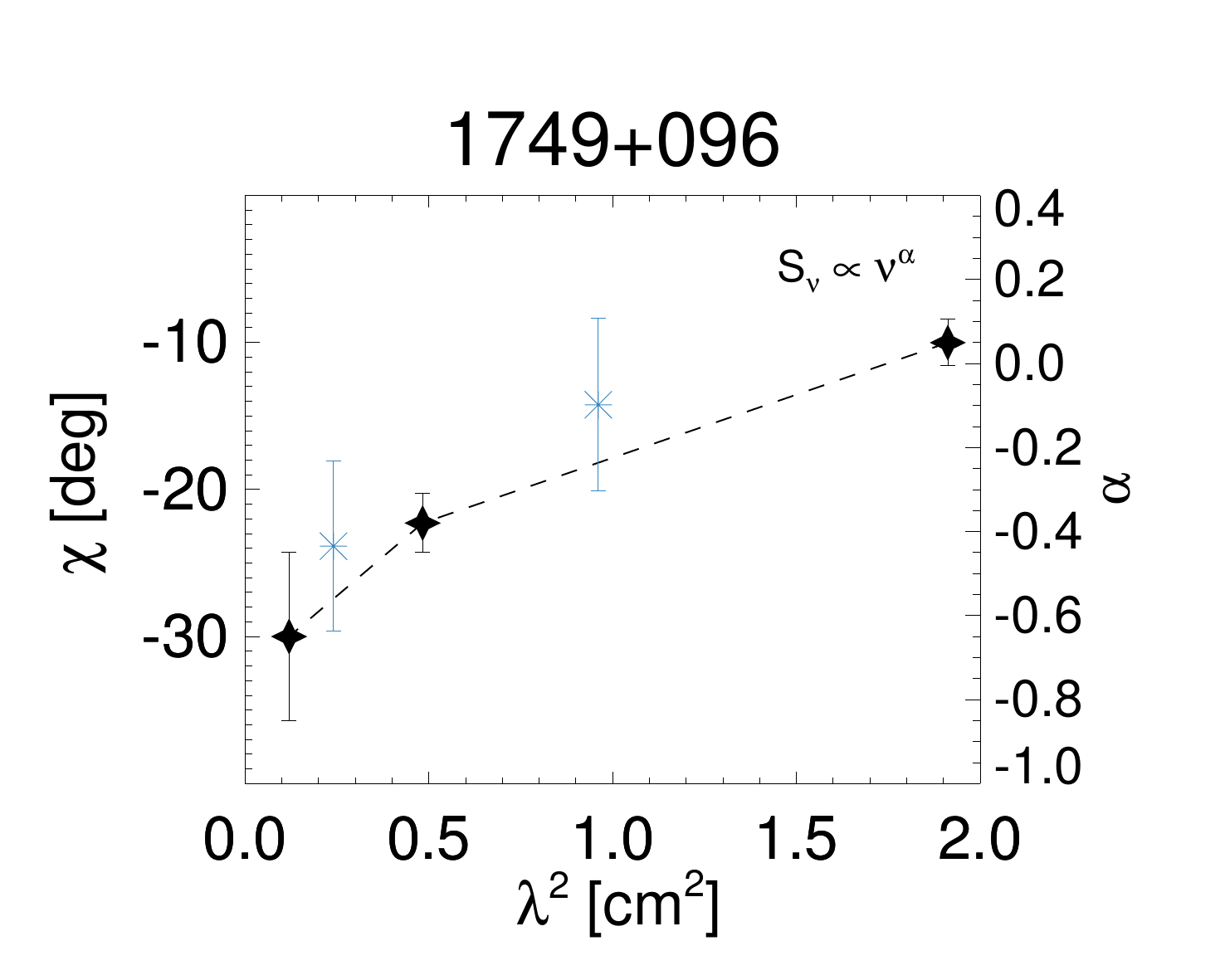}
\includegraphics[trim=5mm 4mm 24mm 13mm, clip, width = 47mm]{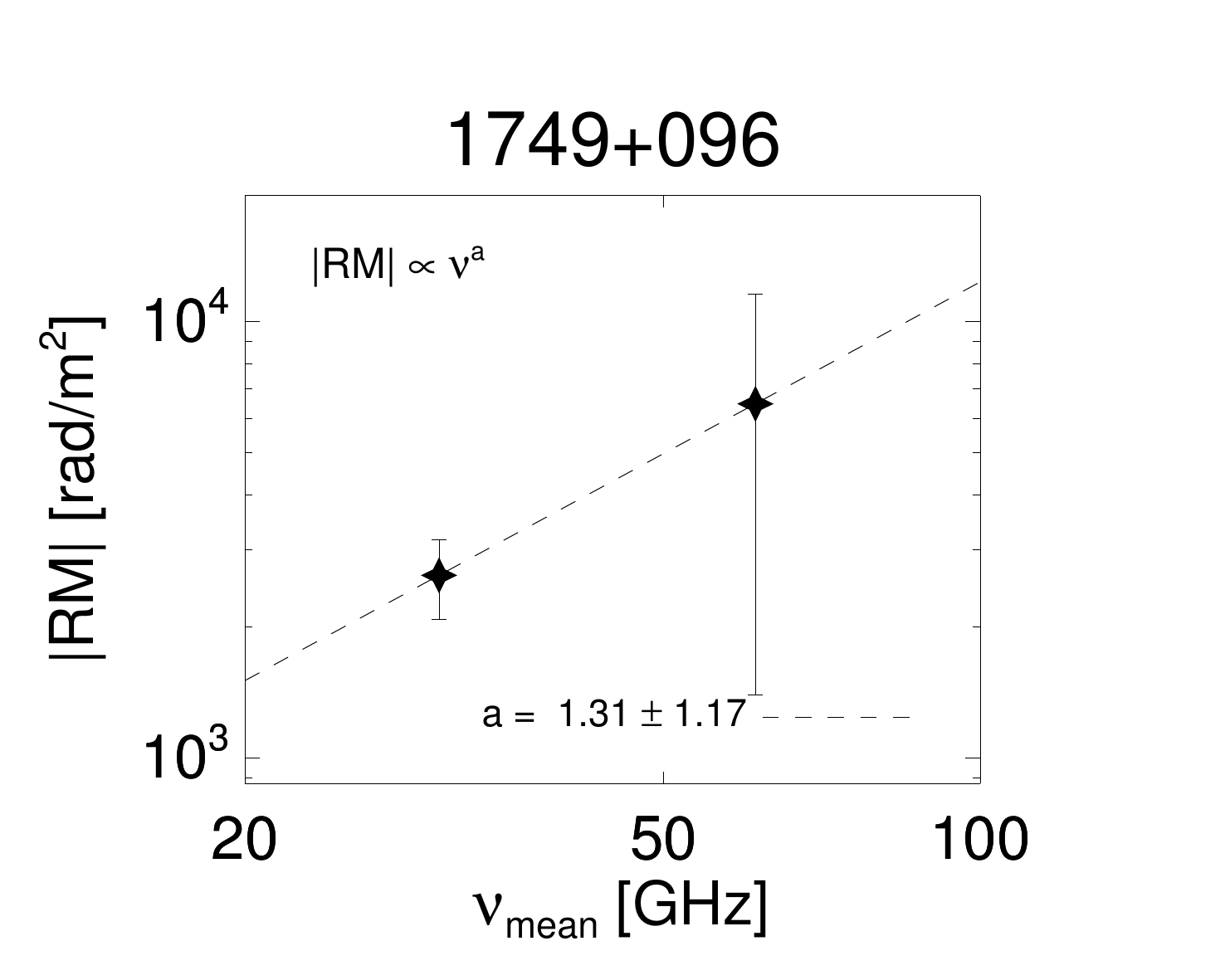}
\caption{\emph{Continued.}}
\end{center}
\end{figure*}

\addtocounter{figure}{-1}

\begin{figure*}[!t]
\begin{center}
\includegraphics[trim=7mm 0mm 6mm 0mm, clip, width = 80mm]{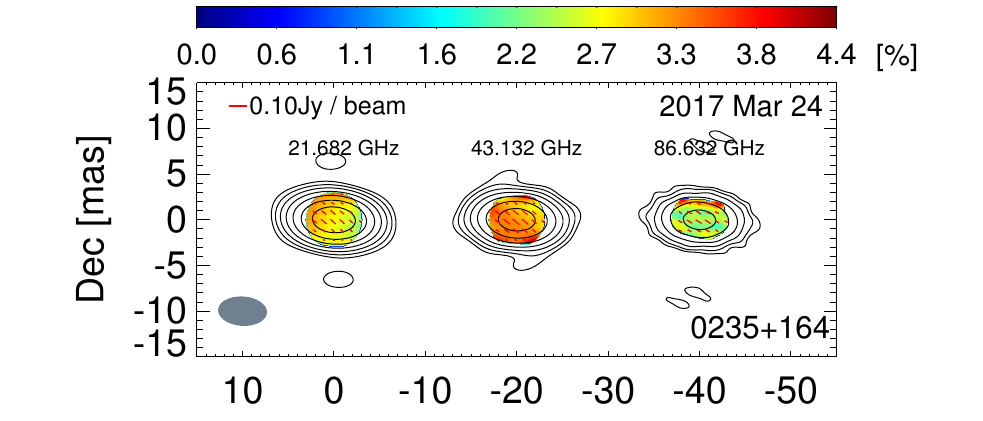}
\includegraphics[trim=7mm 4mm 10mm 13mm, clip, width = 51mm]{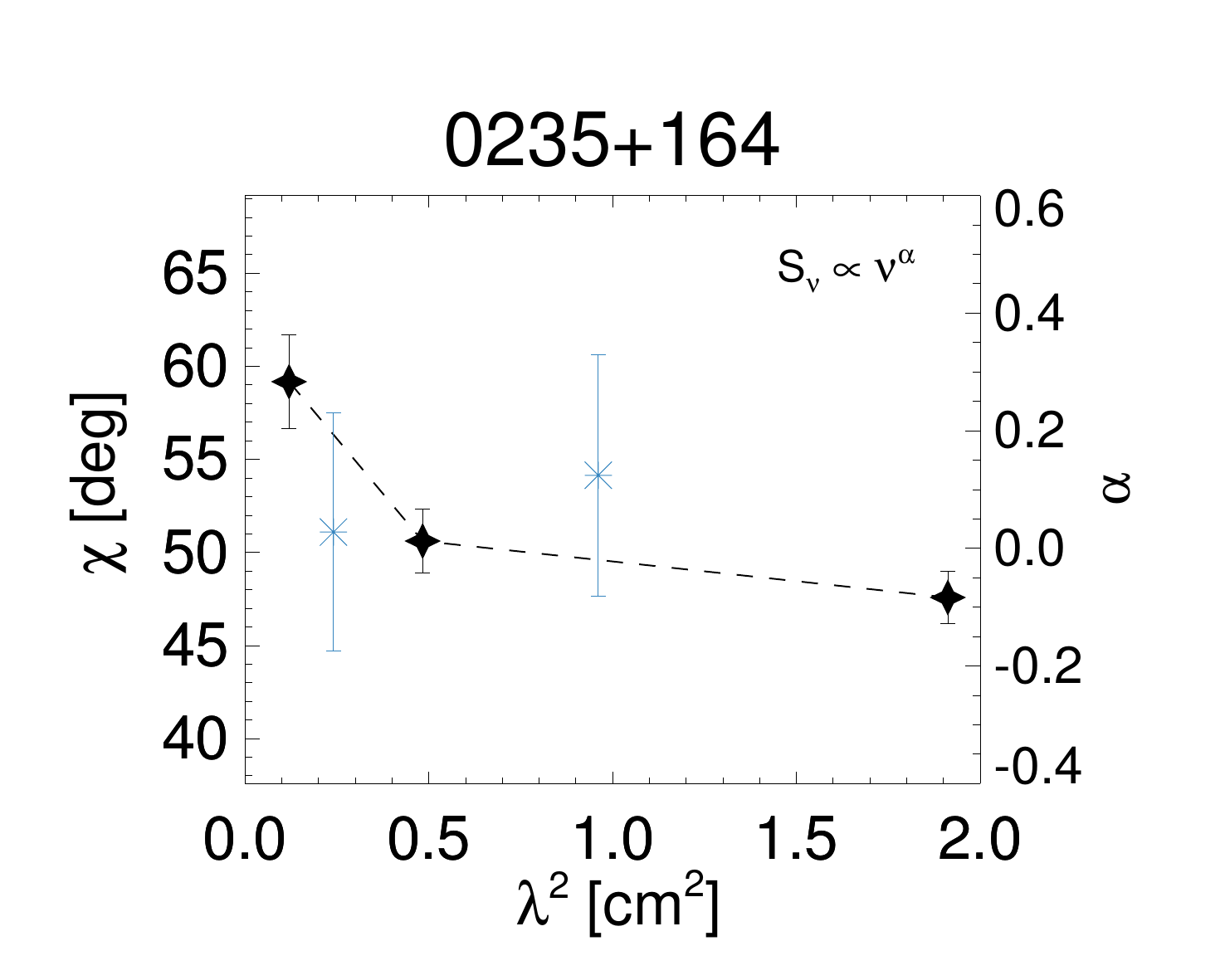}
\includegraphics[trim=5mm 4mm 24mm 13mm, clip, width = 47mm]{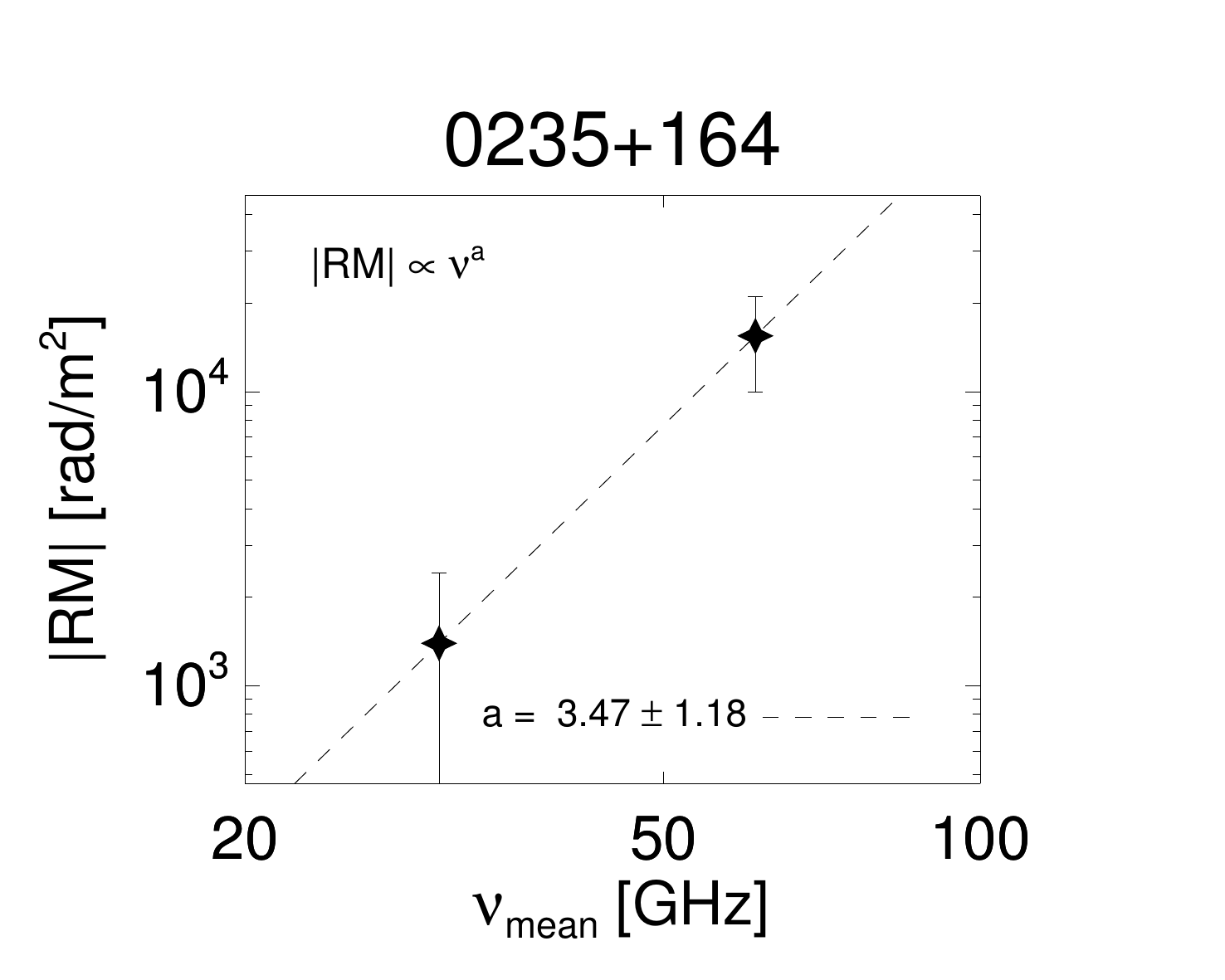}
\includegraphics[trim=7mm 0mm 6mm 0mm, clip, width = 80mm]{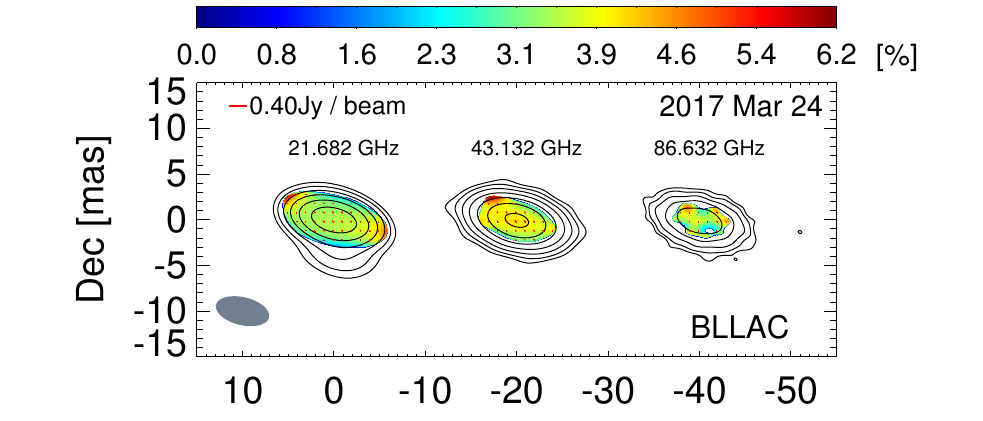}
\includegraphics[trim=7mm 4mm 10mm 13mm, clip, width = 51mm]{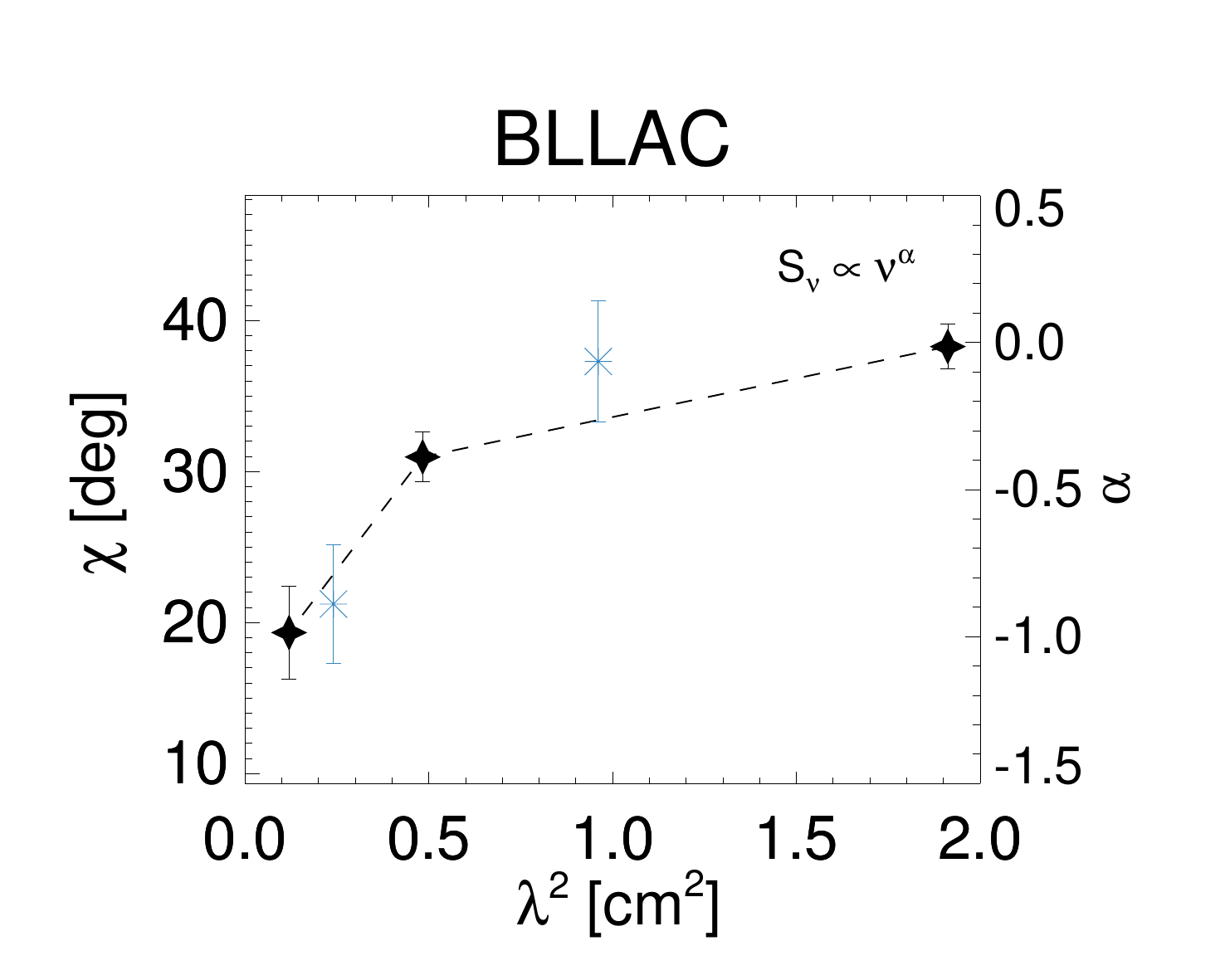}
\includegraphics[trim=5mm 4mm 24mm 13mm, clip, width = 47mm]{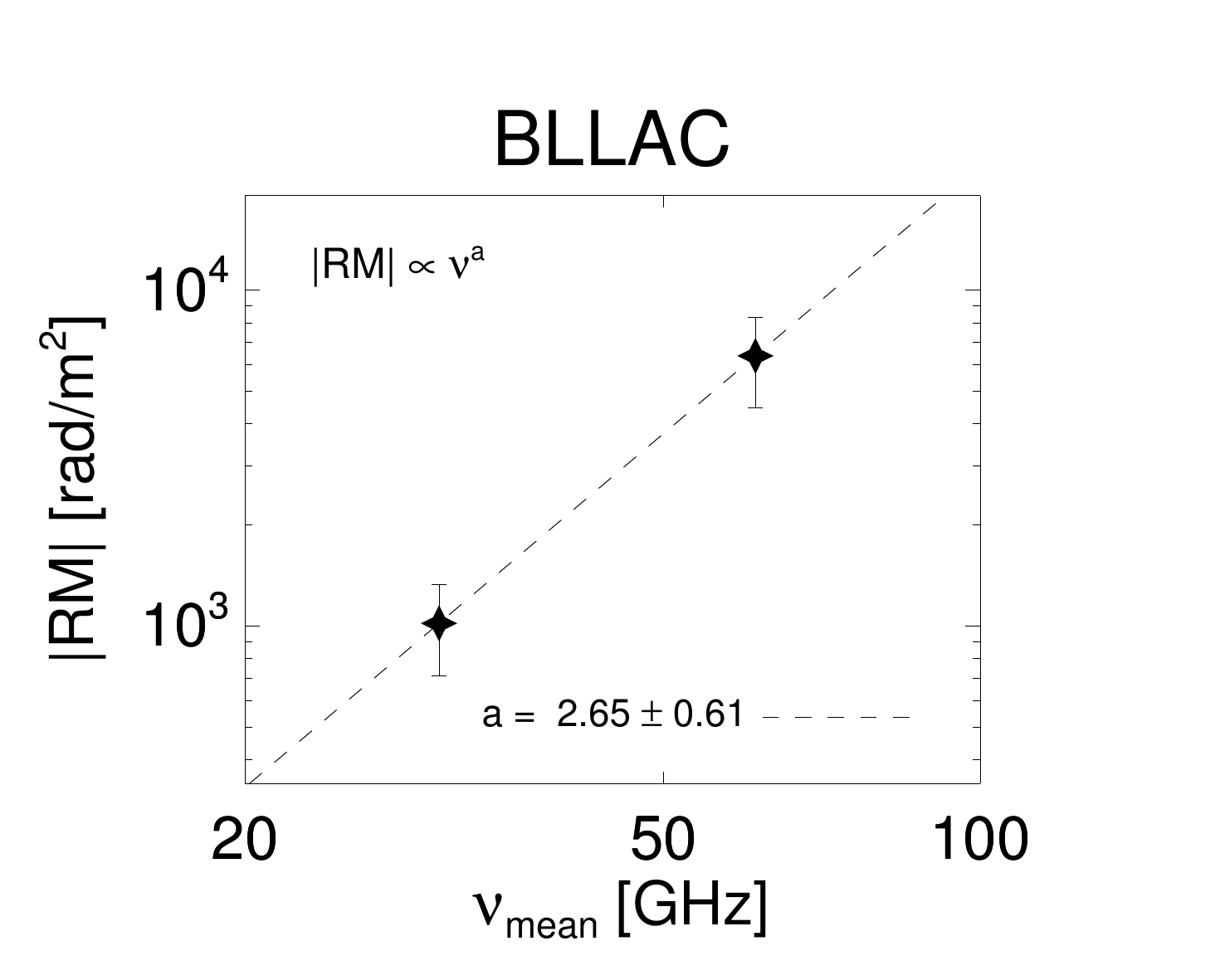}
\includegraphics[trim=7mm 0mm 6mm 0mm, clip, width = 80mm]{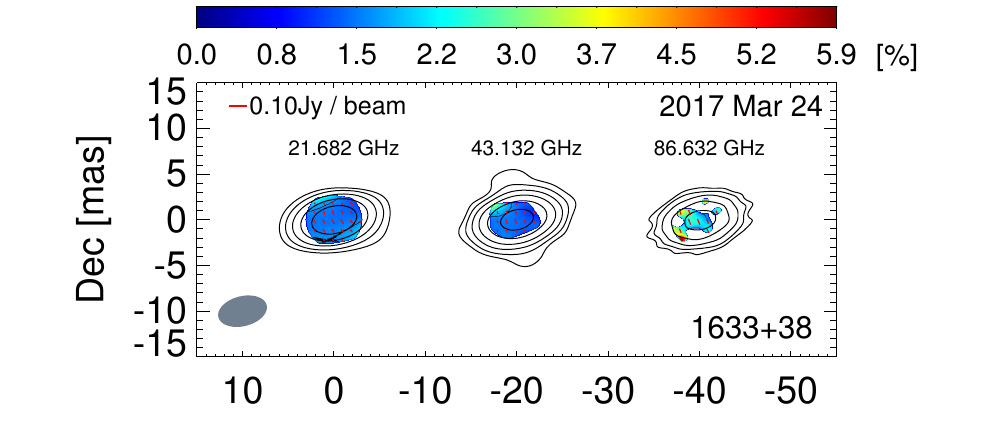}
\includegraphics[trim=7mm 4mm 10mm 13mm, clip, width = 51mm]{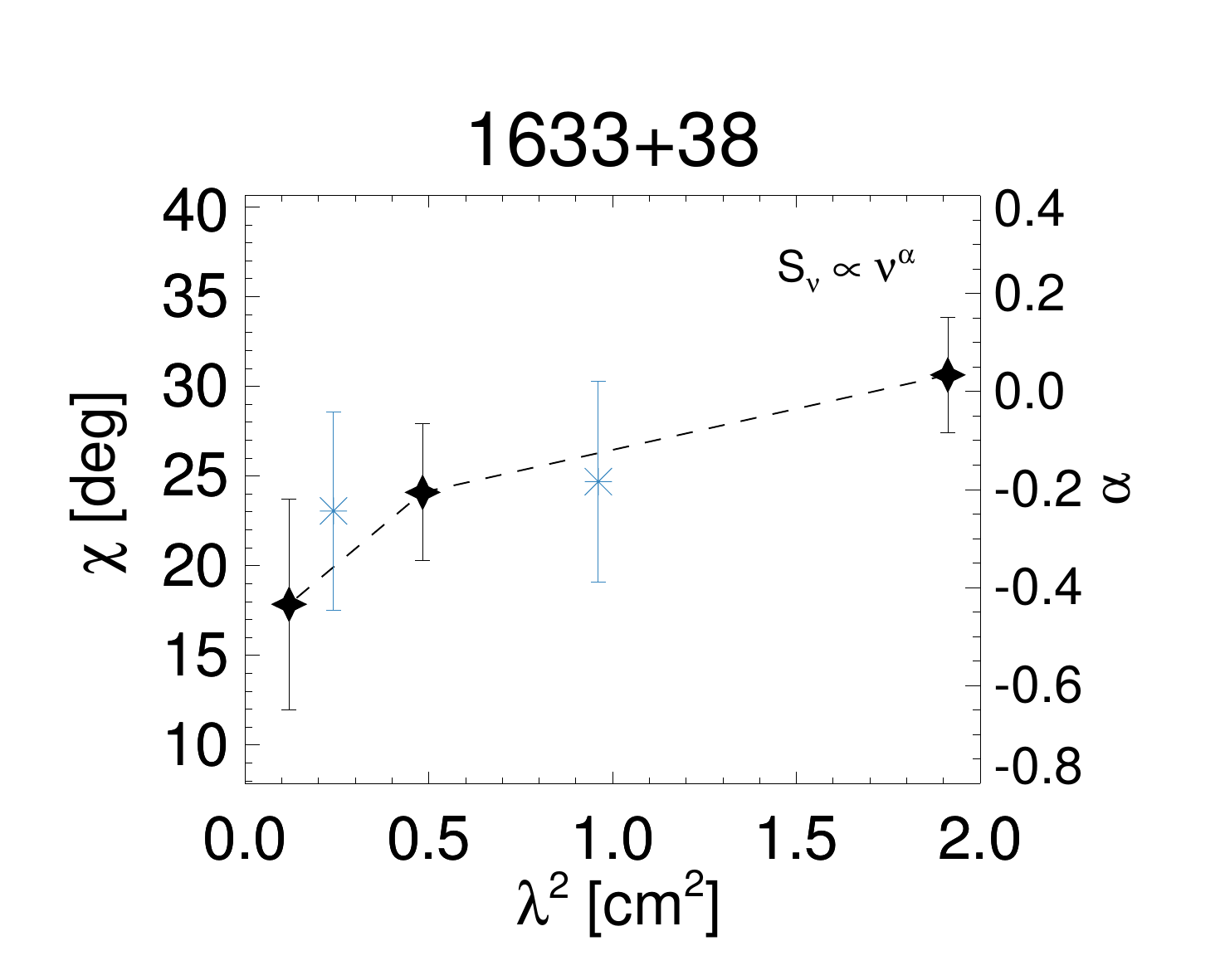}
\includegraphics[trim=5mm 4mm 24mm 13mm, clip, width = 47mm]{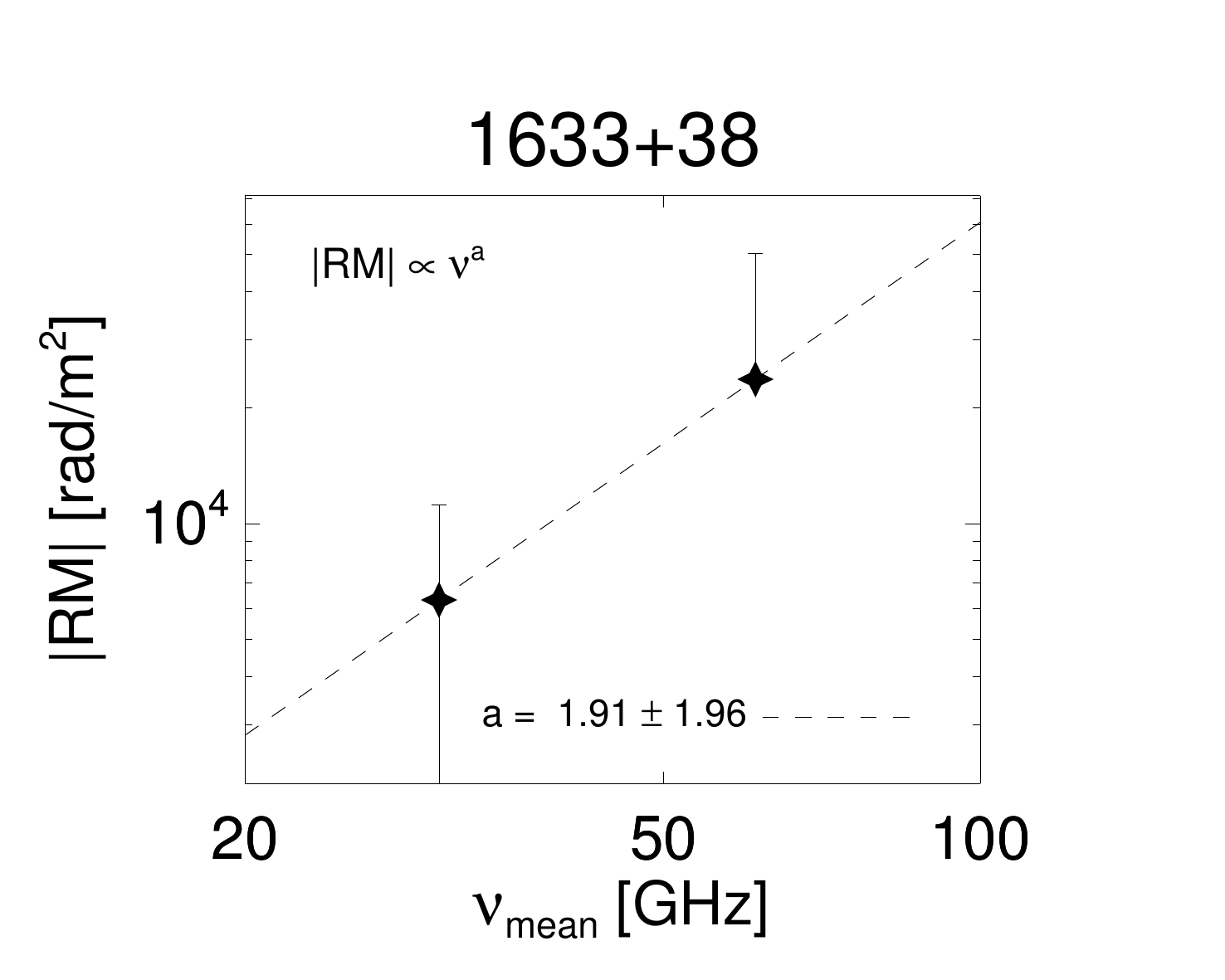}
\caption{\emph{Continued.}}
\end{center}
\end{figure*}

\section{results}
\label{sect3}

\subsection{RM at radio wavelengths}

In Figure~\ref{result}, we present polarization maps for multiple frequencies convolved with the KVN 22 GHz beam size (left panels), EVPAs at the core as a function of $\lambda^2$ (central panels), and RMs as a function of geometric mean observing frequency (right panels). We obtained one RM value from each adjacent data pair in the EVPA--$\lambda^2$ plots because we could not obtain good $\lambda^2$ fits across the three bands in most cases. We have to rotate the 22 and 43 GHz EVPAs by more than $720^{\circ}$ and $180^{\circ}$ to explain this behavior with $n\pi$ ambiguity, which translates into $\rm RM \gtrsim 10^5\ rad/m^2$. This RM value is too large especially at 22 GHz because almost all of the detected core RM of blazars are less than $1000\rm\ rad/m^2$ at $\lesssim15$ GHz \citep{hovatta2012}. Alternatively, different optical depths of the cores at different frequencies might be responsible for the non-$\lambda^2$ fits. Especially, EVPA rotations by $90^\circ$ are expected in case of a transition of the core from optically thick to optically thin \citep{pacholzcyzk1970}. We provide the spectral index, $\alpha$ in $S_{\nu} \propto \nu^{\alpha}$, between adjacent frequencies in columns (3) and (4) in Table~\ref{result} and in the center panels of Figure~\ref{result} (blue asterisks). We found that values of $\alpha$ measured at different frequency pairs differ from each other by more than $1\sigma$ in only four cases, 3C 279 in 2017 January, OJ 287 in 2017 January, 3C 345 in 2017 January, and BL Lac in 2017 March. When we rotate the EVPAs at the lowest frequency by $90^\circ$ for these cases, we are left with even worse $\lambda^2$ fits (higher $\chi^2$ values) compared to the case of no rotation. In addition, the degree of polarization is not much different, less than a factor of $\approx2$, at different frequencies (Section~\ref{fracpol}), while it should decrease by a factor of $\approx7$ if there were a $90^{\circ}$ flip \citep{pacholzcyzk1970}.  Therefore, the progressively steeper EVPA rotations at higher frequencies are more likely to be due to the core shift effect, as shown in the numerical simulations of special relativistic magnetohydrodynamic jets \citep{porth2011}. 

We identified the origin (0,0) of the maps with the location of the cores. This might not be always the case exactly; however, the beam size is quite large and thus the effect of an offset of the core position from the origin would be insignificant. We summarize our sources' basic information and their observed polarization quantities in Table~\ref{table:source}. We note that we did not consider the effect of the integrated (Galactic) RM because it is a few hundreds $\rm rad/m^2$ at most (e.g., \citealt{taylor2009}); this is much smaller than the typical RM error we obtain, and at the KVN frequencies the EVPA rotation due to the integrated RM is negligible. 

Since the beam sizes are quite different at different frequencies, we need to quantify the errors in polarization quantities introduced by the convolution of all maps with the 22 GHz beam. We compared $m$ and EVPA at the core found with and without using convolution for each source at each frequency and added the differences quadratically to the errors of $m$ and EVPA found from convolved maps, respectively. Using contemporaneous BU VLBA maps, We identified and excluded sources which have complex polarization structure near the core that cannot be resolved with the KVN; this leaves us with eight sources. We briefly describe the results for individual sources below.

\subsubsection{3C 279}

This source is characterized by longitudinal (i.e., parallel to the jet direction) EVPAs that show a smooth distribution from the core to the inner jet\footnote{The term `inner jet' denotes any polarized component in the jet that can be resolved from the core by instruments with higher angular resolution than the KVN but cannot be (well) resolved by the KVN itself, like the extended linear polarization structure of 3C 279 at $\approx1$ mas from the core seen in the BU map of 14 January 2017 (see \url{https://www.bu.edu/blazars/VLBA_GLAST/3c279/3C279jan17\_map.jpg})} (e.g., \citealt{jorstad2005}). Similarly, our KVN maps show basically longitudinal EVPAs but rotated by up to $\approx20^{\circ}$ as function of frequency. RMs between adjacent frequency pairs range from $\approx10^3$ to $\approx10^4\rm\ rad/m^2$. We fitted a power law function to the RMs as a function of geometrical mean observing frequency and obtained the power law index $a$ in the relation ${\rm|RM|} \propto \nu^a$. Since we calculate each RM value from only two data points, the RM errors are relatively large, which results in relatively large errors in $a$. However, $a$ values in all three epochs show a good agreement with $a=2$--3, which is quite consistent with the results of previous KVN single-dish polarization monitoring of 3C 279 \citep{kang2015}. We note that the core EVPAs of this source might be contaminated by polarization from the inner jet. However, EVPA rotation of the inner jet region is expected to be very small at $\gtrsim 22\rm\ GHz$ because of a relatively small RM in that region ($\lesssim250\rm\ rad/m^2$; \citealt{hovatta2012}). Therefore, we conclude that the observed EVPA rotation over frequency of this source is dominated by the core polarization.

\subsubsection{OJ 287\label{oj287}}

For this source, we could find contemporaneous VLBA data at 15~GHz from the MOJAVE program\footnote{\url{http://www.physics.purdue.edu/astro/MOJAVE/index.html}}, observed on 2017 January 28 and March 11, with our KVN data being obtained on 2017 January 17--18 and March 22--24, respectively. We included those data in our analysis after convolving the 15~GHz maps with the KVN 22~GHz beam (because the KVN beam is larger than the VLBA one even though it is at a higher frequency.) This source has shown slow and gradual EVPA variation in time and thus a potential variability during the time gap between the MOJAVE and our KVN observations ($\lesssim2$ weeks) would not be significant (see the AGN monitoring database of the 26-meter University of Michigan Radio Astronomy Observatory\footnote{\url{https://dept.astro.lsa.umich.edu/obs/radiotel/gif/0851\_202.gif}}). In addition, OJ~287 has been known for relatively small RMs at cm wavelengths (e.g., \citealt{hovatta2012}). Our KVN maps are consistent with zero RM (within errors) in 2016 December. However, in the 2017 January data, EVPA rotations from 15 to 86 GHz in the same direction being steeper at higher frequency pairs are observed, which results in $a = 0.98\pm0.66$. In the 2017 March data, the EVPA rotations between 15 and 43 GHz were almost zero within errors but a relatively large rotation with $\rm|RM|\approx5\times10^3\rm rad/m^2$ was detected at 43/86 GHz.

\subsubsection{CTA 102\label{cta102}}

This source shows a high degree of polarization, up to $\approx 40\%$, in its jet at 22 GHz but becomes compact at higher frequencies. The EVPAs rotate rapidly as function of frequency, with different slopes in the EVPA--$\lambda^2$ diagram between 22/43 GHz and 43/86 GHz. RMs at 43/86 GHz are a few times $10^4\rm\ rad/m^2$ in the source rest frame in both epochs. However, $a$ value in the 2016 December data is much larger than that in the 2017 January data. The signs of RMs are different in our two epochs, while their absolute values are of the same order of magnitude. This behavior suggests that the line-of-sight component of the jet magnetic field changed its direction within $\approx1$ month, while magnetic field strength and electron density (or at least their product) did not vary substantially. This sign change might be related to a strong flare that occurred during our KVN observations \citep{raiteri2017}. We discuss possible reasons for the sign flip in CTA 102 in Section~\ref{signchange}.

\subsubsection{3C 345}
\label{3c345}

This source shows almost the same RMs at 22/43 and 43/86 GHz, $\rm |RM|\approx10^4\ rad/m^2$ in the 2016 December data, with $a$ being consistent with zero within errors. However, the RM at 43/86 GHz is much larger than that at 22/43 GHz about one month later, resulting in $a=1.86\pm0.3$. These results indicate that there is substantial time variability in this source. Similarly to the case of CTA 102, this source shows a flare during our KVN observations and the flux density in early 2017 is almost three times higher than that in mid 2016 at 1 mm.\footnote{\url{http://sma1.sma.hawaii.edu/callist/callist.html?plot=1642\%2B398} This might be related to a substantial change in $a$ within $\approx1$ month for this source, though the sign of RM does not change during the period.}

\subsubsection{1749+096}

This source displays a rather compact jet geometry at all frequencies and in both epochs (2016 December and 2017 March). Interestingly, the degree of linear polarization is larger at lower frequencies, which is not usually seen in other sources (Section~\ref{fracpol}). This might suggest that we are looking at a mixture of different polarization components with different EVPAs and/or RMs or that internal Faraday rotation occurs in this source (Section~\ref{sectscreen}). The values of RM range from $\approx10^3$ to $\approx10^4\rm\ rad/m^2$ in both epochs data. The values of $a$ are consistent within errors. Therefore, the sign, absolute value, and frequency dependence of RM appear to be stable over $\approx3$ months.

\subsubsection{0235+164}

This source shows a rather compact jet geometry. A systematic rotation of EVPAs in the same sense as function of frequency can be seen from 22 to 86 GHz. RMs range from $\approx10^3$ to $\approx2\times10^4\rm\ rad/m^2$, with a substantially larger RM at a higher frequency pair, resulting in $a = 3.47\pm1.18$.

\subsubsection{BL LAC}

The EVPA--$\lambda^2$ diagram shows a steeper slope between 43 and 86 GHz than between 22 and 43 GHz, providing $a = 2.65\pm 0.61$. RM values range from $\approx10^3$ to $\approx6\times10^3\rm\ rad/m^2$. This is consistent with previous measurements of the core RM between 15 and 43 GHz \citep{OG2009a, gomez2016}. However, the high resolution \emph{RadioAstron} space VLBI image shows a complex RM structure in the core region including a sign change, indicating the presence of helical magnetic fields there \citep{gomez2016}. Therefore, we may be looking at a blend of those structures in our KVN images (see Section~\ref{multipleshocks}).

\subsubsection{1633+38}

This source is relatively faint and shows a low degree of linear polarization ($\approx2\%$) which leads to relatively large EVPA errors. Thus, the RM at 43/86 GHz is comparable to its error and the obtained $a=1.91\pm1.96$ has also a large error. When fitting a single linear function to the EVPAs at the three frequencies available, we obtained $\rm RM = 974\pm509\ rad/m^2$. This is surprisingly low because a very high RM, $\approx2.2\times10^4\rm\ rad/m^2$, was reported previously for the core of this source using six different frequencies from 12 GHz to optical wavelength \citep{algaba2012}. We found that EVPAs at the core of this source in the MOJAVE program are $44^\circ$ and $32^\circ$ in 2016 November and 2017 April, respectively. If there is no substantial EVPA variability in between these two epochs, then a simple $\lambda^2$ fit can explain the data from 15 to 86 GHz, suggesting that there is no $n\pi$ ambiguity in our data and that the core RM of this source is indeed quite small. The observations in \cite{algaba2012} were performed in 2008 November which indicates that there is substantial temporal variability of the EVPA rotation in 1633+38. Four years before their observations, core EVPAs could not be fitted with a single $\lambda^2$ law and the obtained RM value was much smaller, $\rm RM = -570\pm430\ rad/m^2$ \citep{algaba2011}. This also suggests substantial temporal RM variability in 1633+38.

\begin{figure*}[!t]
\begin{center}
\includegraphics[trim=2mm 3mm 10mm 0mm, clip, width = 88mm]{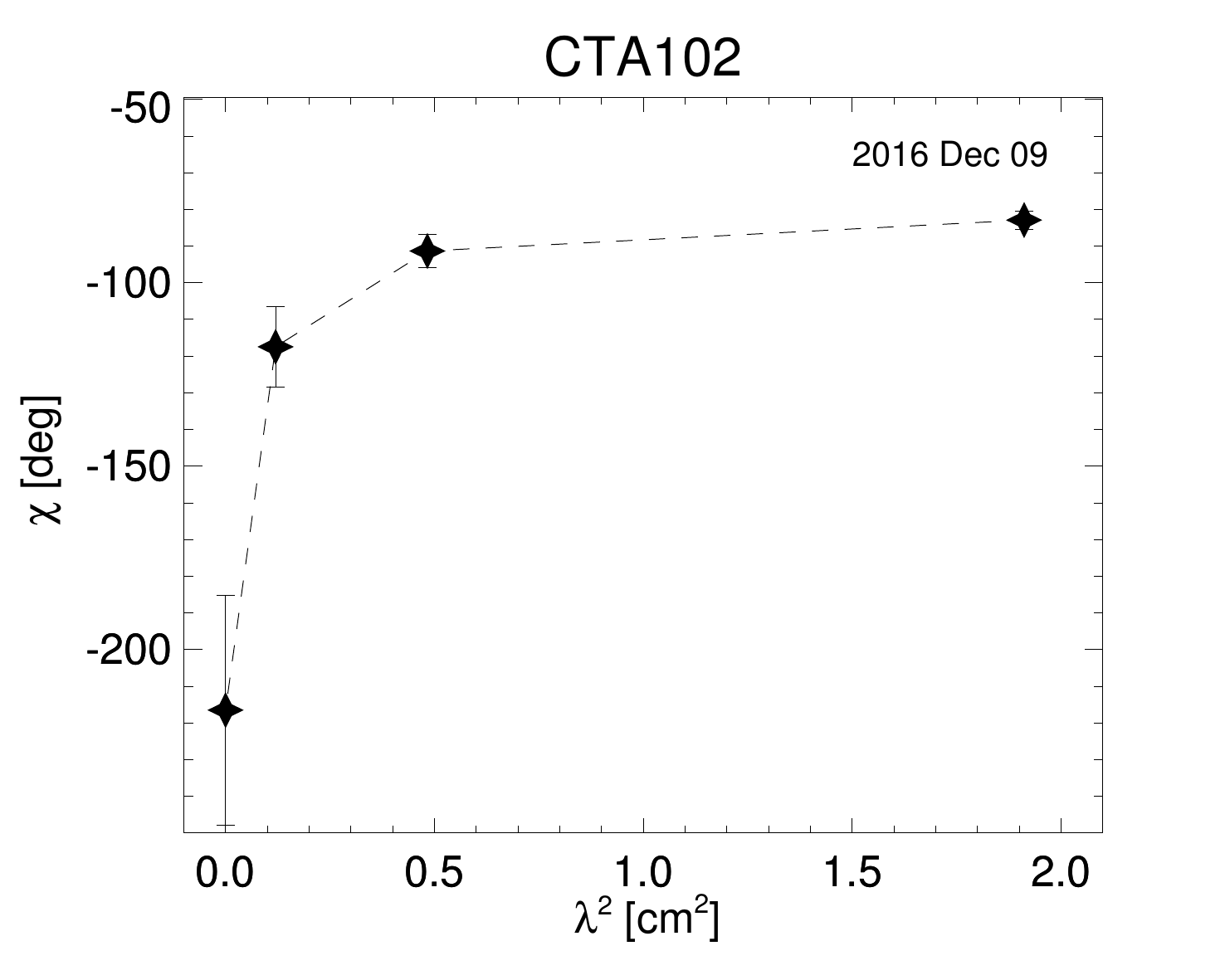}
\includegraphics[trim=2mm 3mm 10mm 0mm, clip, width = 88mm]{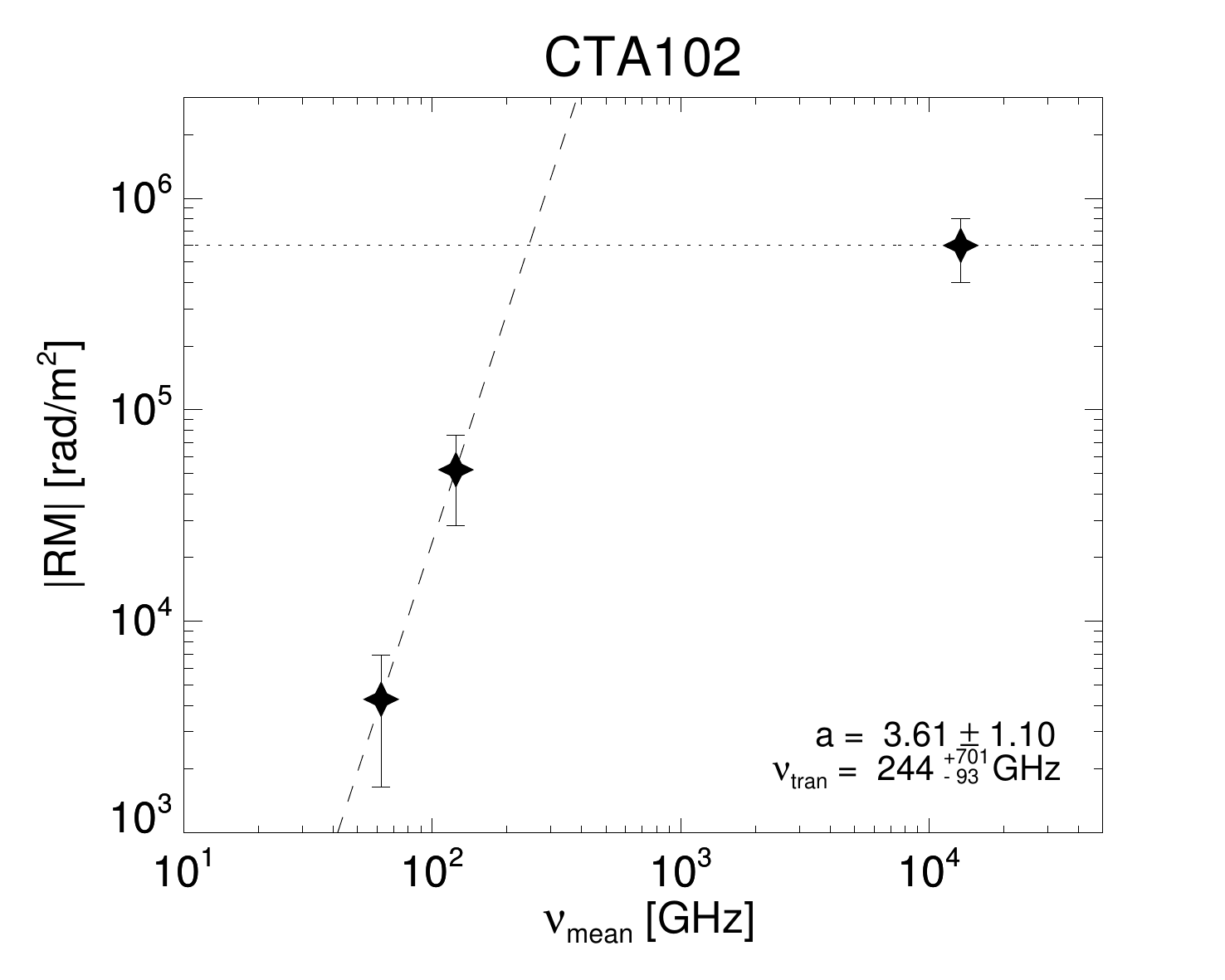}
\includegraphics[trim=2mm 3mm 10mm 0mm, clip, width = 88mm]{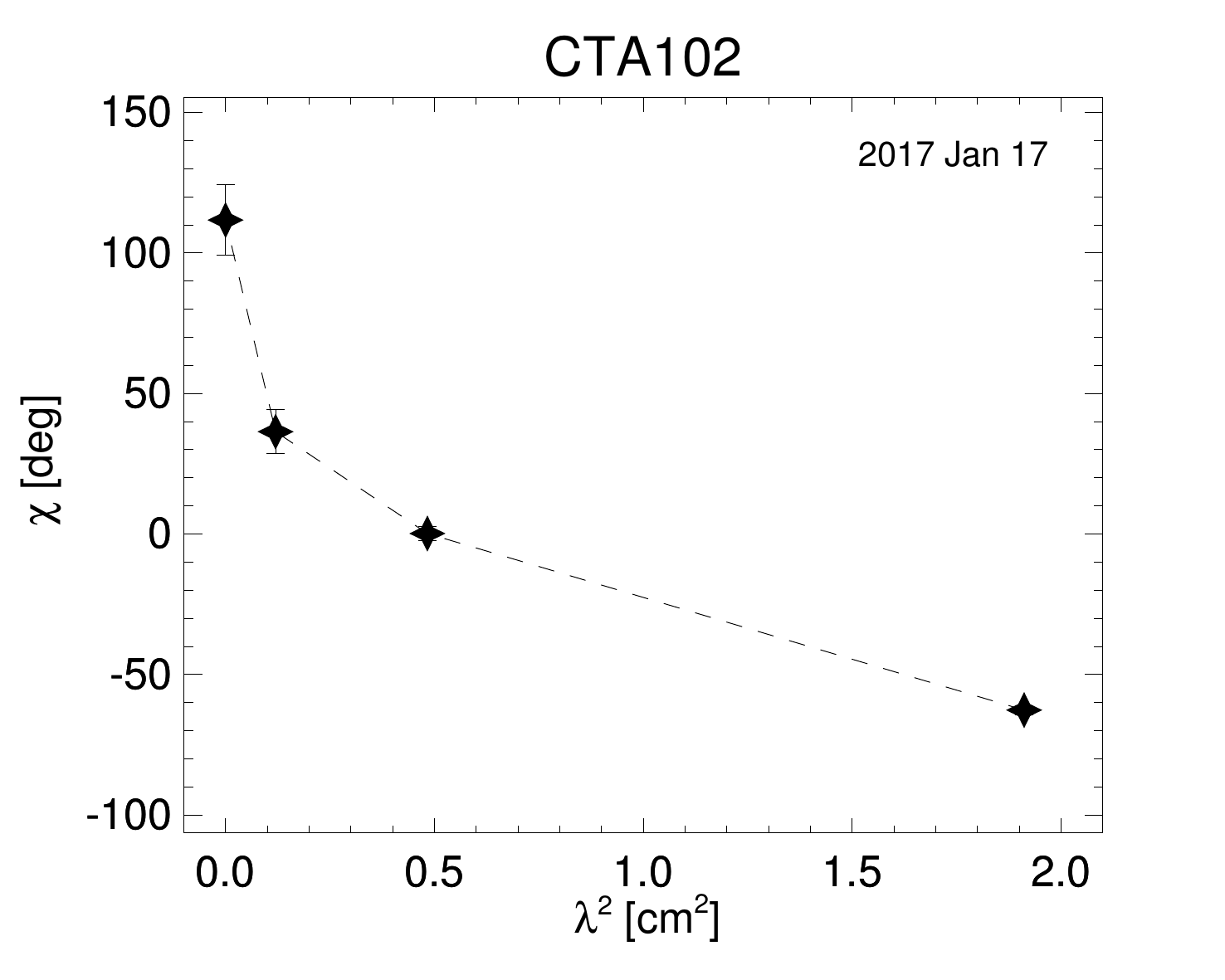}
\includegraphics[trim=2mm 3mm 10mm 0mm, clip, width = 88mm]{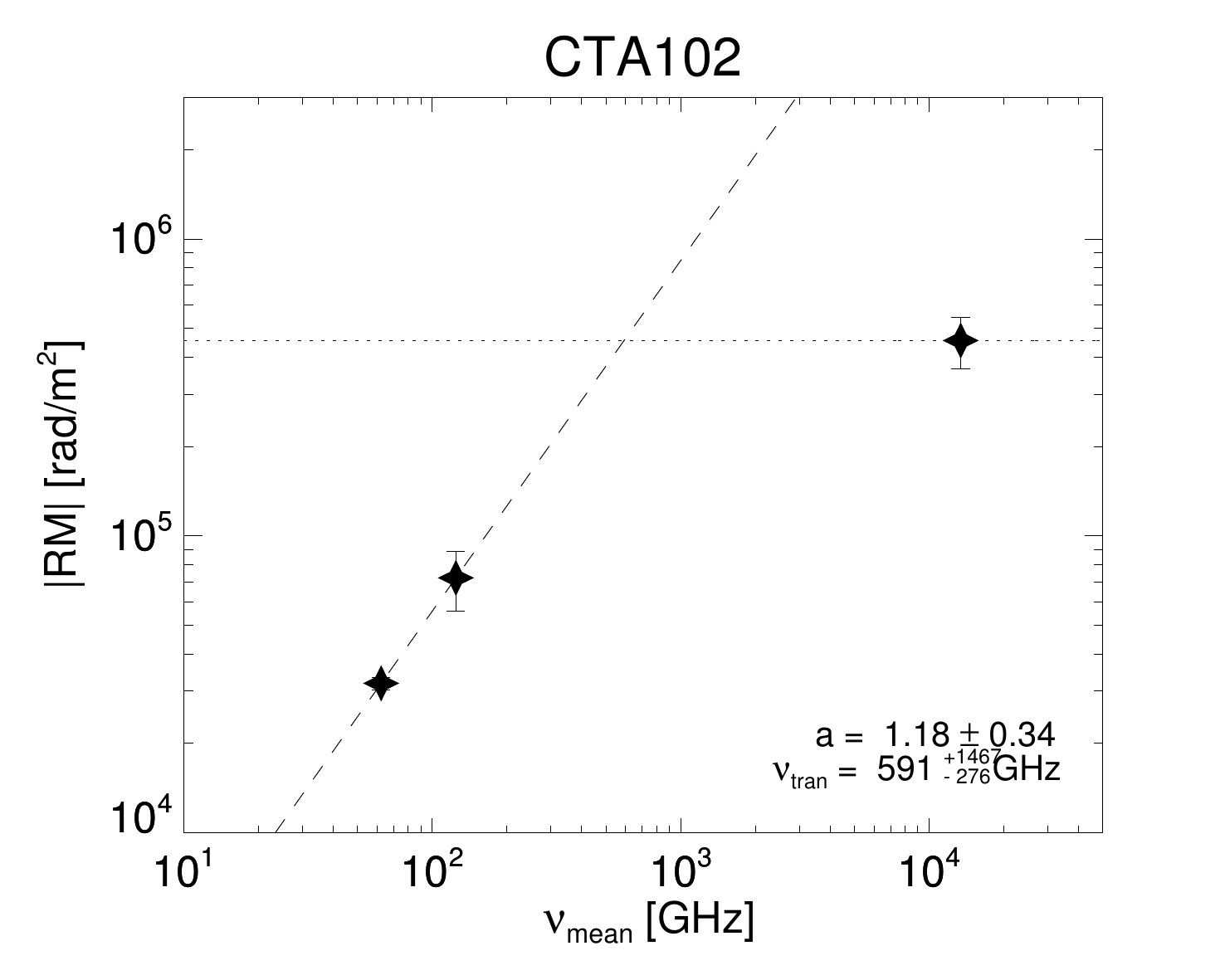}
\includegraphics[trim=2mm 3mm 10mm 0mm, clip, width = 88mm]{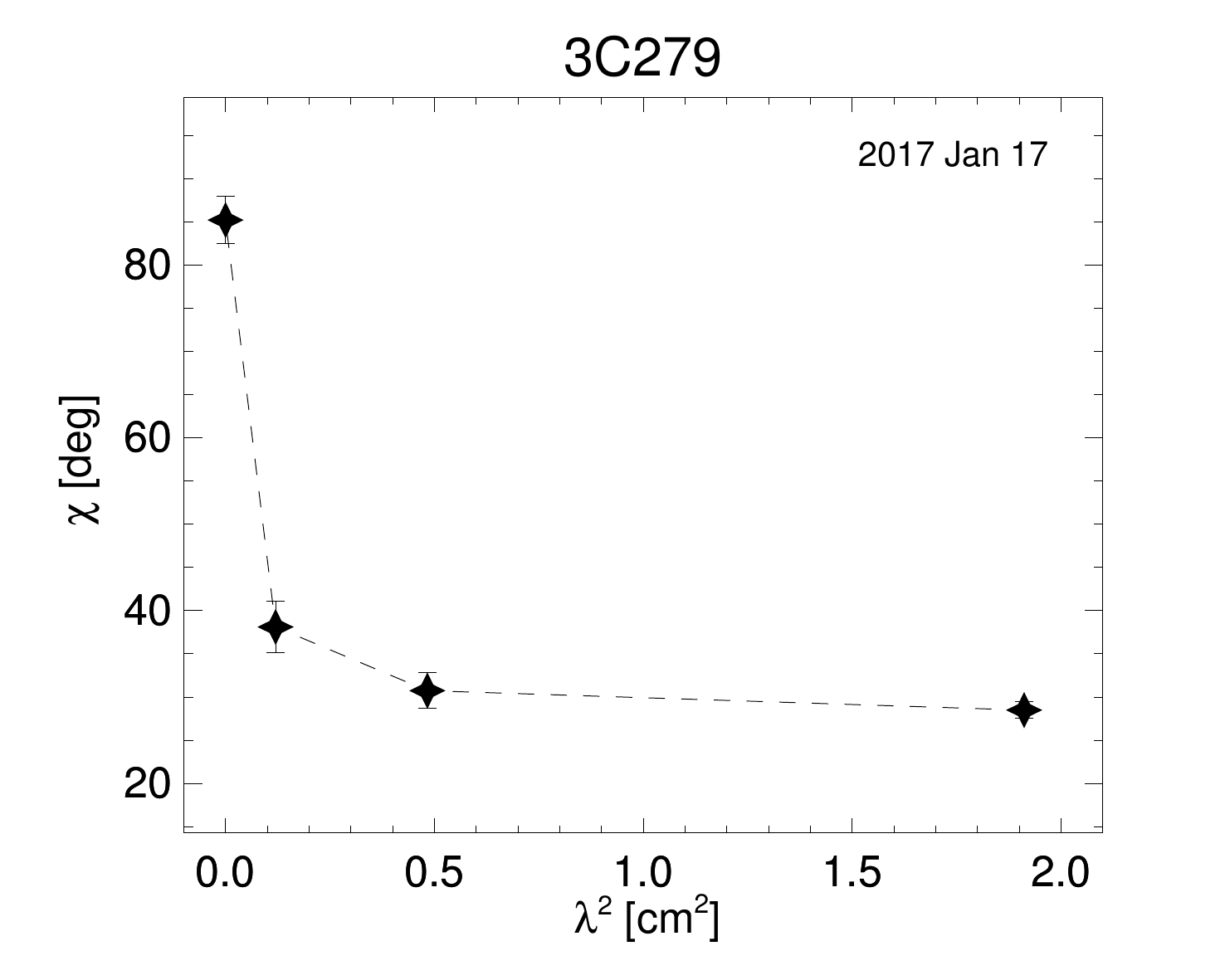}
\includegraphics[trim=2mm 3mm 10mm 0mm, clip, width = 88mm]{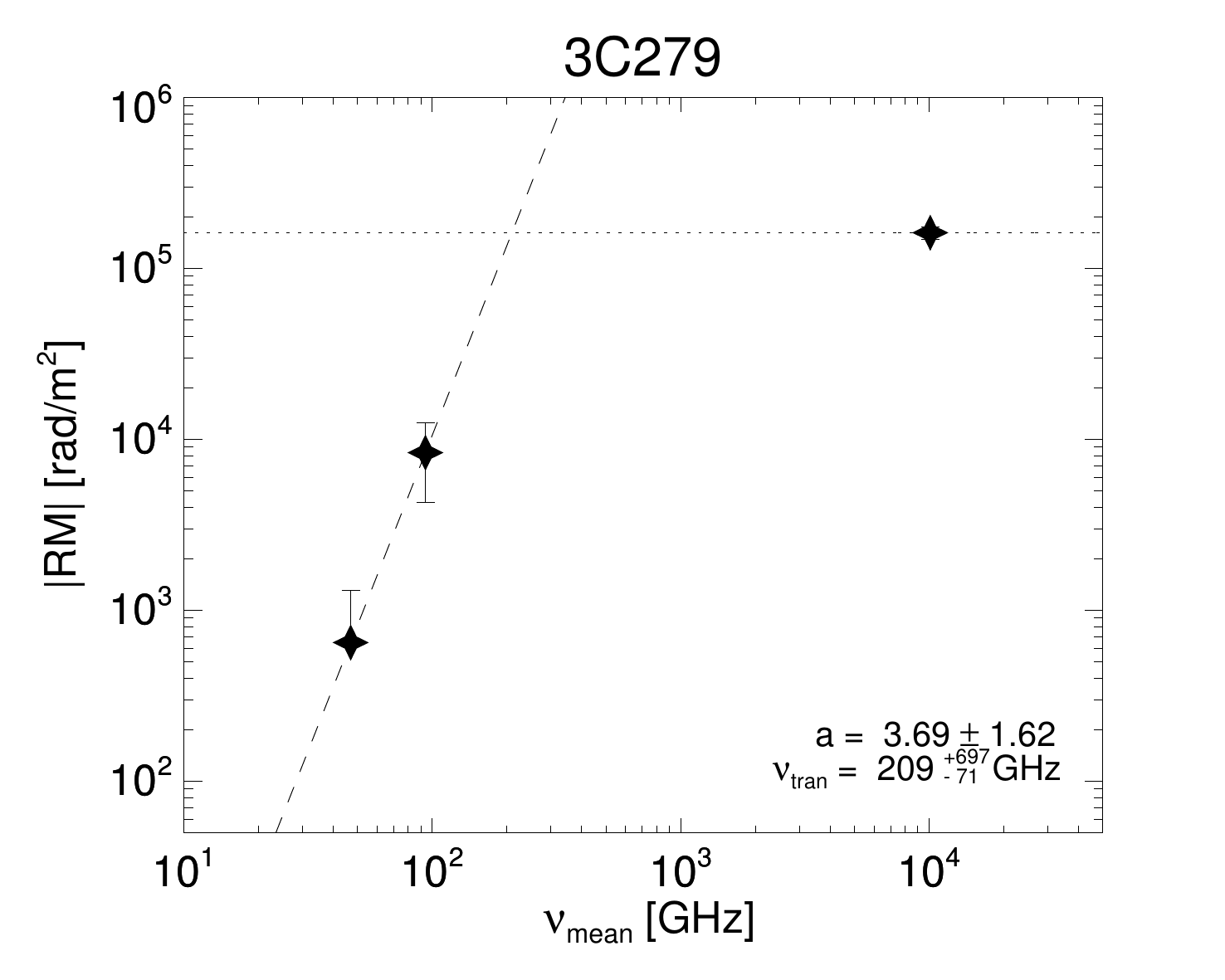}
\caption{Same as the central and right panels of Figure~\ref{result} but combined with contemporaneous optical data (see Table~\ref{Steward}). We obtained the frequency $\nu_{\rm trans}$ where the power-law increase of the RM measured at radio frequencies intersects with the RM obtained with the optical data points (horizontal dotted lines). The values of the power-law index $a$ and $\nu_{\rm trans}$ are given at the bottom right of the right panels. All RM and frequency values are in the source rest frame. \label{paopt}}
\end{center}
\end{figure*}

\addtocounter{figure}{-1}

\begin{figure*}[!t]
\begin{center}
\includegraphics[trim=2mm 3mm 10mm 3mm, clip, width = 88mm]{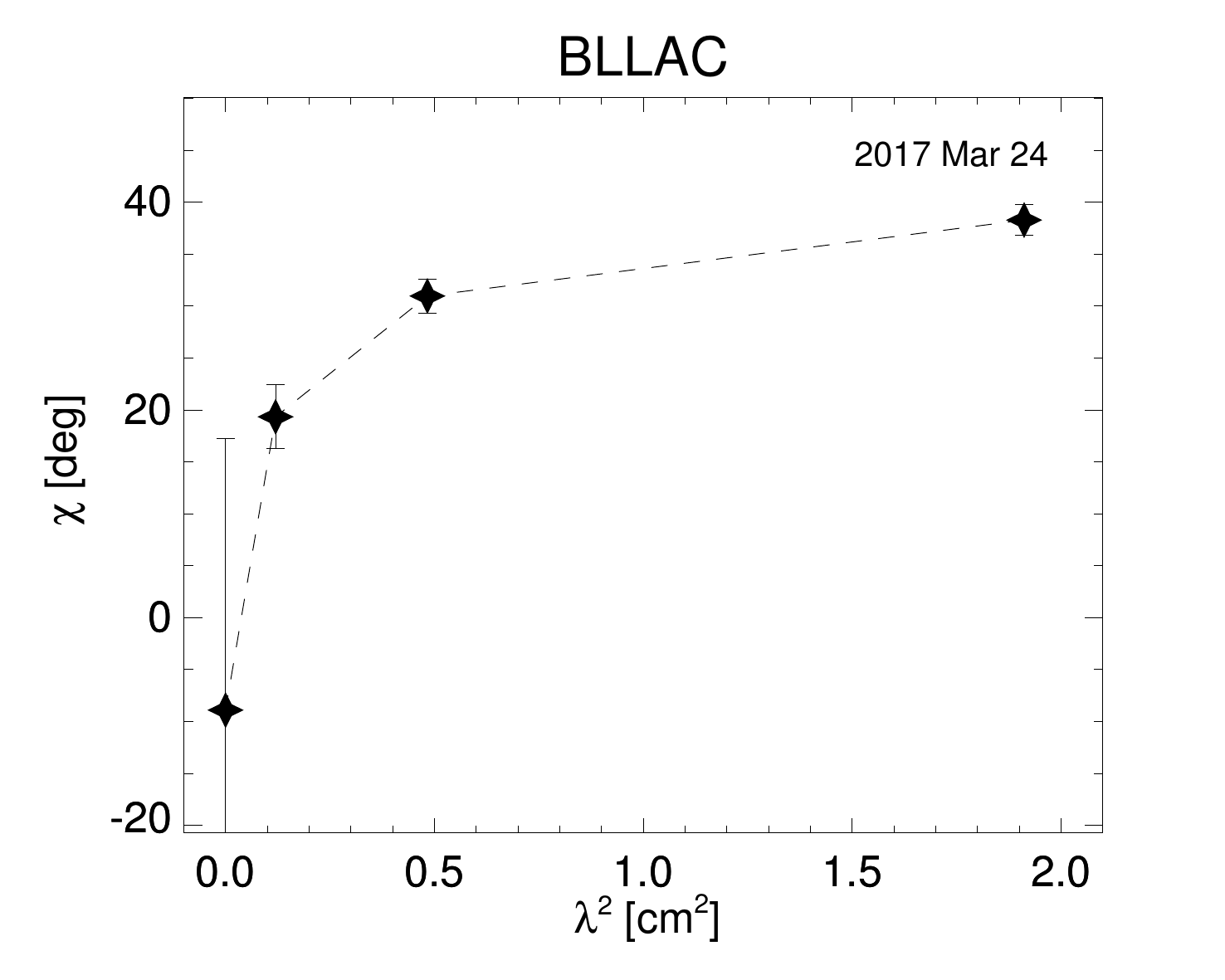}
\includegraphics[trim=2mm 3mm 10mm 3mm, clip, width = 88mm]{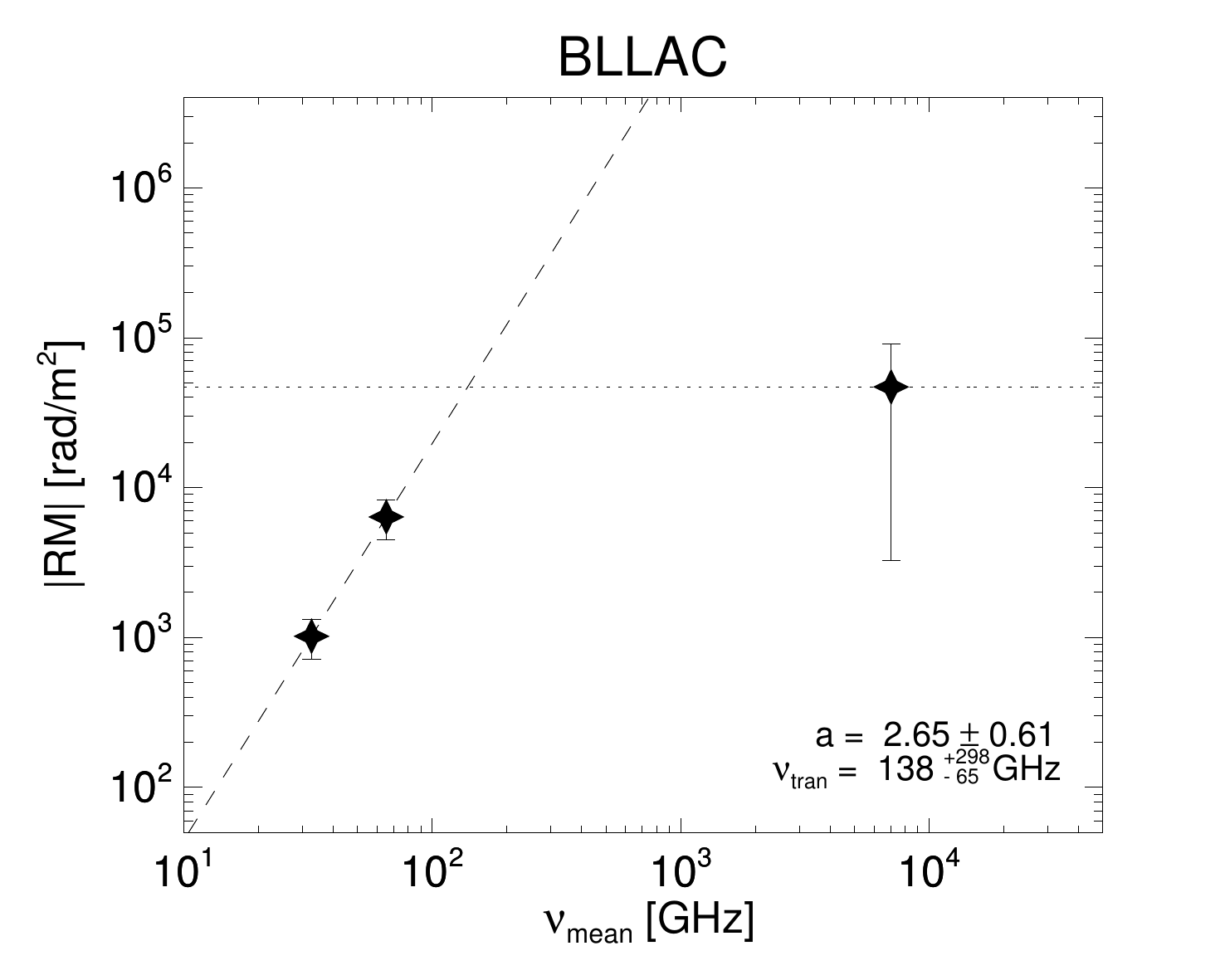}
\caption{\emph{Continued.}}
\end{center}
\end{figure*}

\subsection{Optical EVPAs from the Steward observatory\label{optdata}}

For a few sources, we obtained quasi-contemporaneous (taken within $\lesssim$1 week) optical polarization data from the Steward Observatory blazar monitoring program\footnote{\url{http://james.as.arizona.edu/~psmith/Fermi}} (see \citealt{smith2009} for details) for some epochs. We summarize the optical data we used in Table~\ref{Steward}. (We excluded some additional datasets due to their large errors.) The optical polarimetry errors are usually quite small, $\lesssim1^{\circ}$, unless sources are very weakly polarized. However, optical polarization of blazars often show rapid variability on short time scales (e.g., \citealt{jorstad2013}) presumably due to a smaller size of the emission region at higher frequencies \citep{marscher1996}. In order to take into account the uncertainty arising from the time gap between optical and radio observations we estimated errors from source variability as follows.

The Steward Observatory blazar monitoring program usually observes each source multiple times for a specified period spanning a few days in a broad optical band from 500 to 700~nm. We selected all data in the periods that are close to our KVN observations and used the mean and standard deviation of the data points as representatives of optical EVPA and typical error, respectively. We assumed that their optical EVPAs show random-walk type variations with time. (In addition to statistical variability, many blazars occasionally show smooth, systematic optical EVPA rotations that might be associated with high energy flares; e.g., \citealt{blinov2015}). We multiplied the observed standard deviation by the square root of the ratio of the time gap between optical and radio observations to the duration of the period in which a set of optical data was obtained. Under this assumption, the longer the time separation, the larger the formal uncertainty of an optical EVPA at the time of the corresponding radio observation.

\begin{deluxetable}{clcl}[t!]

\tablecaption{Optical Data from the Steward Observatory\label{Steward}}
\tablehead{
\colhead{Source} & \colhead{Obs. date} &
\colhead{$\chi_{\rm opt}$ [$^{\circ}$]} & \colhead{$\nu_{\rm trans}$ [GHz]}
}
\startdata
\multirow{2}{0.2\columnwidth}{\centering CTA 102} & 11/30/16--12/02/16 & $-216.5 \pm 31.4$ & $244^{+701}_{-93}$\\
& 01/10/17--01/13/17 & $111.7 \pm 12.6$ & $591^{+1467}_{-276}$\\
\centering 3C 279 & 01/10/17--01/14/17 & $85.2 \pm 2.7$ & $209^{+697}_{-71}$\\
\centering BL LAC & 03/29/17--03/30/17 & $-8.9 \pm 26.1$ & $138^{+298}_{-65}$
\enddata
\tablecomments{$\chi_{\rm opt}$ is the EVPA at optical wavelengths taken from the blazar monitoring program of the Steward Observatory. We used the mean value of the EVPAs observed in the noted period as $\chi_{\rm opt}$ (see text for explanation of $n\pi$ ambiguity and error estimation). $\nu_{\rm trans}$ shows the frequency on the source's rest frame at which the power-law increase of RM at radio frequencies is expected to stop (see the right panel of Figure~\ref{paopt}).}

\end{deluxetable}

\begin{figure*}[!t]
\begin{center}
\includegraphics[trim=10mm 8mm 18mm 4mm, clip, width = 52mm]{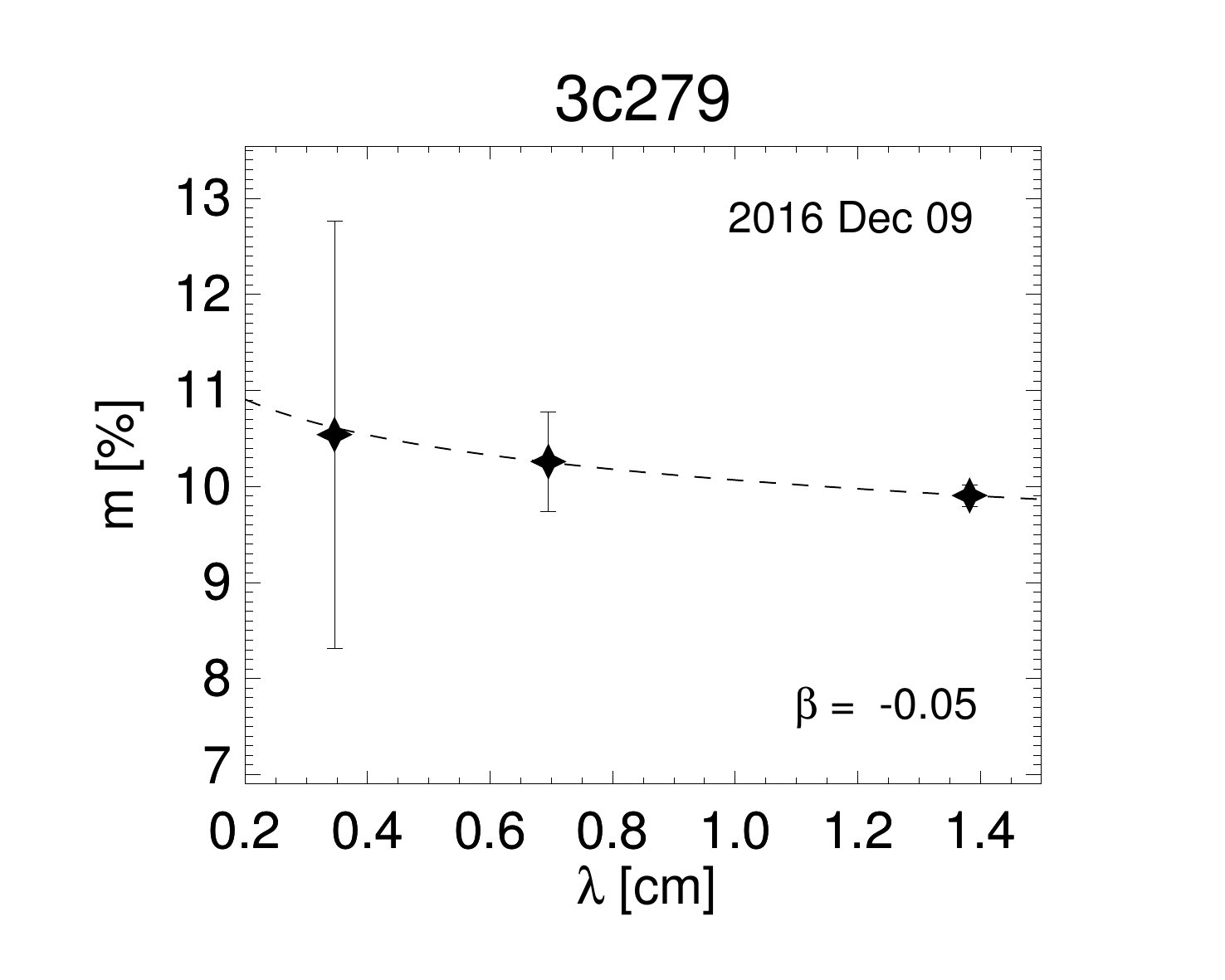}
\includegraphics[trim=10mm 8mm 18mm 4mm, clip, width = 52mm]{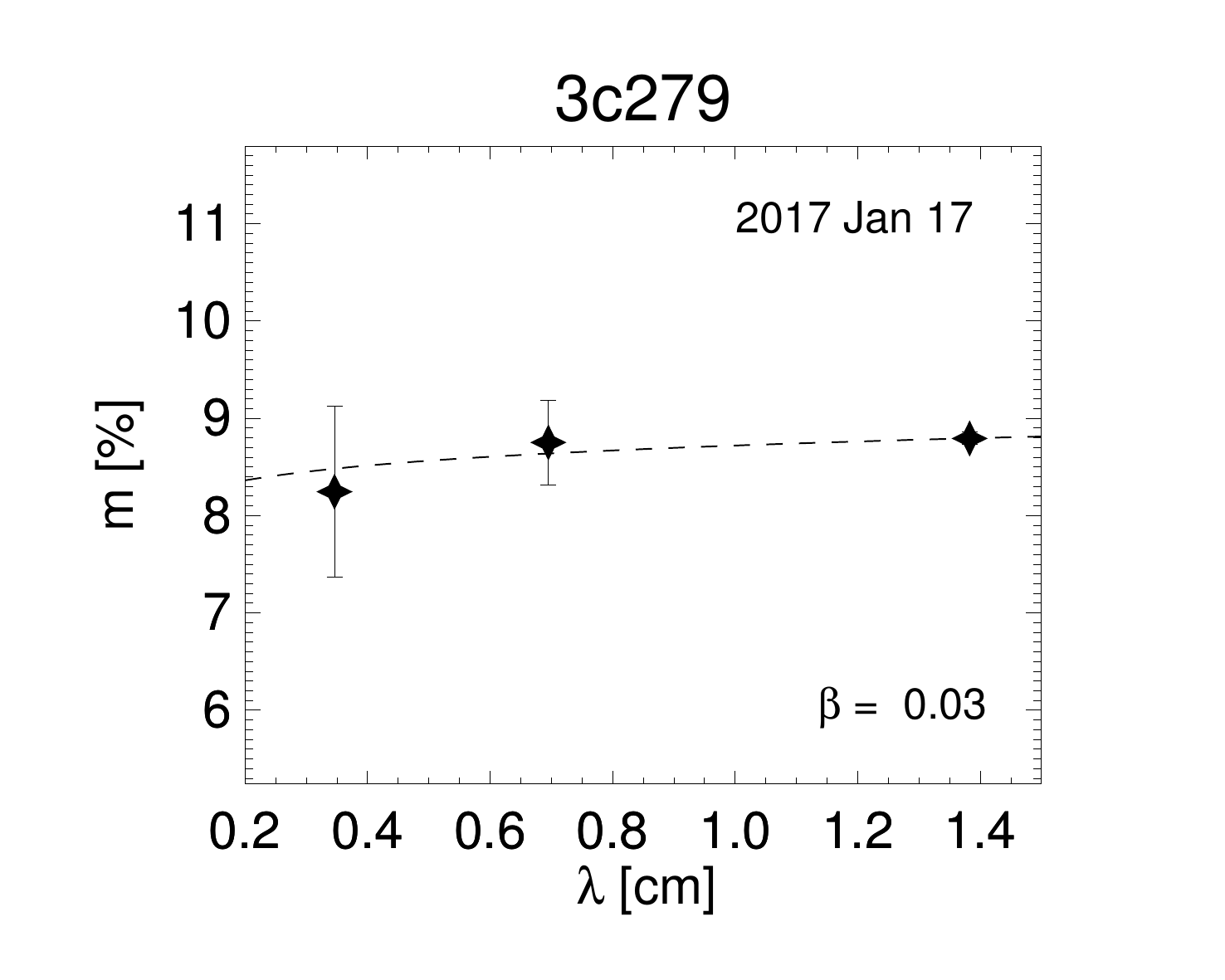}
\includegraphics[trim=10mm 8mm 18mm 4mm, clip, width = 52mm]{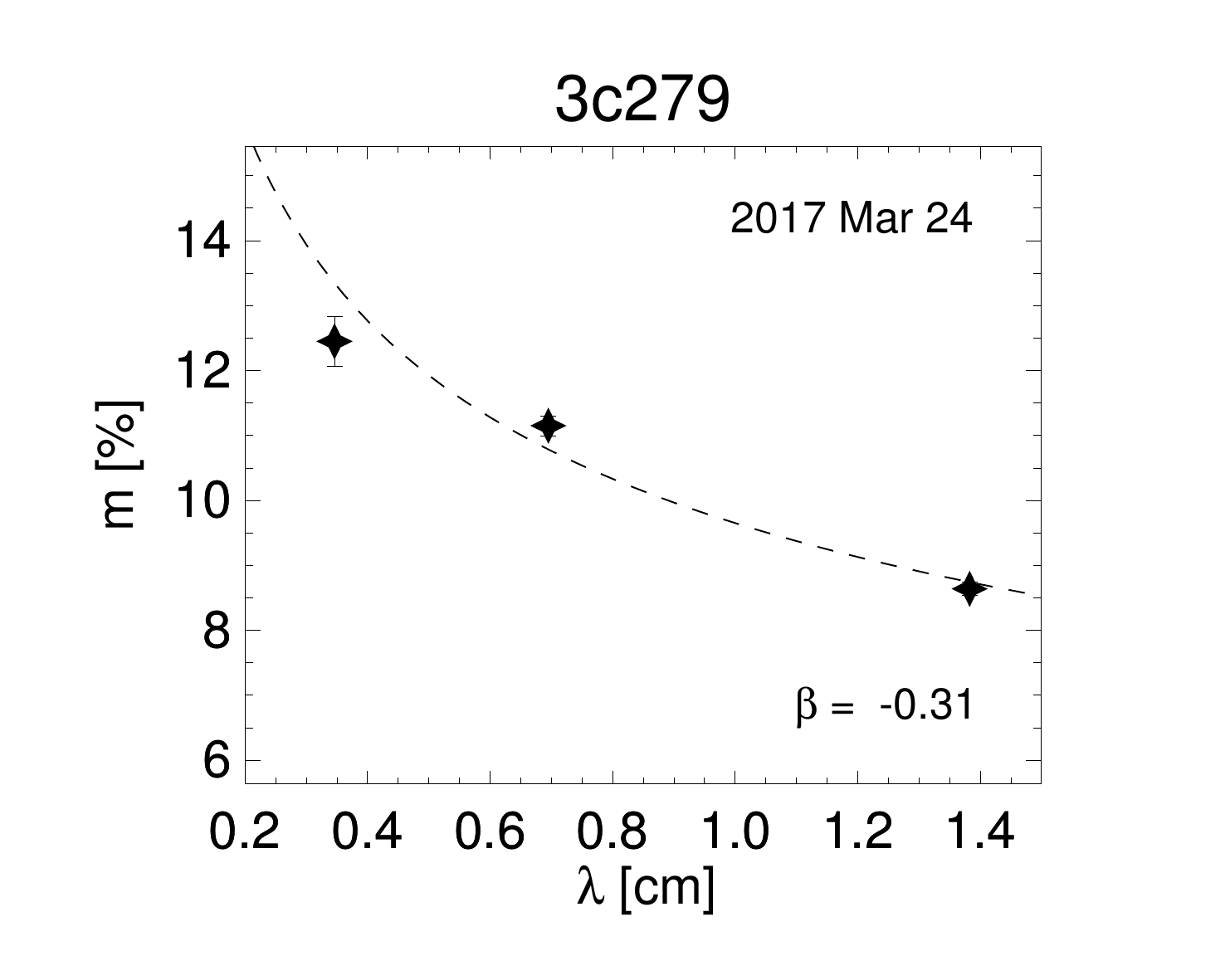}
\includegraphics[trim=10mm 8mm 18mm 4mm, clip, width = 52mm]{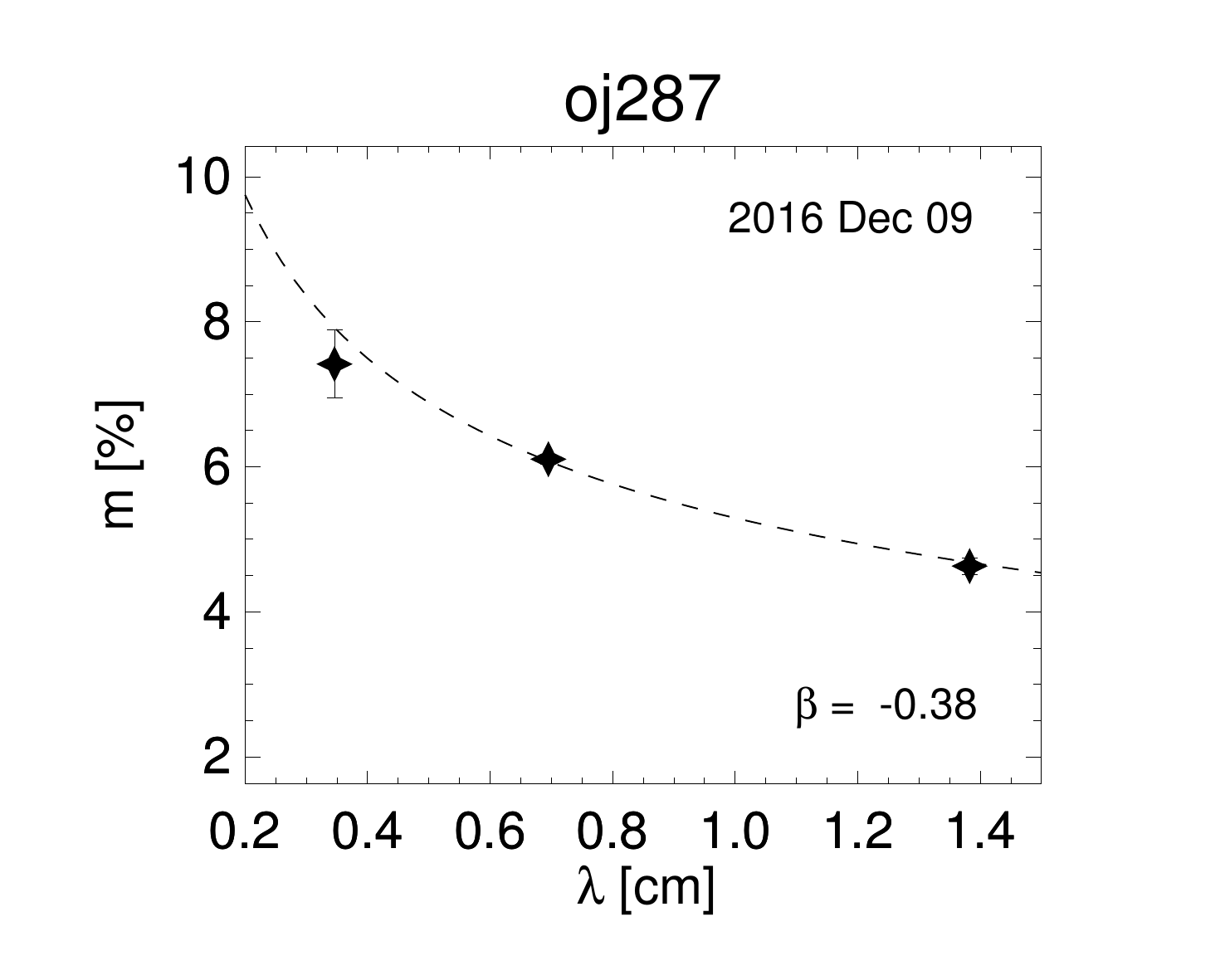}
\includegraphics[trim=10mm 8mm 18mm 4mm, clip, width = 52mm]{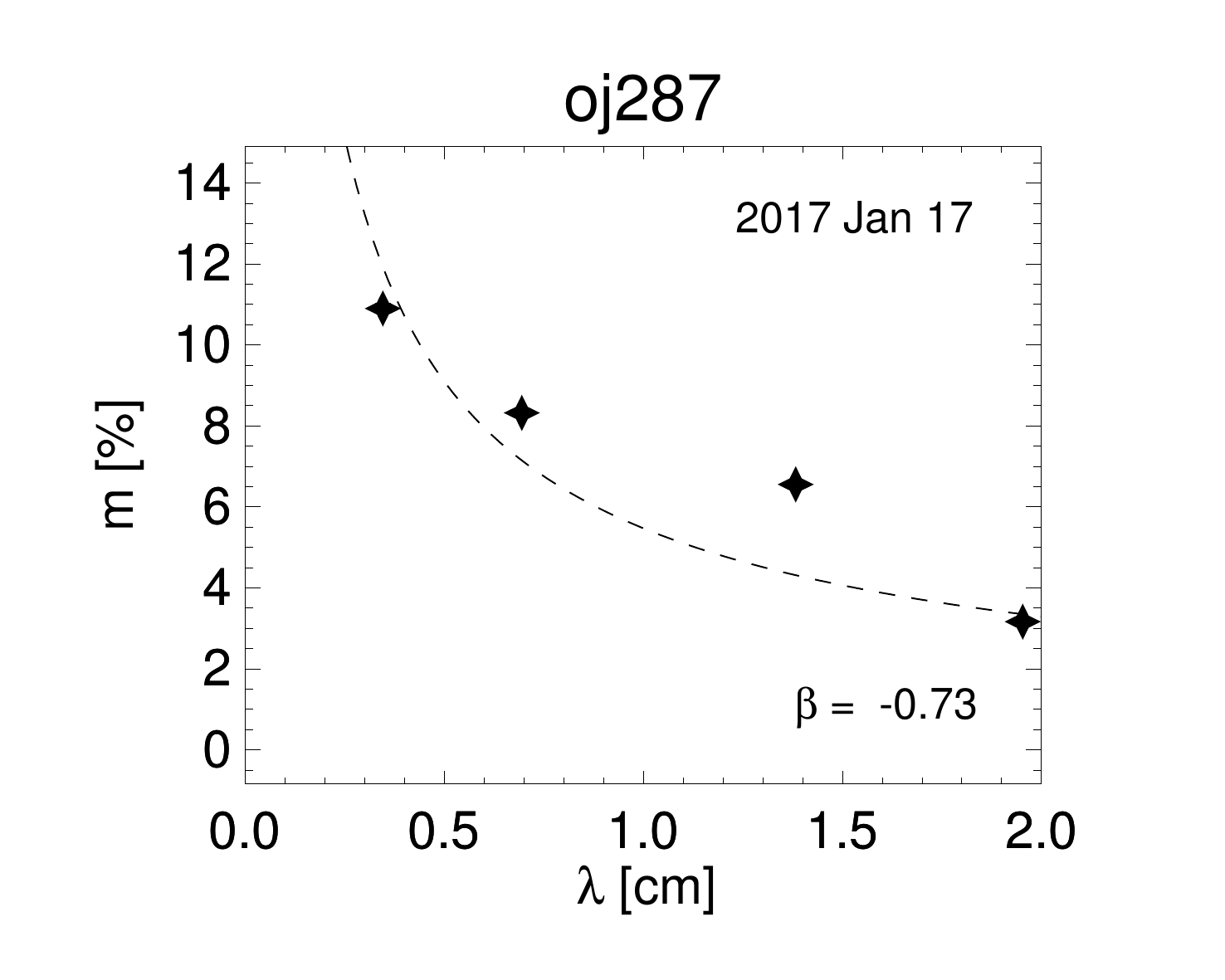}
\includegraphics[trim=10mm 8mm 18mm 4mm, clip, width = 52mm]{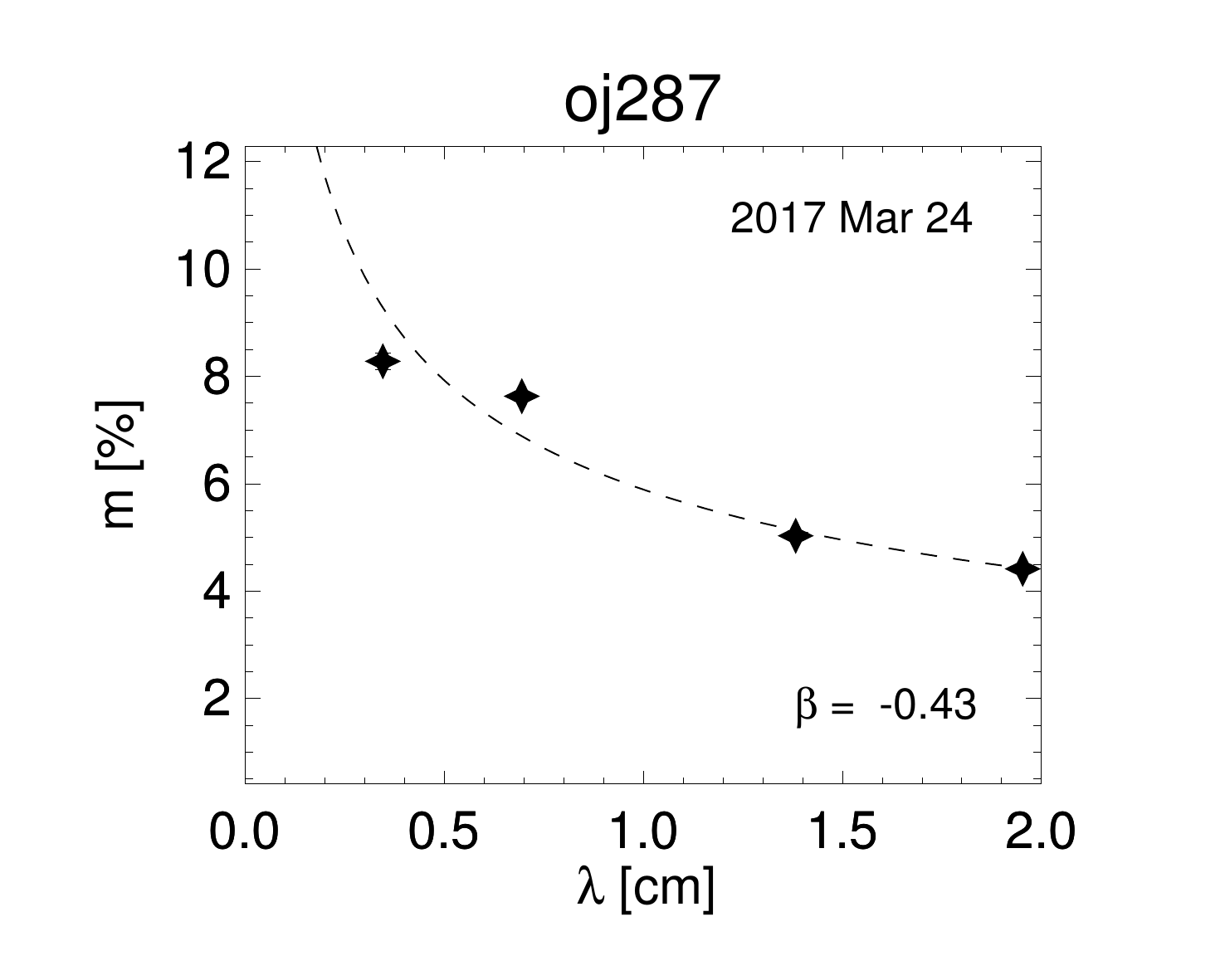}
\includegraphics[trim=10mm 8mm 18mm 4mm, clip, width = 52mm]{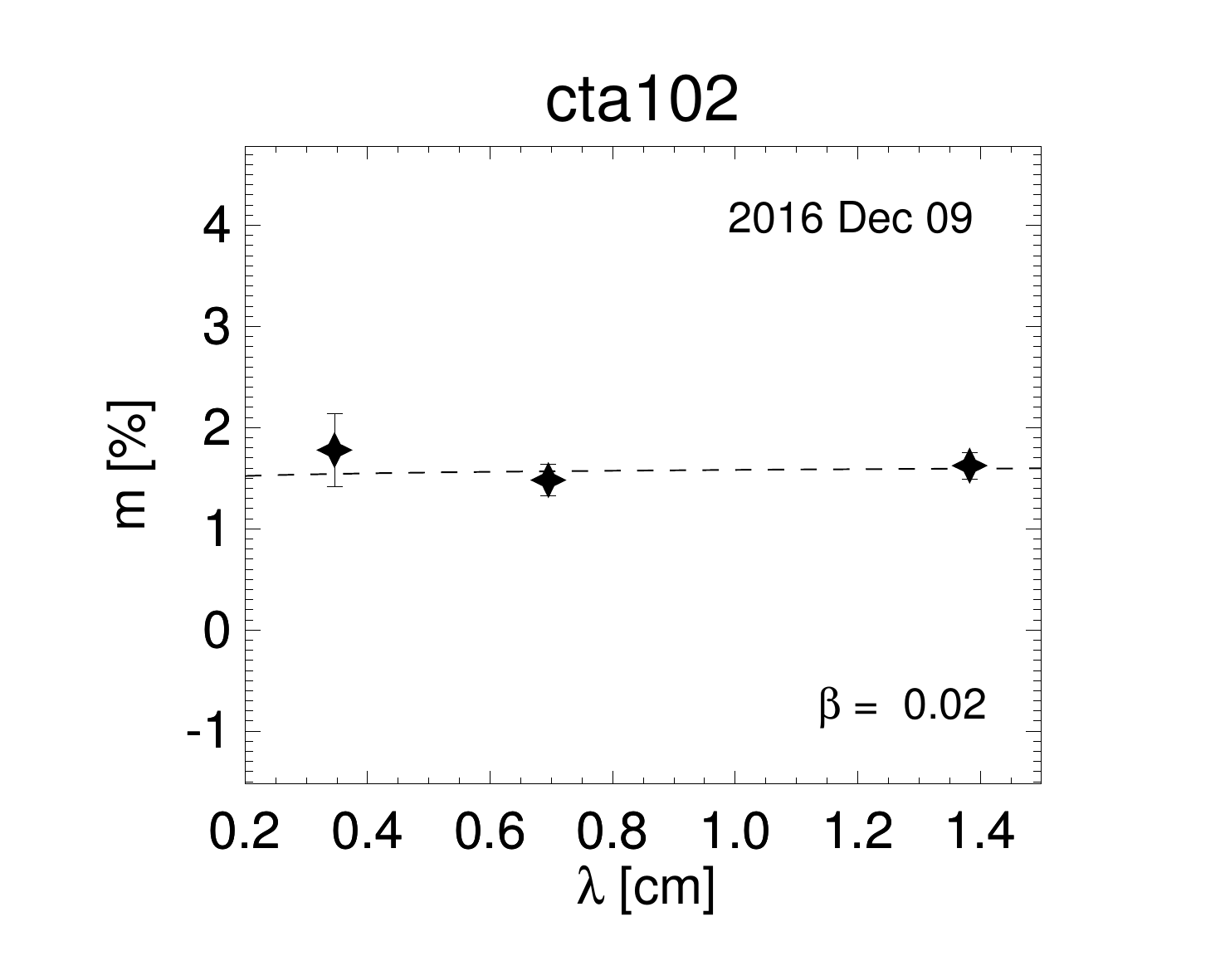}
\includegraphics[trim=10mm 8mm 18mm 4mm, clip, width = 52mm]{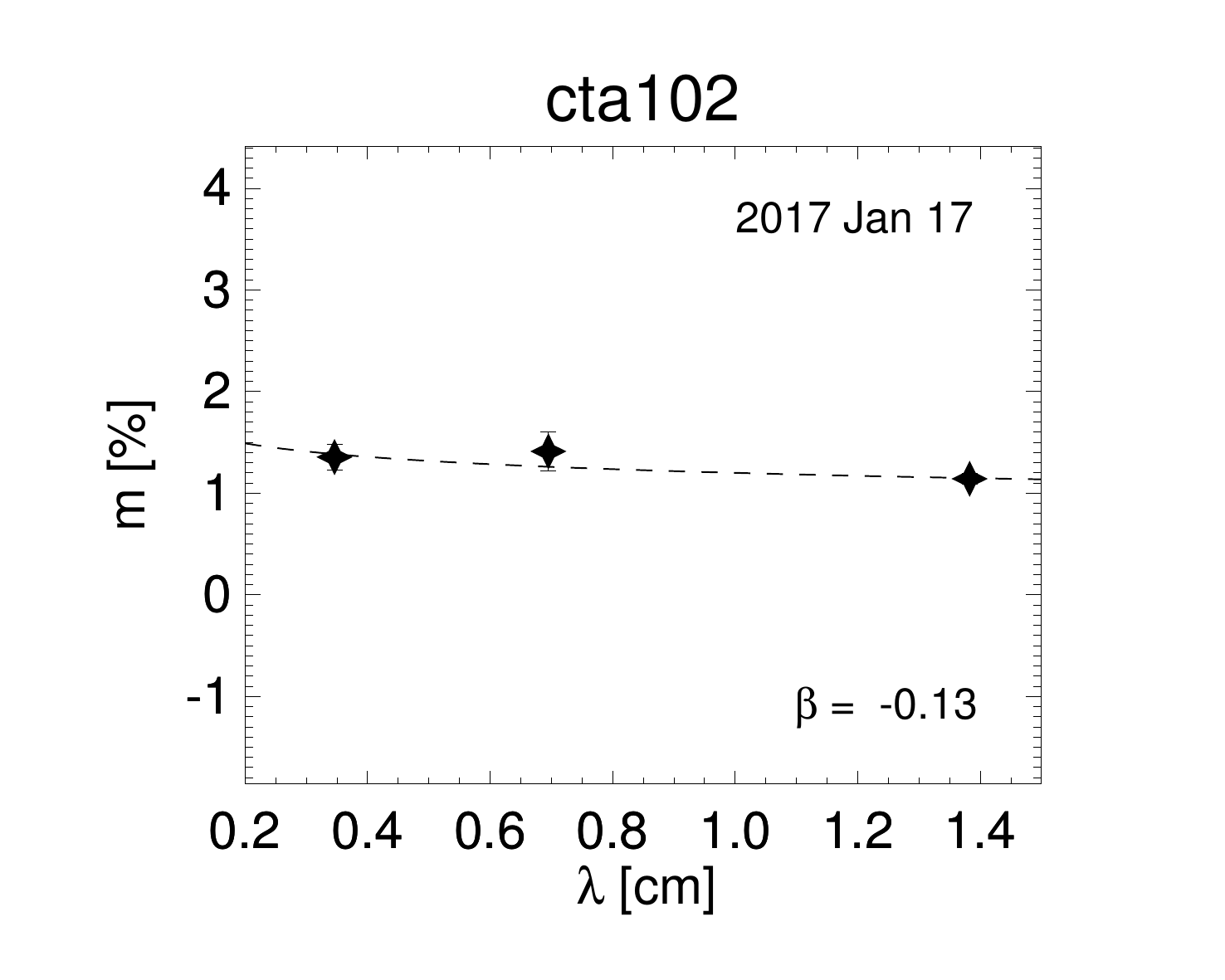}
\includegraphics[trim=10mm 8mm 18mm 4mm, clip, width = 52mm]{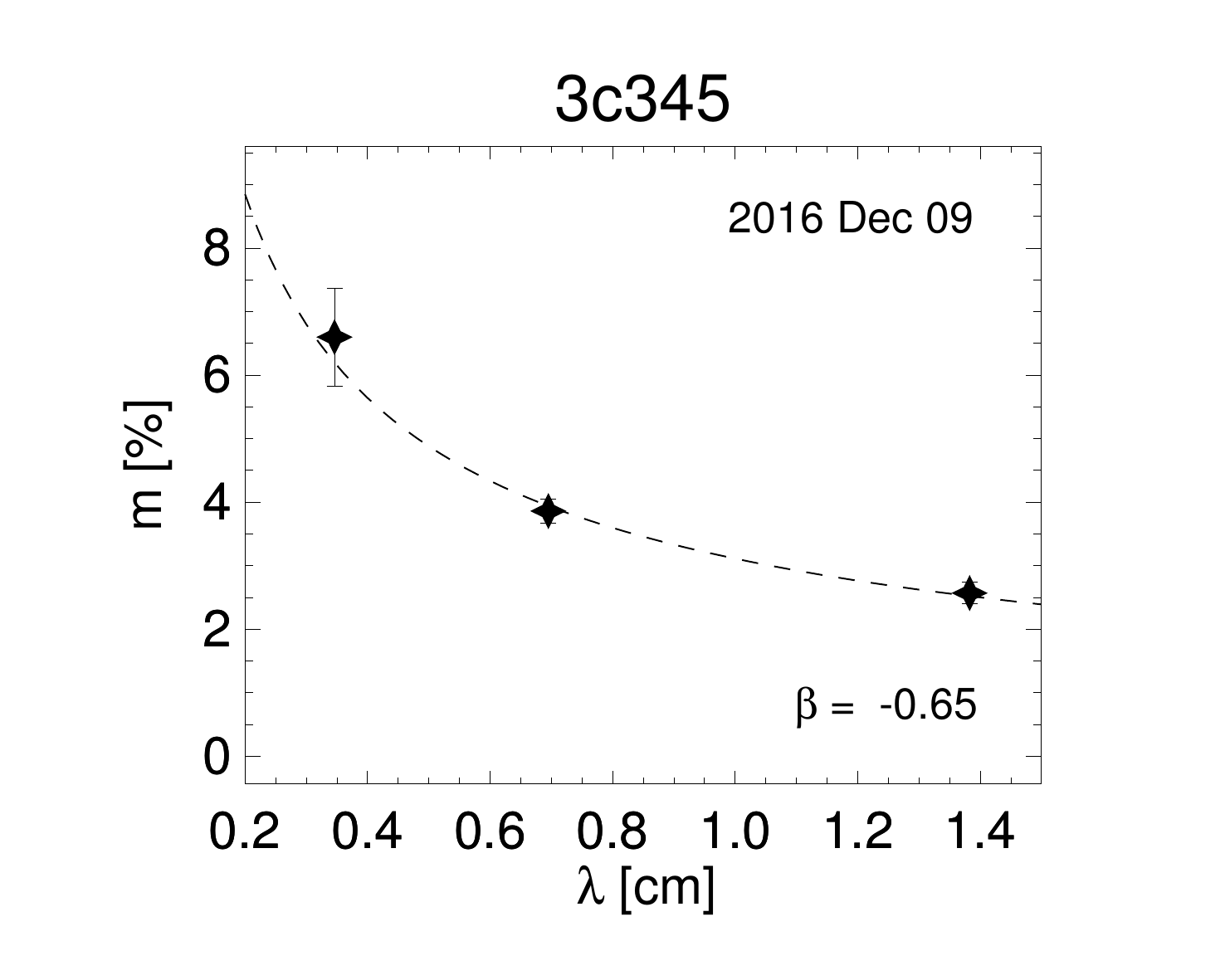}
\includegraphics[trim=10mm 8mm 18mm 4mm, clip, width = 52mm]{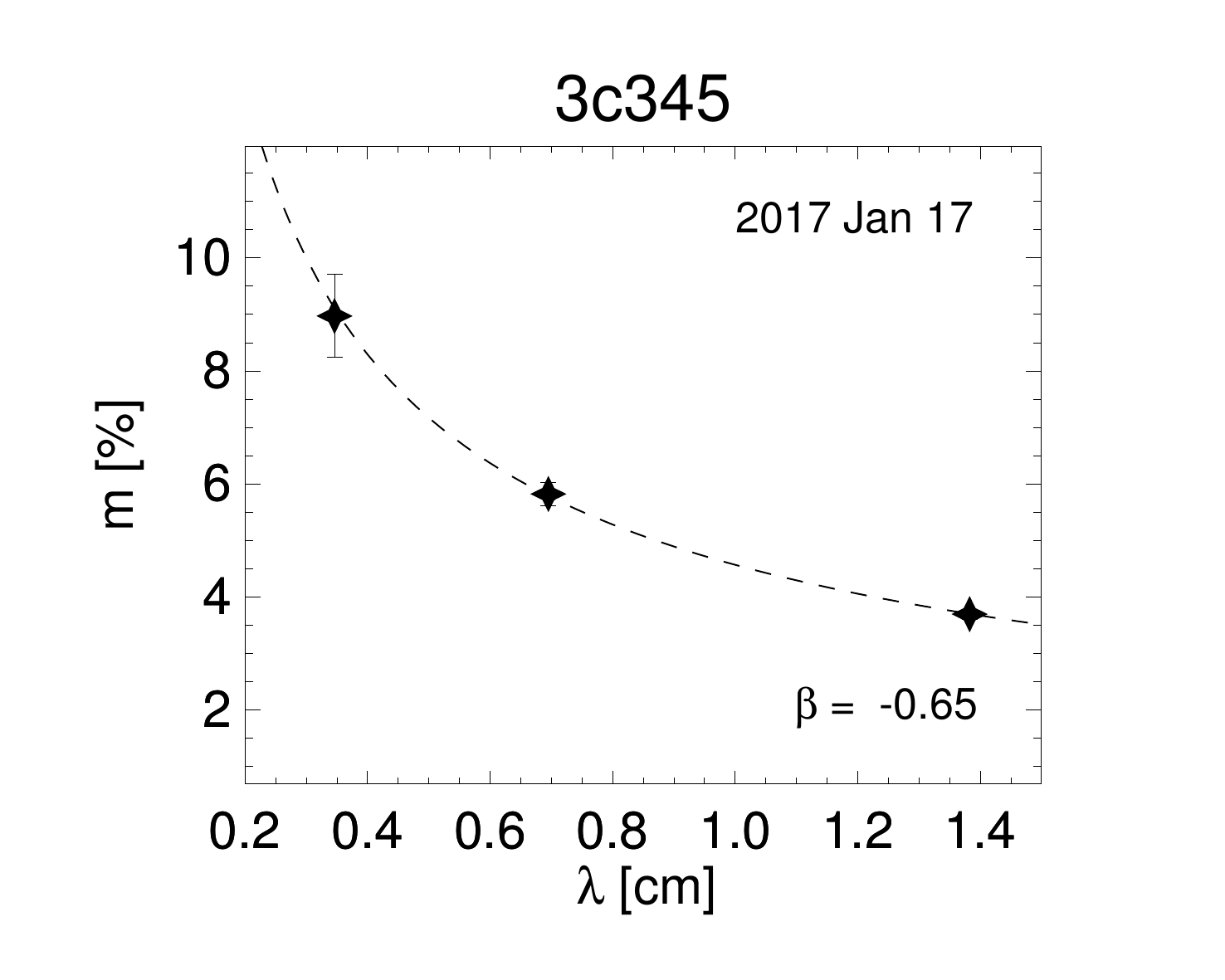}
\includegraphics[trim=10mm 8mm 18mm 4mm, clip, width = 52mm]{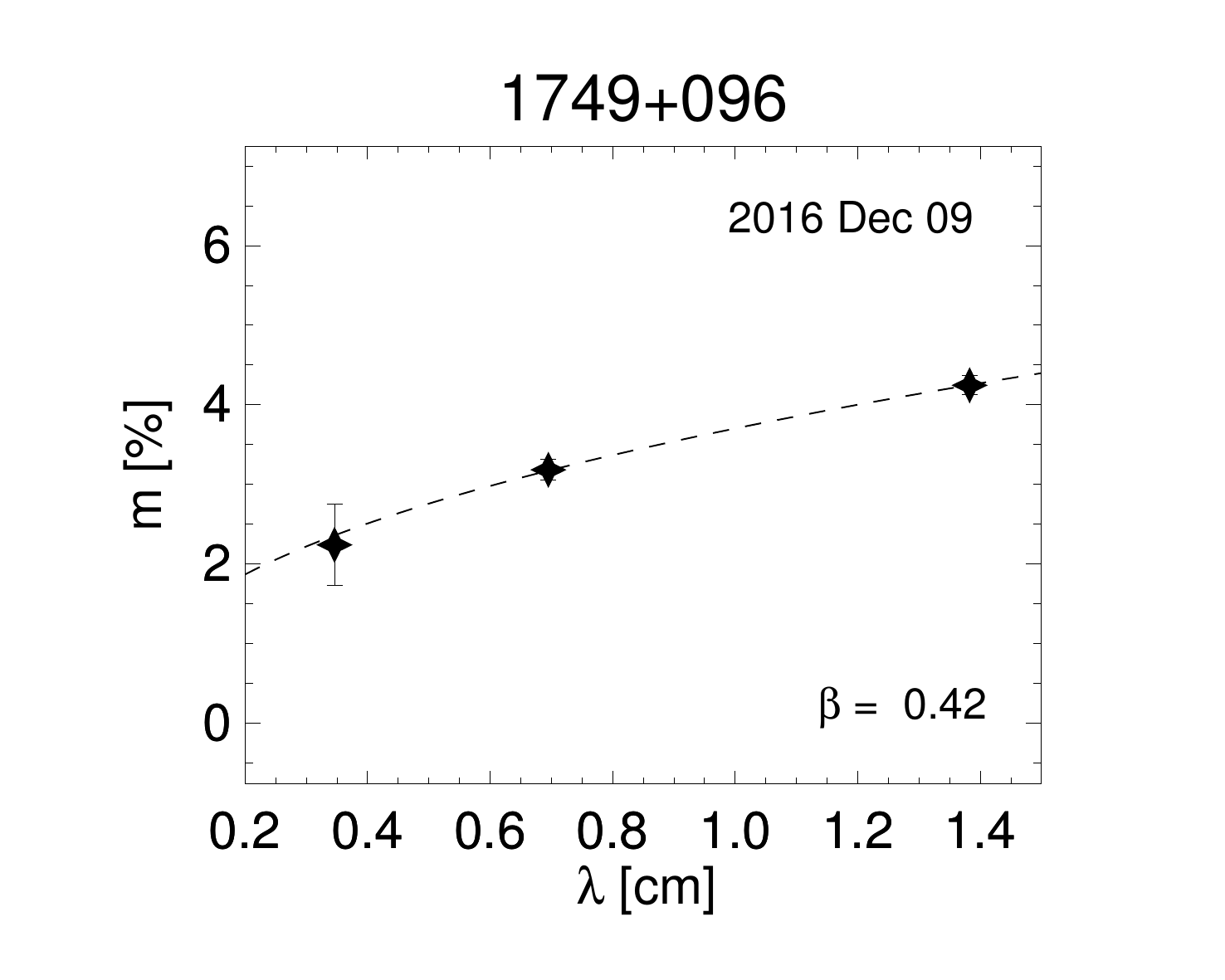}
\includegraphics[trim=10mm 8mm 18mm 4mm, clip, width = 52mm]{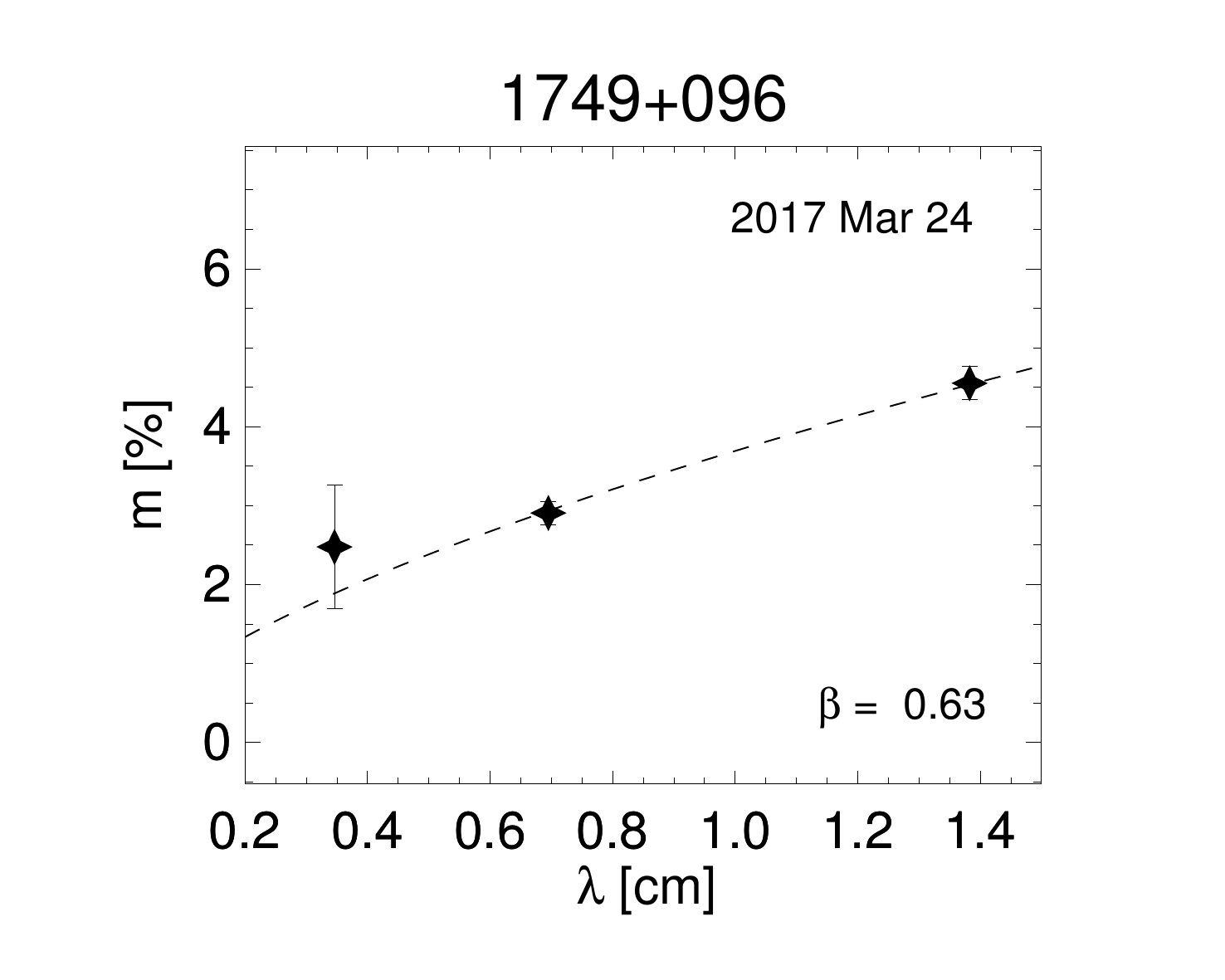}
\includegraphics[trim=10mm 8mm 18mm 4mm, clip, width = 52mm]{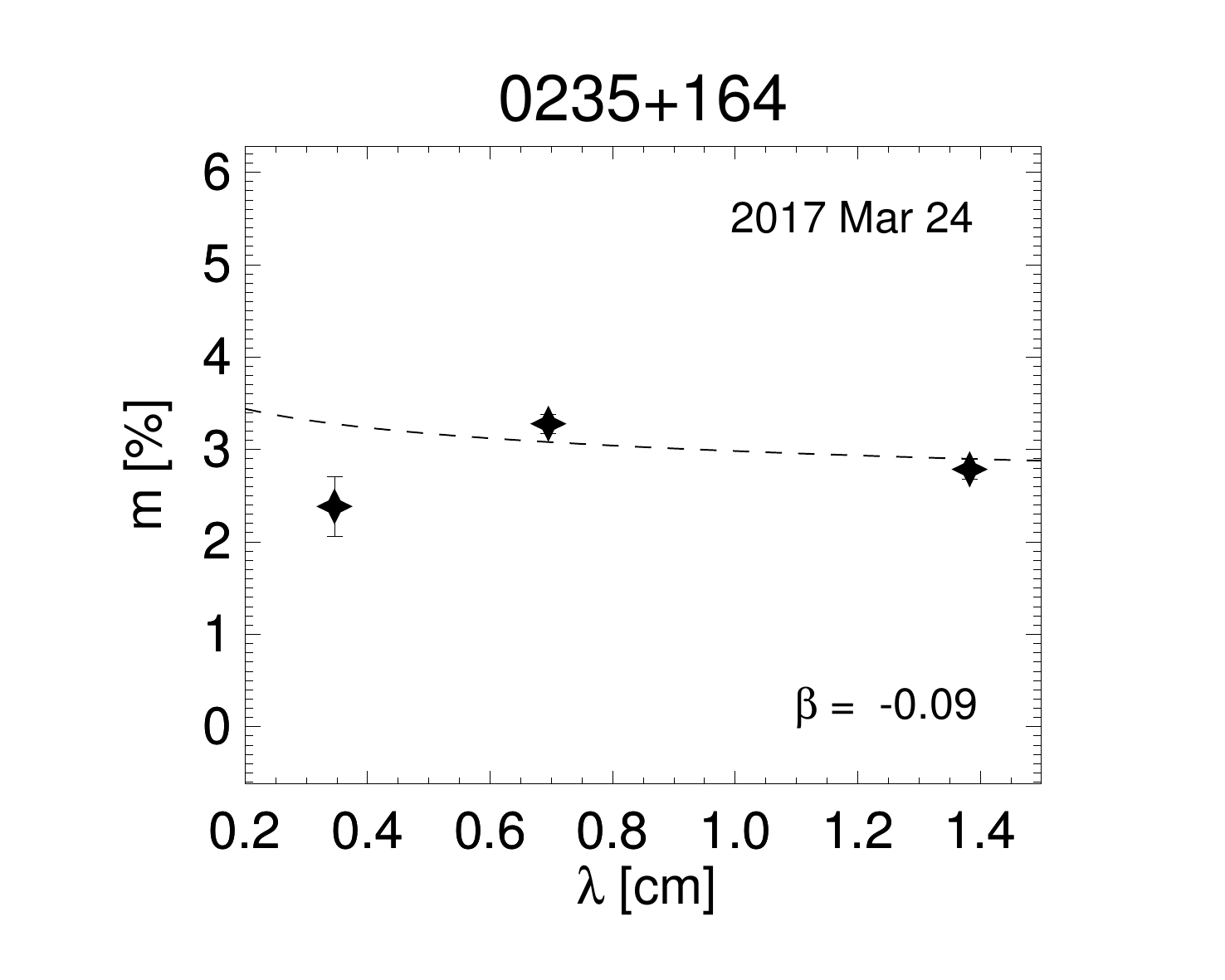}
\includegraphics[trim=10mm 8mm 18mm 4mm, clip, width = 52mm]{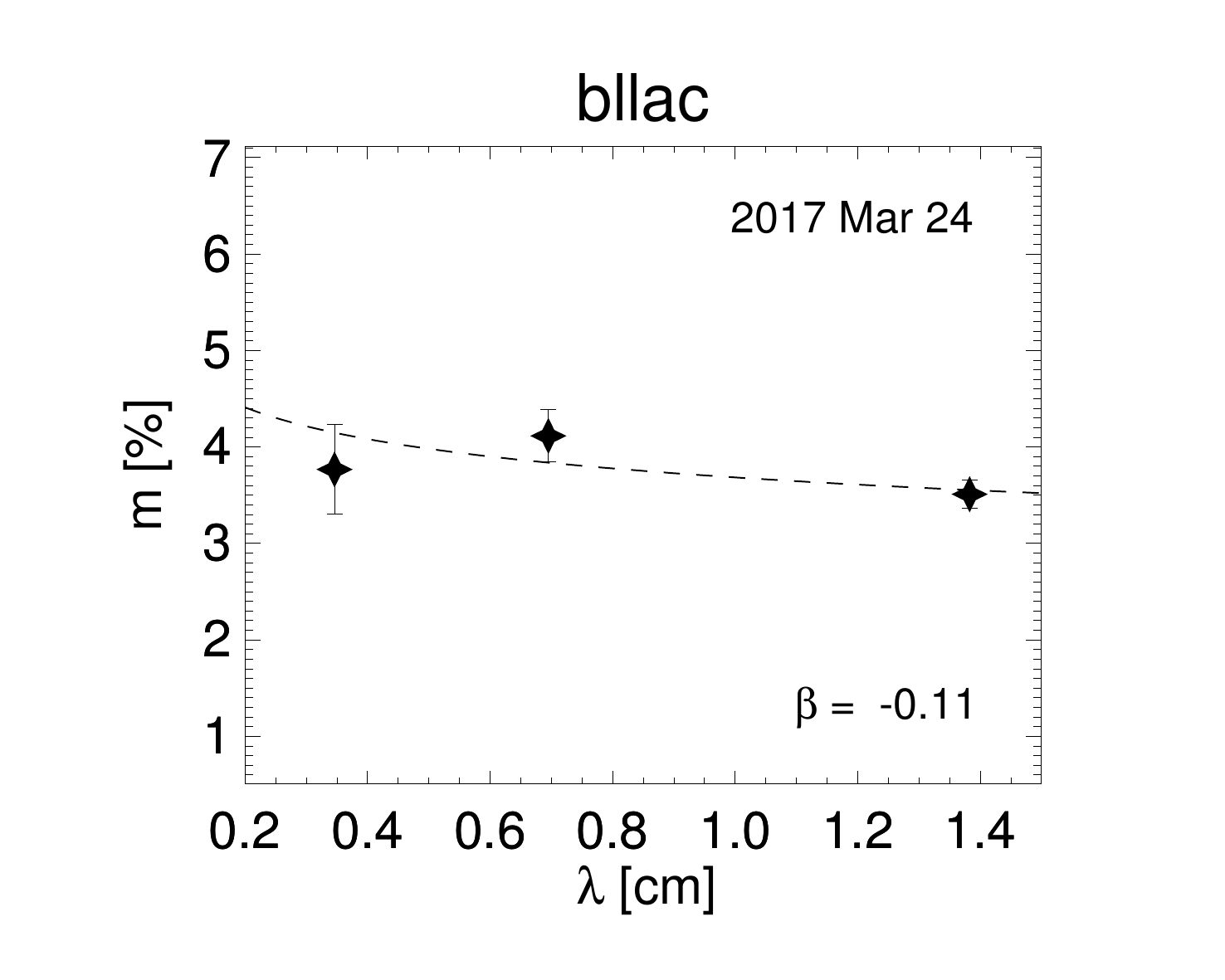}
\includegraphics[trim=10mm 8mm 18mm 4mm, clip, width = 52mm]{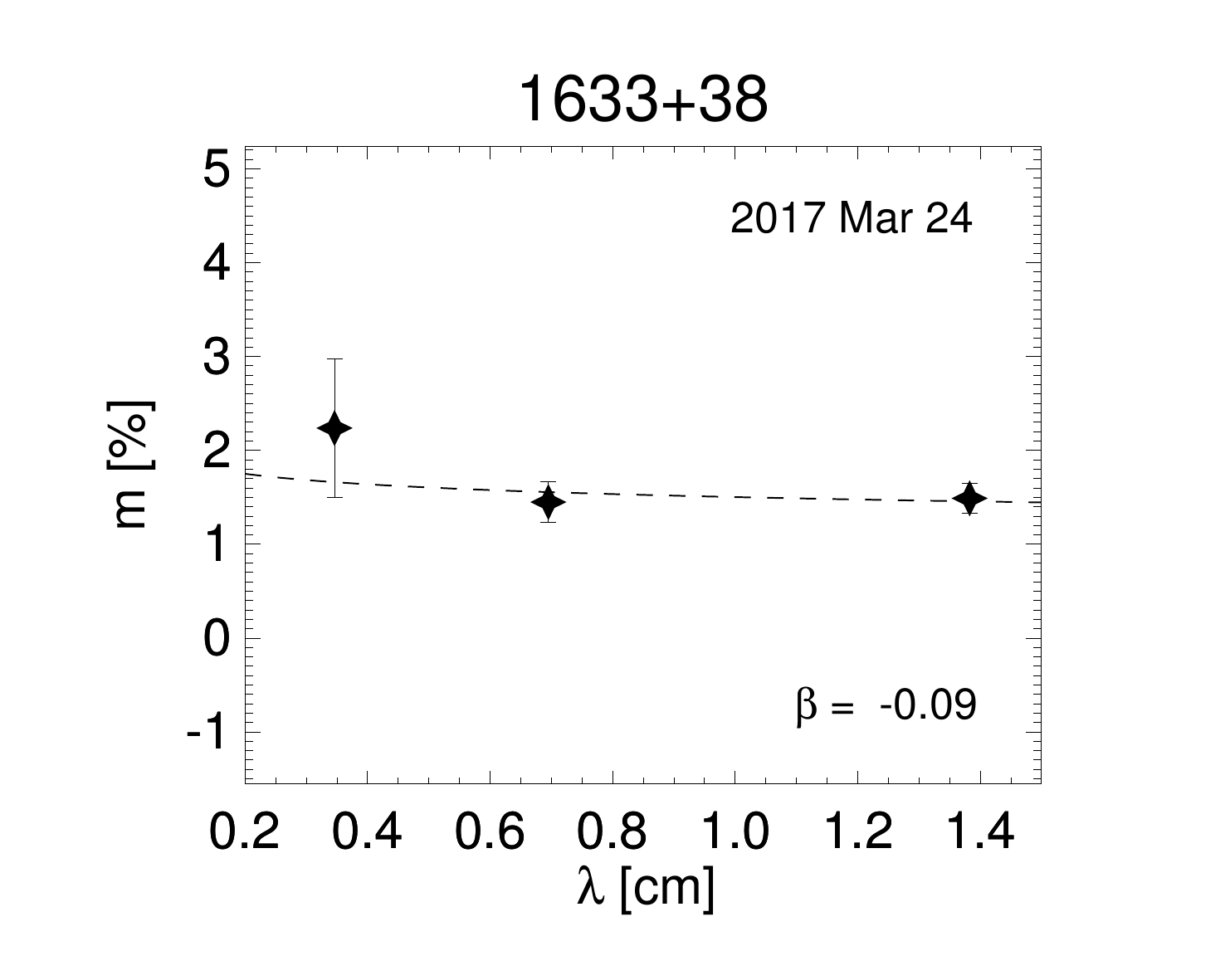}
\caption{Degree of core linear polarization as function of $\lambda$. The dashed lines are the best-fit power-law functions, $m\propto\lambda^{\beta}$, to the data points. The  polarization spectral indices, $\beta$, are noted in each panel. \label{frac}}
\end{center}
\end{figure*}

\begin{figure*}[!t]
\begin{center}
\includegraphics[trim=4mm 2mm 10mm 3mm, clip, width = 88mm]{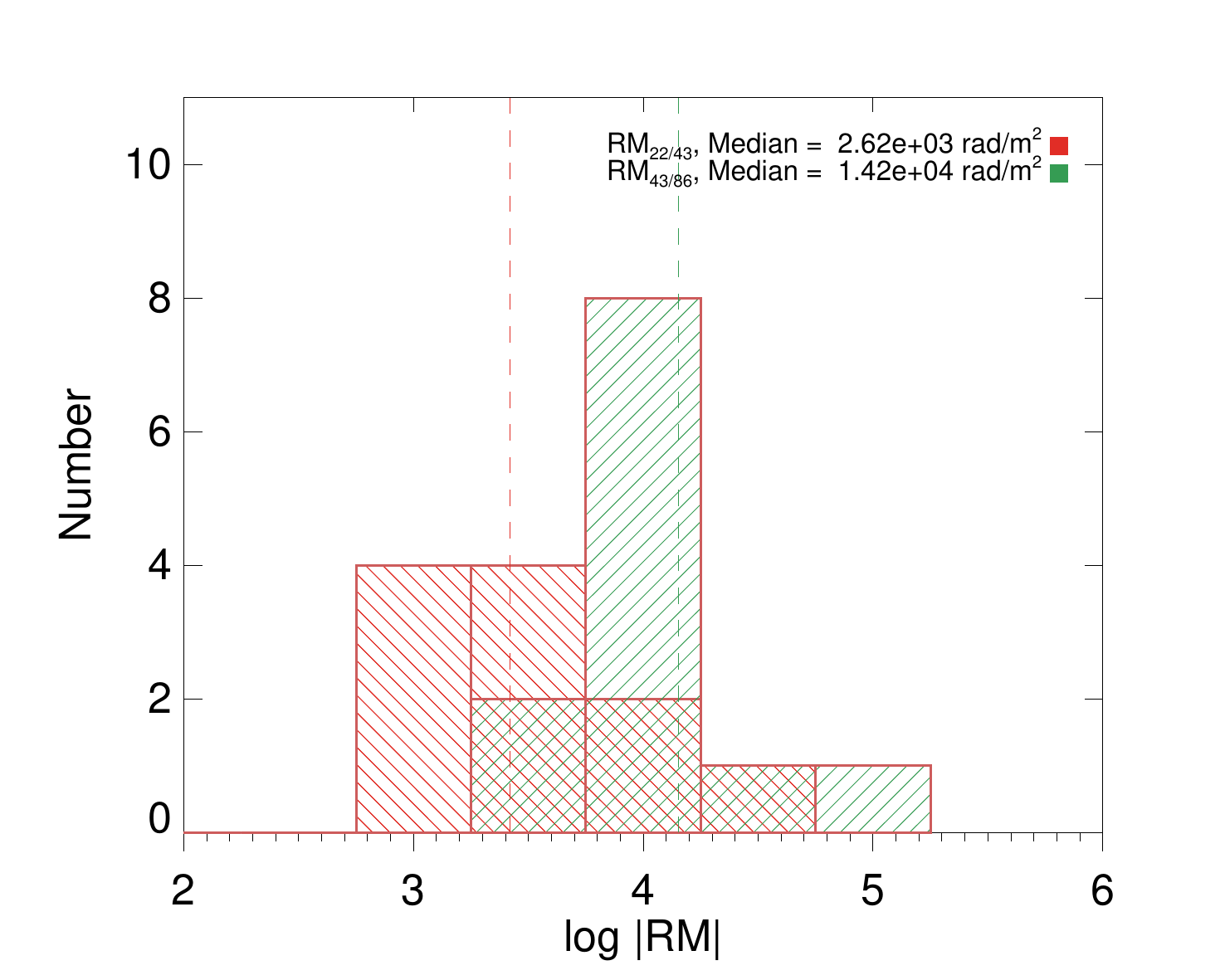}
\includegraphics[trim=4mm 2mm 10mm 3mm, clip, width = 88mm]{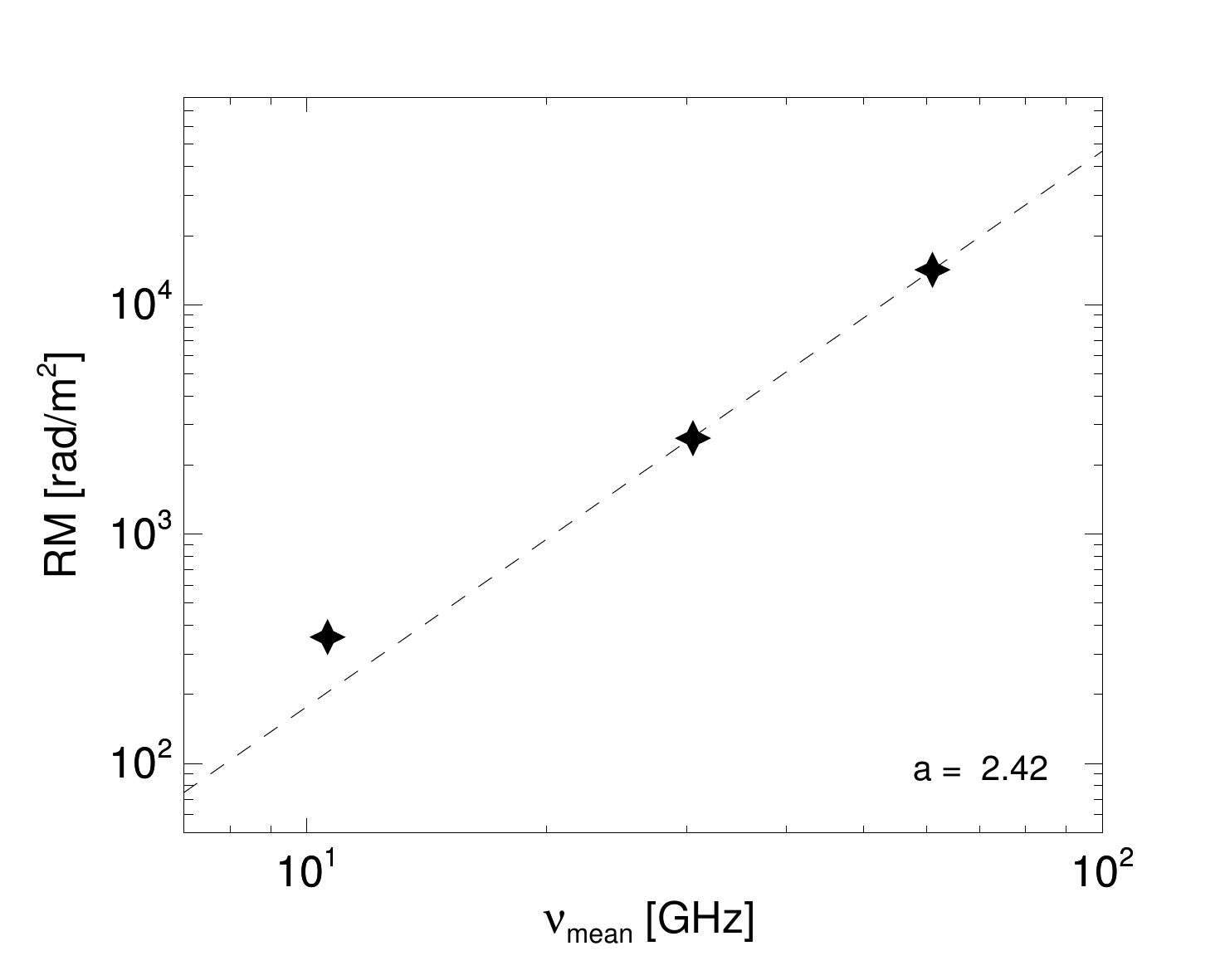}
\caption{\emph{Left:} Histrograms of the logarithms of absolute RM values between 22 and 43 GHz (red) and 43 and 86 GHz (green). Values smaller than their 1$\sigma$ errors are not included. The median values for each frequency pair are noted at the top right and are marked with vertical dashed lines. All RM values are rest frame values. \emph{Right:} Median RM values as function of geometric mean observing frequency, combined with that obtained at 8.1--15.4 GHz for the same sources used for the left histogram by \cite{hovatta2012}. The dashed line is an un-weighted power-law fit with a best-fit power-law index $a=2.42$. \label{rmhisto}}
\end{center}
\end{figure*}

Due to lack of frequency coverage between 86 GHz and optical wavelengths, we suffered from potential $n\pi$ rotation of the optical EVPAs. Therefore, we assumed that the optical EVPAs of our sources rotate in the same direction as the ones at mm wavelengths and that the EVPA rotation between 86 GHz and the optical band does not exceed $\pi$. We present the optical EVPA values of three sources in Table~\ref{Steward} and plot them with the core EVPAs from our KVN observations in the left panel of Figure~\ref{paopt}. We also show the RMs obtained from each adjacent frequency pair in the RM--frequency diagram (the right panels of Figure~\ref{paopt}). Our assumption on $n\pi$ rotation appears reasonable because the optical EVPAs follow the trend of EVPA rotation established at radio frequencies, although we cannot rule out the possibility of coincidence because of the low number of sample. The RMs obtained from the EVPA difference between 86 GHz and the optical frequencies are about an order of magnitude higher than the values obtained from the frequency pair 43/86 GHz. We note that the observed RMs between 86 GHz and optical light exceed the minimum possible measurable RM by an order of magnitude except for BL Lac for which the observed RM is about two times the minimum measurable RM.


The power-law increase of RM as a function of frequency does not continue to optical wavelengths but saturates at a certain frequency (right panel of Figure~\ref{paopt}). We used the term \emph{transition frequency}, $\nu_{\rm trans}$, for this frequency. We calculated asymmetric errors on $\nu_{\rm trans}$ via Monte-Carlo simulations by adding Gaussian random numbers to the best-fit parameters of the radio RM--$\nu$ power-law relation with standard deviations identical to their $1\sigma$ errors. The obtained $\nu_{\rm trans}$ are distributed from 138 to 591 GHz in the source rest frame for different sources and in different epochs (Table~\ref{Steward}). We note that $\nu_{\rm trans}$ for BL Lac is consistent with the observed frequency of 86 GHz within $1\sigma$ because of the relatively large minimum measurable RM.

\subsection{fractional polarization\label{fracpol}}

We present the degree of linear polarization $m$ as function of $\lambda$ in Figure~\ref{frac}. Various de-polarization models are available to explain the evolution of $m$ with wavelength (see \citealt{osullivan2012, farnes2014} for summaries). In principle, $m-\lambda$ scalings can be used to determine whether the emitting region and the Faraday screen are co-spatial or not, whether magnetic fields in the screen are regular or turbulent, whether there are multiple components with different polarization properties on scales smaller than the spatial resolution, and so on \citep{burn1966, conway1974, tribble1991, sokoloff1998}. However, we did not try to apply those models to our data because (i) our data provide sparse frequency sampling over a limited frequency range, (ii) the models are mostly appropriate for optically thin emitters while we are dealing with (partially) optically thick cores, and (iii) different observing frequencies might probe different physical regions, as suggested by the complicated $\chi-\lambda^2$ scalings of the EVPAs. Instead, we obtain a polarization spectral index $\beta$ by fitting $m\propto\lambda^{\beta}$ to our data (Table~\ref{result}, see \citealt{farnes2014}), which could be used for future theoretical studies (e.g., \citealt{porth2011}) and for comparison with observations at lower frequencies (e.g., \citealt{farnes2014}). We refer the readers to detailed studies of degree of linear polarization at different wavelengths of AGNs using broadband radio spectro-polarimetric observations (e.g., \citealt{osullivan2012, osullivan2017, hovatta2018, pasetto2018}) and investigating spatially resolved optically thin emitting regions with multi-frequency VLBI observations (e.g., \citealt{hovatta2012, kravchenko2017}). The median, mean, and standard deviation of $\beta$ are -0.11, -0.17, and 0.38, respectively. All sources show $\beta\lesssim0$ except for 1749+096 which showed $\beta\approx0.5$ in both epochs (2016 December and 2017 March).

\section{discussion}
\label{sect4}

In this section, we interpret the results of the core polarization properties of eight blazars, five flat spectrum radio quasars (FSRQs) and three BL Lac objects (BLOs).

\subsection{RM distributions at different frequencies}
\label{rmdist}

We present the distributions of the absolute RM values for different frequency pairs in the left panel of Figure~\ref{rmhisto}. We excluded RM values whose absolute values are smaller than their $1\sigma$ errors. The histograms show that the median RM increases with frequency. We note that the minimum possible measurable RMs are 440 and 2038 $\rm rad/m^2$ at 22/43 and 43/86 GHz, respectively, assuming a typical EVPA error of $2^\circ$, $3^\circ$, and $3^\circ$ at 22, 43, and 86 GHz, respectively. As is evident in Figure~\ref{rmhisto}, the RM values we found are much larger than these minimum possible measurable RMs. Notably, the trend of increasing RMs with increasing observing frequencies cannot be produced artificially.

We collected the median core RMs at cm wavelengths for our sources from \cite{hovatta2012} and show all median RM values as function of frequency in the right panel of Figure~\ref{rmhisto}. As expected, RMs increase with increasing frequency (355, 2620, and 14200 $\rm rad/m^2$ for 8.1--15.4, 22--43, and 43--86 GHz, respectively). Un-weighted fitting of a power-law function returns a best-fit power-law index $a=2.42$. Although the sample size is small and the standard deviations of the RM distributions are large, the obtained power-law index is quite close to $a = 2$, indicating that Faraday rotating media of blazars core can be represented as conical outflows statistically \citep{jorstad2007}.

\begin{figure}[!t]
\begin{center}
\includegraphics[trim=16mm 6mm 10mm 2mm, clip, width = 88mm]{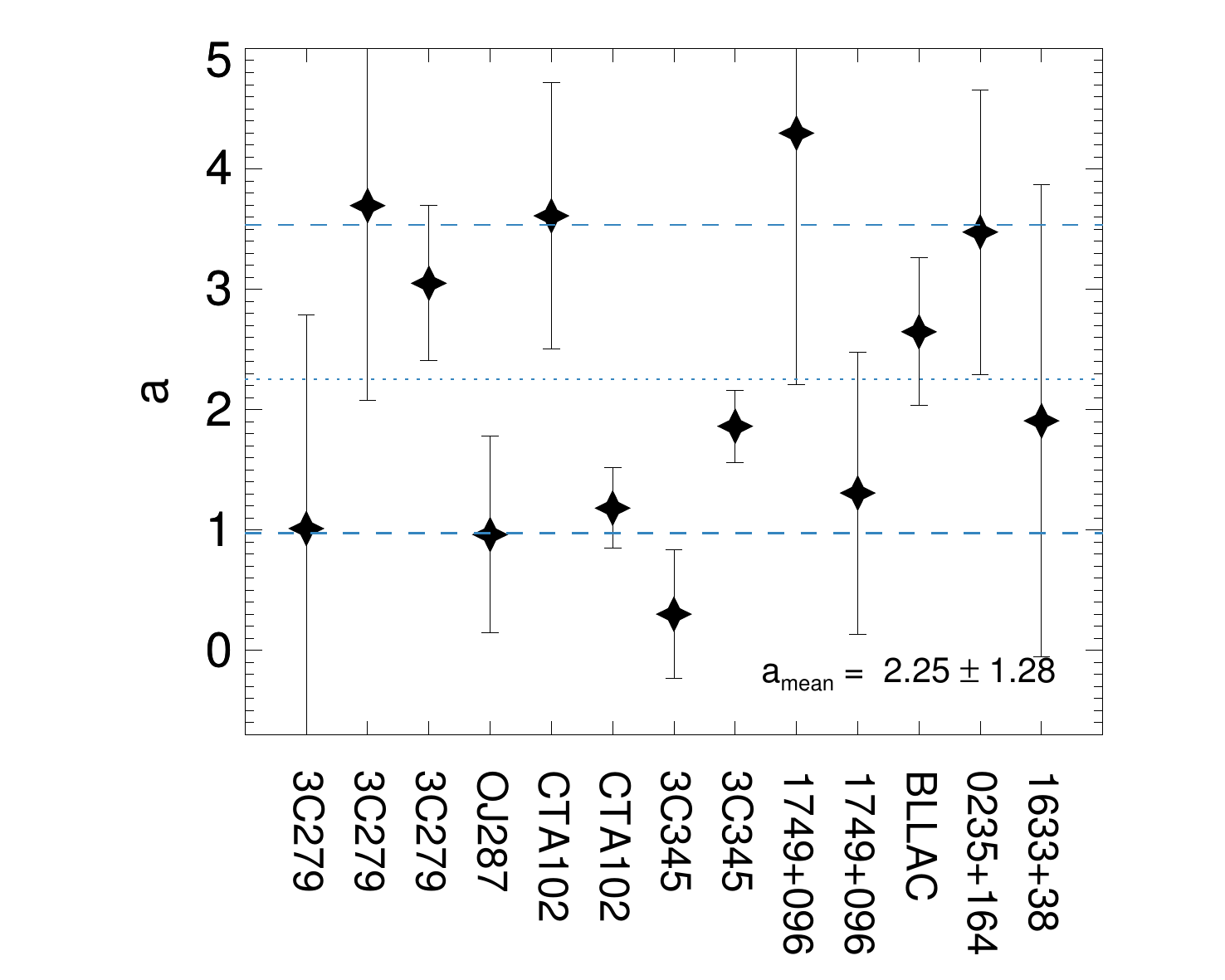}
\caption{$a$ values obtained from fitting ${\rm |RM|}\propto\nu^a$ to the data of each source (right panels of Figure~\ref{result}). Multiple values for the same source correspond to different epochs. The blue dotted and dashed lines mark the mean and standard deviation of the full set of measurements, respectively; the values are noted on the bottom right. \label{aplot}}
\end{center}
\end{figure}

Instead of comparing RM distributions of all sources at different frequencies, we collected the power-law indices $a$ obtained for each source in Figure~\ref{aplot}. We have 13 measurements in total, with some sources having more than one measurement. The mean and standard deviation of all $a$ values are $a = 2.25 \pm 1.28$, which is consistent with $a=2$ and the fitting results for the median values of RM distributions at different frequencies. Our results are also consistent with previous studies of blazars at both cm and mm wavelengths (e.g., \citealt{jorstad2007, OG2009a, algaba2013, kravchenko2017, hovatta2018}). However, many $a$ values are located far from the mean value, which potentially indicates a bimodal distribution. Assuming a power-law electron density distribution as function of jet distance ($d$), $N_e \propto d^{-a}$, toroidal magnetic fields dominant in the Faraday screen, $B\propto d^{-1}$, a conical geometry of the Faraday screen, $dl\propto d$ and $a=2$, and energy equipartition, $d_{\rm core} \propto \nu^{-1}$, one obtains ${\rm RM_{core}} \propto \int N_eBdl \propto \nu^2$ \citep{jorstad2007}. If some of these assumptions are not satisfied, one might expect deviations from a $a=2$ scaling. For example, there is growing evidence for a parabolic geometry of the blazar cores (e.g., \citealt{algaba2017,pushkarev2017}). In some cases, a need for helical magnetic fields instead of dominant toroidal fields in blazars has been pointed out (e.g., \citealt{zamaninasab2013}). Likewise, the assumption of energy equipartition between radiating particles and magnetic fields may not hold for some sources (e.g., \citealt{homan2006}). However, it is difficult to determine accurate $a$ values for each source with the current data only due to source variability and relatively large errors in $a$. For example, the values of $a$ are likely related with source's flaring activity as seen in the case of CTA 102 and 3C 345 (see Section~\ref{cta102} and ~\ref{3c345}). We expect that our monthly monitoring program will allow us to investigate the reason for potential difference in $a$ values for different sources and in different epochs, together with the detailed information of the compact core region provided by the ultra high resolution arrays such as the Global Millimeter-Very Long Baseline Interferometry Array (GMVA, see e.g., \citealt{kim2016}) or the Radioastron space VLBI (e.g., \citealt{gomez2016}).

\subsection{Change of core opacities from optically thick to thin\label{transition}}

In Section~\ref{optdata}, we show that the power-law increase of RM as a function of frequency might not continue to optical wavelengths but flatten out at a certain frequency, $\nu_{\rm trans}$. We suggest that the cores of blazars become fully transparent at $\nu > \nu_{\rm trans}$, meaning no core-shift and thus no more frequency dependence of RM at those frequencies. Accordingly, the radio core may be a standing recollimation shock at $\nu > \nu_{\rm trans}$. For CTA 102, $\nu_{\rm trans}$ increased substantially from $\approx240$ GHz to $\approx590$ GHz within one month, albeit within large errors (Table~\ref{Steward}). This might be related to a strong flare that occurred at the time of our observations (see Section~\ref{cta102}) which ejected a large amount of relativistic electrons into the core, causing it to become optically thick. We obtained $\nu_{\rm trans} \approx 210$ and 140 GHz for 3C 279 and BL Lac, respectively. This result seems to be in line with recent astrometric observations of BL Lac which found a systematic deviation of the amount of core-shift from the one expected for a Blandford--K\"{o}nigl type jet at 22/43 GHz \citep{dodson2017}. Likewise, the scaling of synchrotron cooling time with frequency in BL~Lac matches a standing shock better than an optically thick jet \citep{kim2017}. In summary, one may expect no frequency dependence of RM and no core-shift above $\approx140$ GHz for BL Lac and above $\approx210$ GHz for 3C 279 and CTA 102. However, we stress that the conclusions presented in this section are valid only when the assumptions of (a) no $n\pi$ ambiguity and (b) EVPA rotations in the same sense from mm to optical hold. We will study core opacity evolution and RM saturation further with dedicated upcoming multi-frequency observations at mm and sub-mm, combining data from KVN and ALMA (Park et al., in preparation).

\subsection{The Faraday screen\label{sectscreen}}

Identifying the source of Faraday rotation is very difficult. As the amount of Faraday rotation is inversely proportional to the square of the mass of charged particles, thermal electrons and/or low-energy end of radiating non-thermal electrons would be the dominant source of the observed RM. If those Faraday rotating electrons are mixed in with the emitting plasma in jets, internal Faraday rotation occurs. However, if the rotating medium is located outside the jet, e.g., in a sheath surrounding the jet or the broad/narrow line regions (BLRs/NLRs), then the observed Faraday rotation is external to the jet. Theoretical models assuming an optically thin jet with spherical or slab geometries showed that it is very difficult for internal Faraday rotation to cause EVPA rotations larger than $45^{\circ}$ without severe depolarization \citep{burn1966}. Multiple studies showed that many blazars indeed have EVPA rotations larger than $45^{\circ}$ without significant depolarization, indicating that the source of Faraday rotation is external to the jets usually (e.g., \citealt{ZT2003, ZT2004, jorstad2007, OG2009a, hovatta2012}). A sheath surrounding the jet is considered to be the most viable candidate for an external Faraday rotating medium; in addition, BLRs/NLRs are unlikely sources of RM given the time variability of RMs in jets and volume filling factor arguments \citep{ZT2002, ZT2004, hovatta2012}. Nevertheless, there is indication for potential internal Faraday rotation in some sources \citep{hovatta2012}.

We cannot identify the Faraday screen from our data because of their limitations (Section~\ref{fracpol}). Nevertheless, we note that the observed RM--frequency relations having $a\approx2$ (Section~\ref{rmdist}) and the polarization spectral indices being predominantly negative ($\beta\lesssim0$) for our sources support the conclusion of previous studies that an external jet sheath acts as Faraday screen \citep{ZT2002, ZT2004, hovatta2012}. However, for 1749+096 we observed the degree of fractional polarization at high frequencies to be smaller than the one at lower frequencies,  with $\beta\approx0.5$ in both epochs (2016 December 9 and 2017 March 24) -- which is almost impossible to explain with any standard external depolarization model \citep{hovatta2012}. Such `inverse depolarization' can be due to blending of different polarized inner jet components at different frequencies \citep{conway1974} or internal Faraday rotation in helical or loosely tangled random magnetic field configurations \citep{homan2012}.

\subsection{RM sign change}
\label{signchange}

We observed a RM sign change for CTA 102 within $\approx$1 month, while the absolute values of RM did not change much (Figure~\ref{result}). Previous studies found temporal sign reversals in RMs for other sources (e.g., \citealt{mahmud2009, lico2017}), sign reversals between core and jet (e.g., \citealt{mahmud2013}), and sign reversals in the cores at different frequencies intervals (e.g., \citealt{OG2009a}). Scenarios proposed to explain such RM sign changes include: (i) a reversal of the magnetic pole of the black hole facing the Earth; (ii) torsional oscillations of the jet; (iii) a `nested-helix' magnetic field structure; and (iv) helical magnetic fields in jets seen at different orientations due to relativistic abberation, depending on whether $\theta\Gamma$ is larger or smaller than 1, where $\theta$ is the viewing angle and $\Gamma$ is the bulk Lorentz factor of jets (see \citealt{mahmud2009, mahmud2013} for (i)-(iii) and \citealt{OG2009a} for (iv) for details). Although all scenarios are possible theoretically, we focus on the fact that CTA 102 underwent a relatively strong flare in the period of our KVN observations (Section~\ref{cta102}).

Evidence for the presence of helical magnetic fields in AGN jets has been provided by many studies, starting with the detection of a transverse RM gradient in the jet of 3C 273 \citep{asada2002} which was later confirmed by other studies \citep{ZT2005, hovatta2012}. Similar behaviour has been found in many BL Lac objects (e.g., \citealt{gabuzda2004, gabuzda2015}), radio galaxies (e.g., \citealt{kharb2009}), and quasars (e.g., \citealt{asada2008, algaba2013, gabuzda2015}). Furthermore, general relativistic magnetohydrodynamic simulations of AGN jets showed that the combination of the rotation of the jet base and the outflow leads to the generation of a helical field and associated Faraday rotation gradients \citep{BM2010}. If helical magnetic fields pervade in the jet sheaths and if they are the main contributor of the observed RMs as speculated in Section~\ref{sectscreen}), the sign of RMs would be determined by whether $\theta\Gamma$ is larger or smaller than 1, as explained in \cite{OG2009a}. 

As noted in Section~\ref{cta102}, a strong flare at multiple wavelengths occurred during our KVN observations \citep{raiteri2017}. Flares in blazars are usually associated with new VLBI components emerging from the cores (e.g., \citealt{savolainen2002}). The flare in CTA 102 would then likewise be connected to newly ejected VLBI components. \cite{jorstad2005} found $\theta = 2.6^{\circ}$ and $\Gamma=17.2$ for CTA 102, which yields $\theta\Gamma = 0.78$. If there is bending and/or acceleration or deceleration of the ejected component, changes of $\theta\Gamma$ across the value $\theta\Gamma=1$ can occur and the sign of RM reverses. Assuming scenario (i) or (ii) as mechanism behind the sign reversal requires coincidence with the recent strong flaring activity of this source. Furthermore, scenario (iii) can be related to flaring since a new jet component might lead to temporal increase of the relative contribution of the inner field to the outer field in the magnetic tower model. However, in this case it is difficult to explain the observation of similar RM magnitudes in the two epochs (a few times $10^4\rm\ rad/m^2$ for CTA 102); the relative contributions by the inner and outer magnetic fields to the observed RMs must be almost exactly opposite in different epochs, which, again, would be a coincidence (but see \citealt{lico2017} for the case of Mrk 421 which supports this scenario). Therefore, we conclude that scenario (iv) provides the most natural way to explain the observed sign change in RMs of CTA 102 as it does not require substantial changes in the physical properties of the jets. Our interpretation is also consistent with modelling the multi-wavelength flare in this source in late 2016 with a twisted inhomogeneous jet \citep{raiteri2017}. We note that bending and acceleration/deceleration of blazar jets are quite common indeed (e.g., \citealt{lister2013}).

\subsection{Optical subclasses}
\label{sect:opticalsubclasses}

\begin{figure}[!t]
\begin{center}
\includegraphics[trim=10mm 2mm 10mm 10mm, clip, width = 88mm]{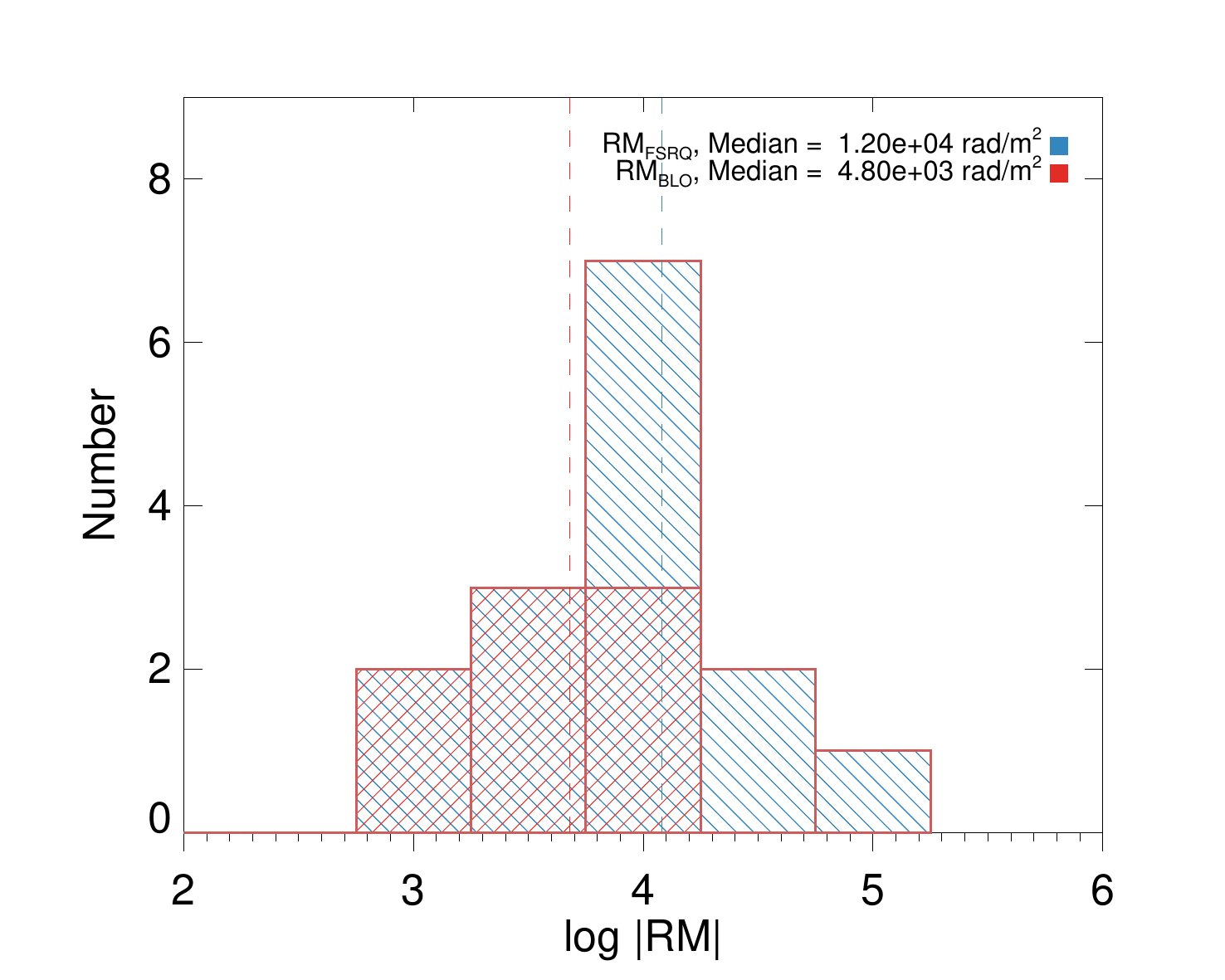}
\caption{Histograms of the logarithmic absolute values of RMs of FSRQs (blue) and BLOs (red). Median values are indicated by vertical dashed lines and noted at the top right. We omitted RM values consistent with zero (within $1\sigma$). All the RM values are rest-frame values.  \label{histoclass}}
\end{center}
\end{figure}

Phenomenologically, blazars can be divided into two classes based on their optical properties: FSRQs and BLOs. Previous studies showed that FSRQs tend to have higher RMs than BLOs \citep{ZT2004, hovatta2012}. We collected all available core RM values from all frequency pairs and present the distributions of the (logarithmic) RMs of FSRQs and BLOs in Figure~\ref{histoclass}. The median RM values are $1.2\times10^4\rm\ rad/m^2$ and $4.8\times10^3\rm\ rad/m^2$ for FSRQs and BLOs, respectively -- the value for FSRQs is higher than that for BLOs by a factor close to three. However, a Kolmogorov-Smirnov test \citep{press1992} finds a probability of 7\% that the FSRQ and BLO values are drawn from the same parent population. Therefore, it is possible that their RM properties are intrinsically the same.

In Section~\ref{sectscreen}, we claimed that the observed Faraday rotation mostly originates from jet sheaths. Relatively slow, possibly non-relativistic winds launched by an accretion disk that surround and confine the highly relativistic jet spine are one of the candidates for a jet sheath (e.g., \citealt{devilliers2005}). A fundamental difference between FSRQs and BLOs is their accretion luminosities relative to their Eddington luminosities, above and below $\approx1\%$, respectively (e.g., \citealt{ghisellini2011}, see also \citealt{PC2015} for further discussion). This suggests that sources in high accretion states tend to have larger RMs. A simple explanation would be that high accretion rates lead to relatively larger amounts of matter in jet sheaths. There is indeed evidence for a relation between the rate of matter injection into the jet and the accretion rate (e.g., \citealt{ghisellini2014, PT2017}), supporting this idea. However, the strength and degree of ordering of core magnetic fields as function of blazar subclass are poorly understood yet; the difference in RM may not be solely due to the difference in particle density.

\subsection{Intrinsic polarization orientation}

Intrinsic EVPAs (projected onto the sky plane) of AGN jets can be obtained by correcting for Faraday rotation. It has been consistently shown that BLOs have intrinsic EVPAs well aligned with their jets, while a wide range of angles between EVPAs and jet orientations, sometimes seen as double-peaked distribution of relative angles, is observed for FSRQs (e.g., \citealt{LH2005, jorstad2007}). The good alignment and the mis-alignment were associated with a transverse or oblique shock and a conical shock, respectively \citep{jorstad2007}. These results, however, used RMs obtained from a single $\lambda^2$ law description of EVPA variation between 7 and 1 mm. The RM values were of the order $10^4\rm\ rad/m^2$.

However, our results show that there is a possibility that the core RMs of blazars can increase up to $\approx10^6\rm\ rad/m^2$ at $\approx250$ GHz (Figure~\ref{paopt}). The possible difference in core RM between FSRQs and BLOs, discussed in Section~\ref{sect:opticalsubclasses}, suggests that it is easier for FSRQs to have high core RMs up to $\approx10^6\rm\ rad/m^2$ than for BLOs -- unless the transition frequency for BLOs is much larger than for FSRQs, which seems not to be the case (Figure~\ref{paopt}). Even at 1 mm, the high RM of $\approx10^6\rm\ rad/m^2$ leads to an EVPA rotation of $\approx57\deg$. Therefore, one needs to take the frequency dependence of RM into account -- especially for FSRQs -- when comparing the intrinsic EVPAs at mm wavelengths with the direction of the inner jet. However, FSRQs are unlikely to have intrinsic EVPAs aligned with their jets (see also \citealt{yuan2001, gabuzda2006, hovatta2016}) even after correcting for Faraday rotation; even their optical EVPAs, which do not suffer from strong Faraday rotation, show bimodal distributions in the angles between jets and EVPAs (\citealt{jorstad2007}, see also \citealt{LS2000}).

We observe RMs to increase with increasing frequency, meaning that intrinsic core EVPAs are different for different observing frequencies. Such a frequency dependence implies that polarized emission observed at higher frequencies comes from regions closer to the jet base. This indicates that intrinsic EVPAs can vary with distance from the jet base. A similar behaviour has been observed for a few sources in other studies. \citet{OG2009a} found that the jet of BL Lac shows EVPAs well aligned to the jet direction in inter-knot regions and even when the jet bends. \cite{gomez2016} showed that their high resolution polarization image of the same source shows smooth but non-negligible variations of EVPA upstream and downstream from the core. Both results were interpreted as the presence of helical magnetic fields in the jet. Similarly, different intrinsic EVPAs at different frequencies might imply the presence of helical magnetic fields in the core regions. However, a firm conclusion requires confirming the frequency dependence of RM at a wide range of observing frequencies with both short and long $\lambda^2$ spacings (see Section~\ref{multipleshocks}).

\subsection{Multiple recollimation shocks in the cores}
\label{multipleshocks}

Theoretical studies have shown that a series of recollimation shocks can form in relativistic jets: in analytic works (e.g., \citealt{DM1988}), in hydrodynamic numerical simulations (e.g., \citealt{gomez1995, gomez1997, agudo2001}), and in magneto-hydrodynamic simulations (e.g., \citealt{mizuno2015, marti2016}). Observationally, the presence of stationary features in AGN jets in addition to their VLBI cores has been verified in many studies (e.g., \citealt{jorstad2005}). Especially, high resolution images of 3C 454.3 \citep{jorstad2010} and BL Lac \citep{gomez2016} revealed that their cores may consist of multiple stationary components. For BL Lac, emission upstream the radio core, leading to multi-wavelength flares when it passes through the core, was observed \citep{marscher2008, gomez2016}; this supports the idea that the core can be identified with one of a series of recollimation shocks (e.g., \citealt{marscher2009}).

We found that the EVPA--$\lambda^2$ relations of our sources are usually non-linear, instead showing breaks in their slopes. We obtained the RMs for pairs of adjacent frequencies and discovered that the core RMs systematically increase with observing frequency.
Based on VLBA observations at 8 different frequencies from 4.6 to 43 GHz, \cite{OG2009a} showed that breaks in RM appear frequently, with the best-fit lines in the EVPA--$\lambda^2$ diagram connecting smoothly over a wide range of frequencies (though not for BL Lac in their sample). In contrast, \cite{kravchenko2017} presented large discontinuities between the different EVPA--$\lambda^2$ fits at much lower frequencies (between 2 and 5 GHz). Therefore, one might expect that potential discontinuities in EVPA rotations might not be substantial at mm wavelengths and the assumption underlying our analysis -- no RM discontinuities -- might be justified. 

Furthermore, these studies showed that core EVPA rotations could be fitted well by a $\lambda^2$ law in some frequency ranges, then breaks, and then shows another good $\lambda^2$ fit in other frequency ranges (e.g., \citealt{OG2009a, kravchenko2017}). Other studies obtained good $\lambda^2$ fits for the core EVPA rotations in most cases when they used relatively small frequency intervals, e.g., 8--15 GHz (e.g., \citealt{ZT2004, hovatta2012}). This indicates that polarized emission from a single emission region is dominant over relatively small frequency intervals, without showing a systematic increase of RMs as a function of frequency. However, over a wide range of frequencies, the RM--frequency relations appear to show multiple breaks; this implies that $\rm |RM| \propto \nu^a$ predicted by \cite{jorstad2007} assuming a continuous core-shift effect might not hold for narrow frequency intervals. One possible explanation is that blazar cores actually consist of multiple recollimation shocks and we observe polarized emission from one of the shocks in a given narrow frequency interval. As one goes to higher frequencies, polarized emission from inner shocks close to the jet base becomes more dominant due to lower opacity, leading to another good $\lambda^2$ fit with higher RMs.

\section{Conclusions}
\label{sect5}

We studied polarization properties of 8 blazars -- 5 FSRQs and 3 BLOs with multi-frequency simultaneous observations with the KVN at 22, 43, and 86 GHz. We investigated the nature of blazar radio cores by means of measuring Faraday rotation measures at different observing frequencies. Our work leads us to the following principal conclusions:
\begin{enumerate} 
\item We found that RMs increase with frequency, with median values of $2.62\times10^3\rm\ rad/m^2$ and $1.42\times10^4\rm\ rad/m^2$ for the frequency pairs 22/43 GHz and 43/86 GHz, respectively. These values are also higher than those obtained by \cite{hovatta2012} at 8.1--15.4 GHz for the same sources. The median values are described well by a power-law function with $\rm|RM|\propto\nu^{a}$ with $a=2.42$. When $a$ values are obtained separately for each source, they are distributed around $a=2$ with mean and standard deviation of $a=2.25\pm1.28$. This agrees with the expectation from core-shift \citep{jorstad2007} for many blazars at the KVN frequencies. This finding implies that the geometry of Faraday rotating media in blazar cores can be approximated as conical.
\item We compared our KVN data with contemporaneous (within $\approx1$ week) optical polarization data from the Steward Observatory for a few sources. When we assume that the direction of EVPA rotation at radio frequencies is the same at optical wavelengths and that there is no $n\pi$ ambiguity, the optical data show a trend of EVPA rotation similar to that of the radio data. The RM values obtained with the optical data indicate that the power-law increase of RM with frequency continues up to a certain frequency, $\nu_{\rm trans}$, and then saturates, with $\rm |RM|\approx10^{5-6}\rm\ rad/m^2$ at $\approx250\rm\  GHz$, depending on source and flaring activity. We suggest that this saturation is due to the absence of core shift above $\nu_{\rm trans}$; instead, radio cores are standing recollimation shocks. This is in agreement with other studies which concluded that the radio cores of blazars cannot purely be explained as the unity optical depth surface of a continuous conical jet but are physical structures at least in some cases.
\item We detected a sign change in the observed RMs of CTA 102 over $\approx1$ month, while the magnitudes of RM were roughly preserved. Since this source showed strong flaring at the time of our observations, we suggest that new relativistic jet components emerging from the core undergo acceleration/deceleration and/or jet bending, thus leading to a change in the direction of the line-of-sight component of helical magnetic fields in the jet because of relativistic abberation.
\item We found indication that the absolute values of the core RMs of FSRQs are larger than those of BLOs at 22--86~GHz, which is consistent with results found at cm wavelengths. This difference might arise from FSRQs having higher accretion rates than BLOs, resulting in larger amounts of material in the central engine.
\item For those sources which show non-linear EVPAs--$\lambda^2$ relations, the RM-corrected (intrinsic) EVPAs might be different at different frequencies and thus at different locations of the jets. A recent ultra-high resolution image of BL Lac observed with space VLBI shows that its intrinsic EVPAs in the core region vary with different locations indeed.
\item We suggest that the systematic increase of RM as function of observing frequency appears only when covering sufficiently large ranges in frequency, with different $\lambda^2$ laws at different frequency ranges connecting smoothly. Combining this with the fact that linear EVPA--$\lambda^2$ relations are commonly observed over narrow frequency ranges suggests that blazars cores might consist of multiple recollimation shocks such that polarized emission from one of the shocks is dominant in a given narrow frequency range. 
\end{enumerate}

\acknowledgments 
We thank the referee for constructive comments, which helped to improve the paper. The KVN is a facility operated by the Korea Astronomy and Space Science Institute. We are grateful to the staff of the KVN who helped to operate the array and to correlate the data. The KVN and a high-performance computing cluster are facilities operated by the KASI (Korea Astronomy and Space Science Institute). The KVN observations and correlations are supported through the high-speed network connections among the KVN sites provided by the KREONET (Korea Research Environment Open NETwork), which is managed and operated by the KISTI (Korea Institute of Science and Technology Information). This research has made use of the NASA/IPAC Extragalactic Database (NED) which is operated by the Jet Propulsion Laboratory, California Institute of Technology, under contract with the National Aeronautics and Space Administration. Data from the Steward Observatory spectropolarimetric monitoring project were used. This program is supported by Fermi Guest Investigator grants NNX08AW56G, NNX09AU10G, NNX12AO93G, and NNX15AU81G. This study makes use of 43 GHz VLBA data from the VLBA-BU Blazar Monitoring Program (VLBA-BU-BLAZAR; \url{http://www.bu.edu/blazars/VLBAproject.html}), funded by NASA through the Fermi Guest Investigator Program. The VLBA is an instrument of the Long Baseline Observatory. The Long Baseline Observatory is a facility of the National Science Foundation operated by Associated Universities, Inc. This research has made use of data from the MOJAVE database that is maintained by the MOJAVE team \citep{lister2009}. We acknowledge financial support from the Korean National Research Foundation (NRF) via Global Ph.D. Fellowship Grant 2014H1A2A1018695 (J.P.) and Basic Research Grant NRF-2015R1D1A1A01056807 (S.T., J.C.A.). S. S. Lee was supported by NRF grant NRF-2016R1C1B2006697. GYZ is supported by Korea Research Fellowship Program through the NRF (NRF-2015H1D3A1066561). Correspondence should be addressed to S.T.

\appendix

\section{D-Term calibration and evolution \label{appendixa}}
\label{appendixa}

\begin{figure*}[!t]
\centering
\includegraphics[trim=7mm 10mm 7mm 5mm, clip, width = 58mm]{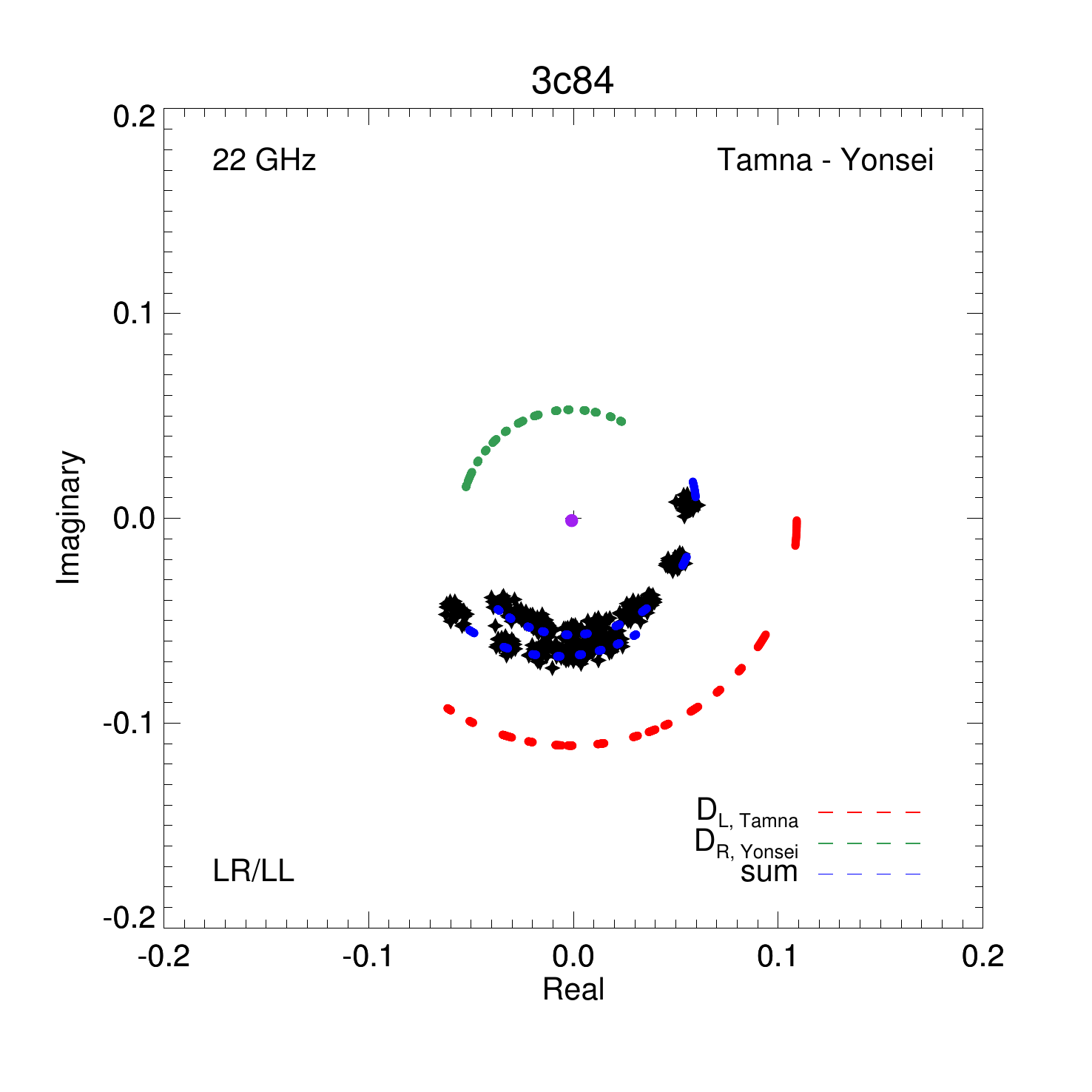}
\includegraphics[trim=7mm 10mm 7mm 5mm, clip, width = 58mm]{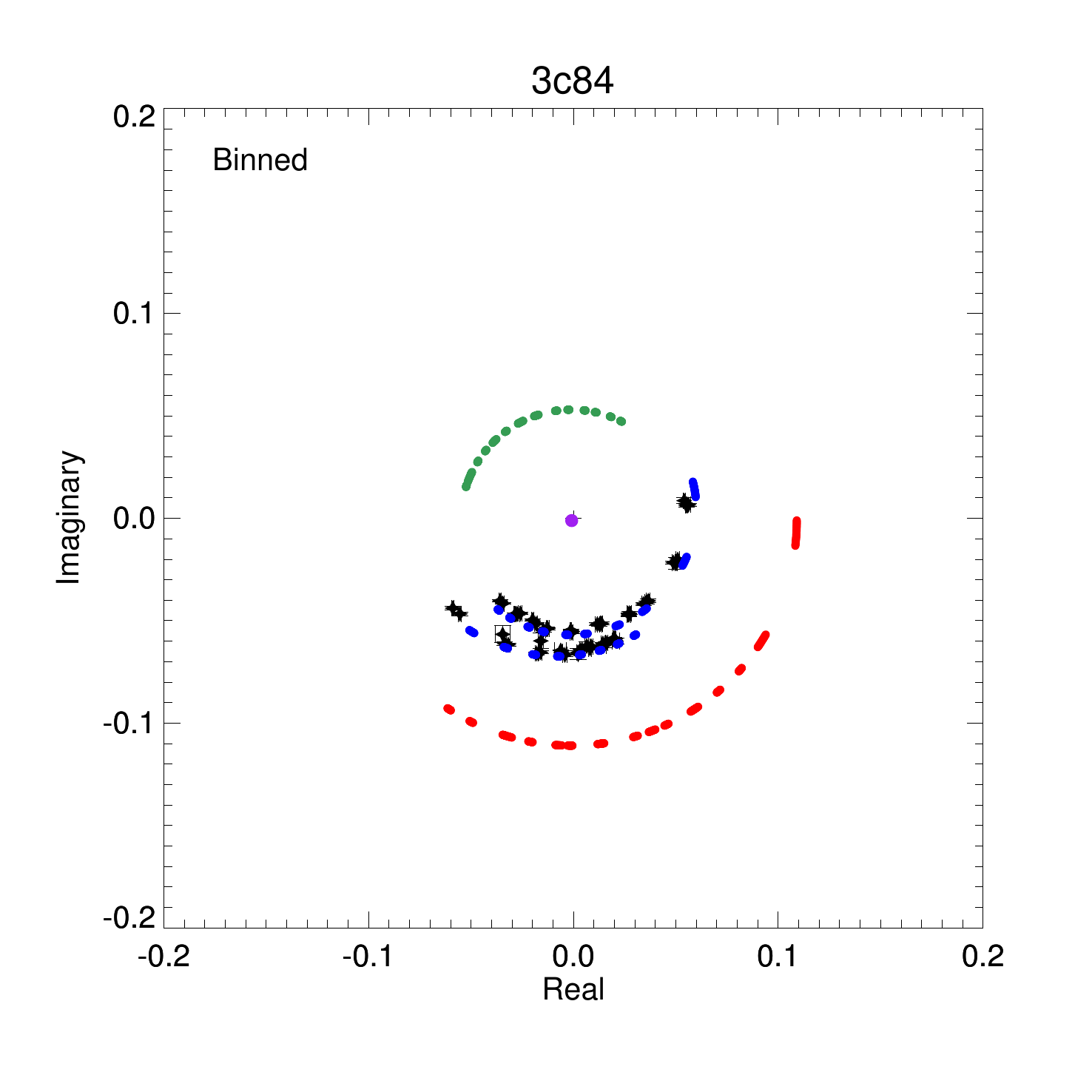}
\includegraphics[trim=7mm 10mm 7mm 5mm, clip, width = 58mm]{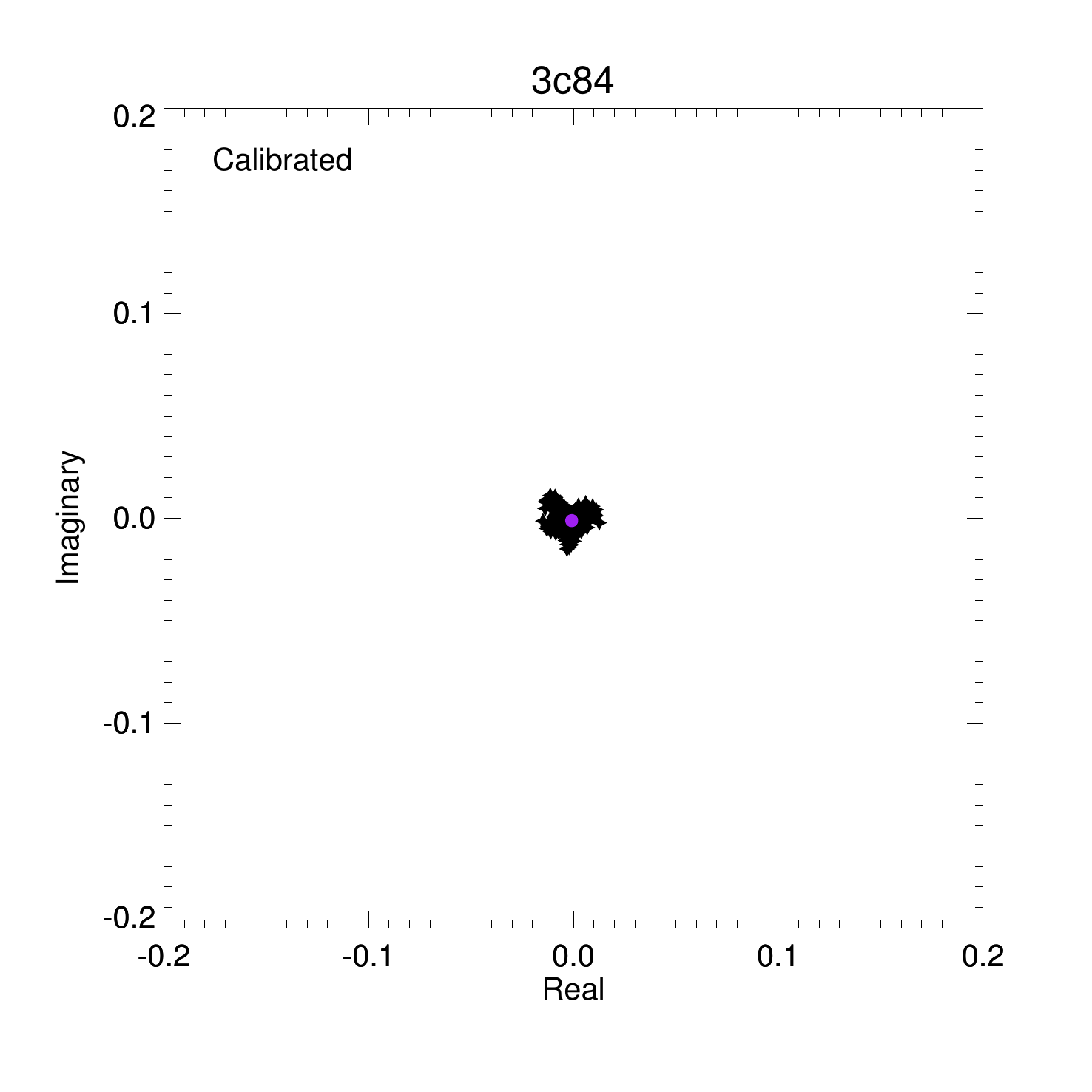}
\includegraphics[trim=7mm 10mm 7mm 5mm, clip, width = 58mm]{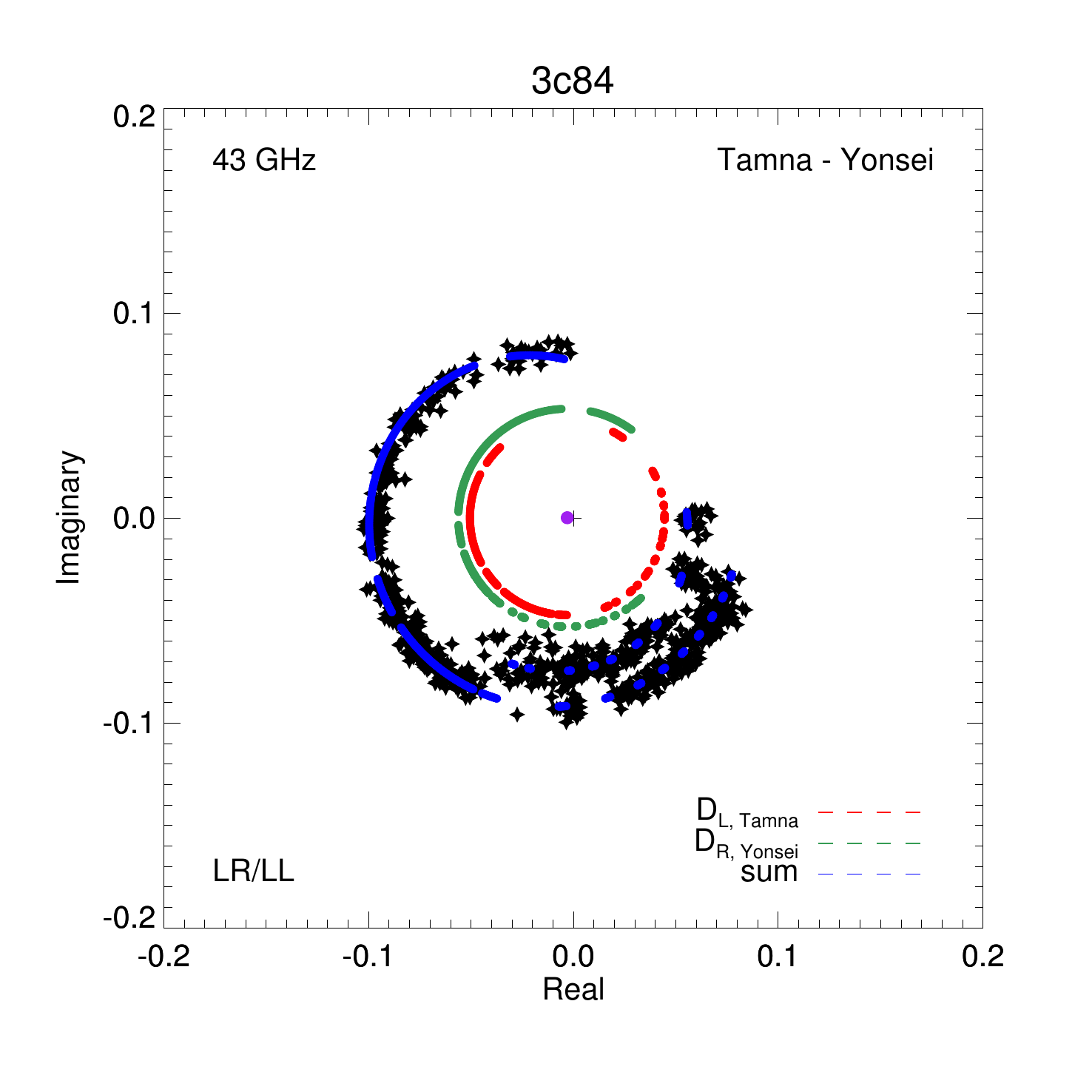}
\includegraphics[trim=7mm 10mm 7mm 5mm, clip, width = 58mm]{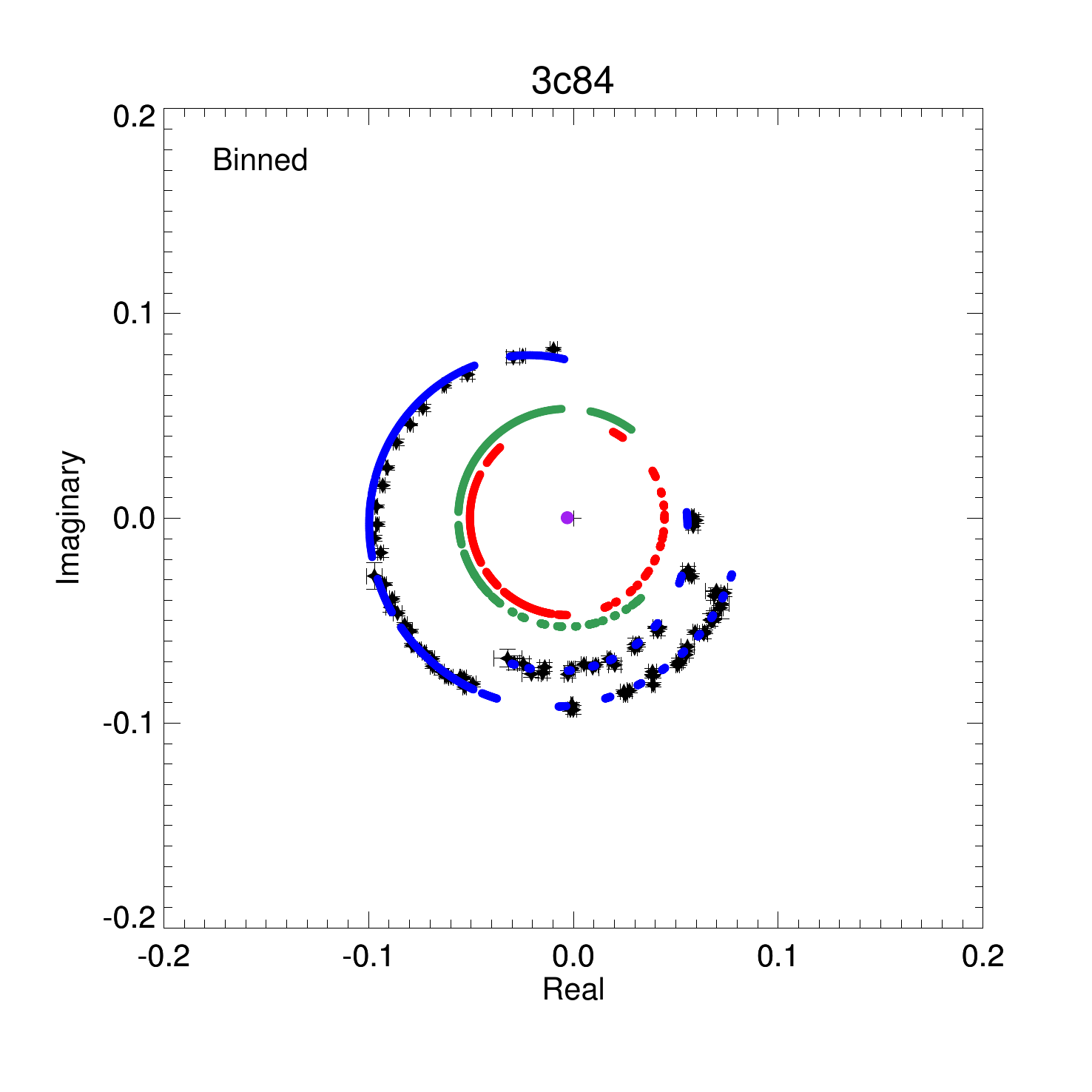}
\includegraphics[trim=7mm 10mm 7mm 5mm, clip, width = 58mm]{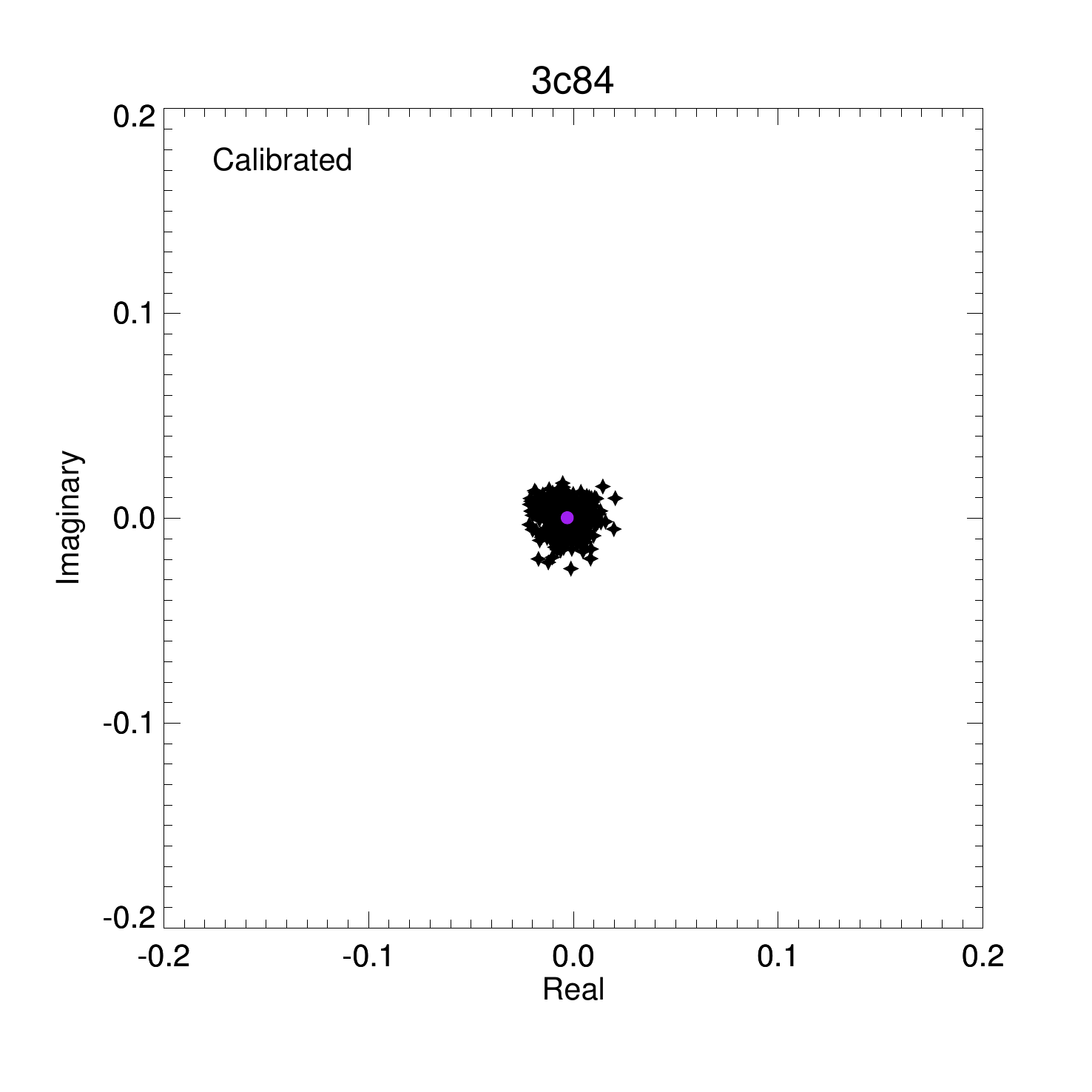}
\includegraphics[trim=7mm 10mm 7mm 5mm, clip, width = 58mm]{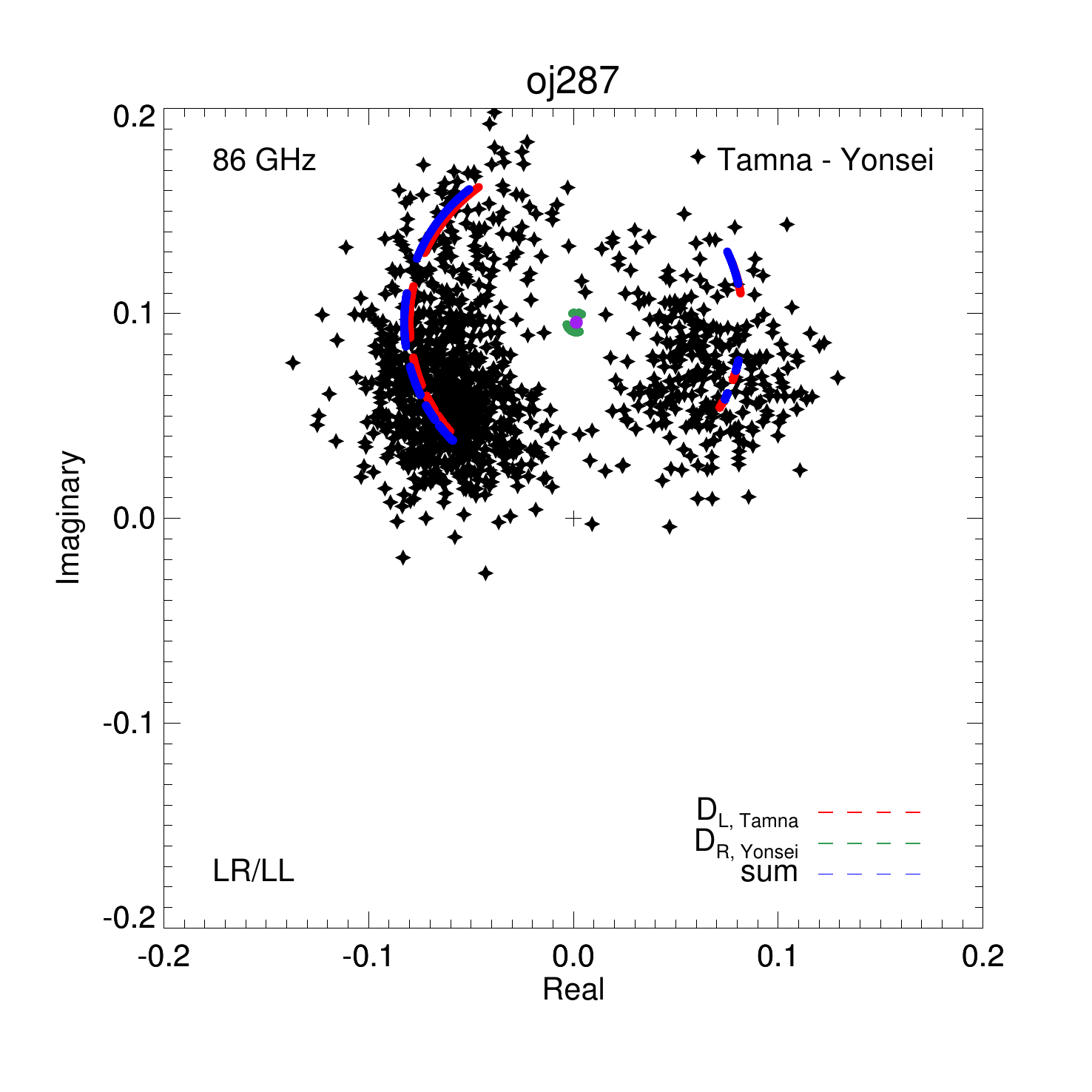}
\includegraphics[trim=7mm 10mm 7mm 5mm, clip, width = 58mm]{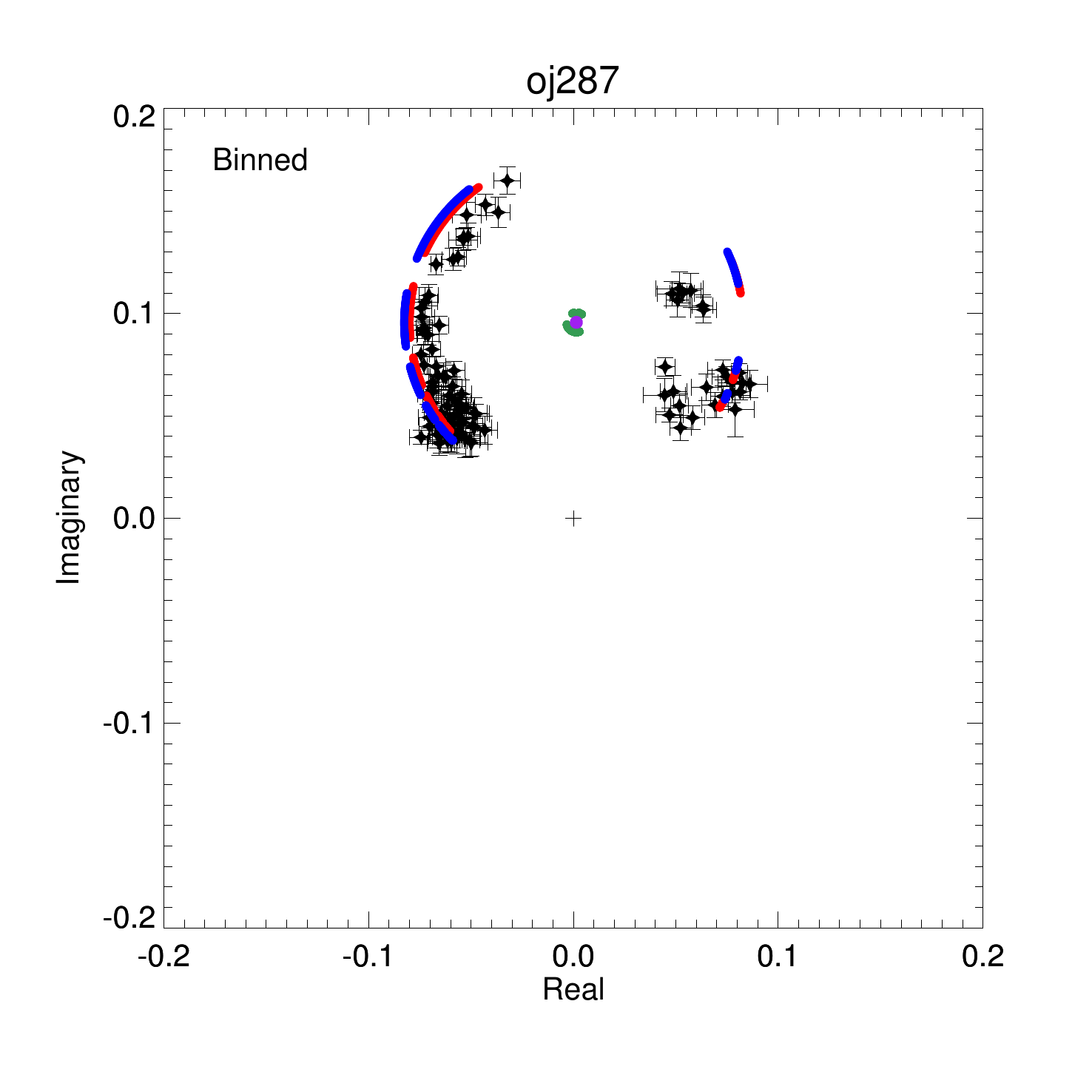}
\includegraphics[trim=7mm 10mm 7mm 5mm, clip, width = 58mm]{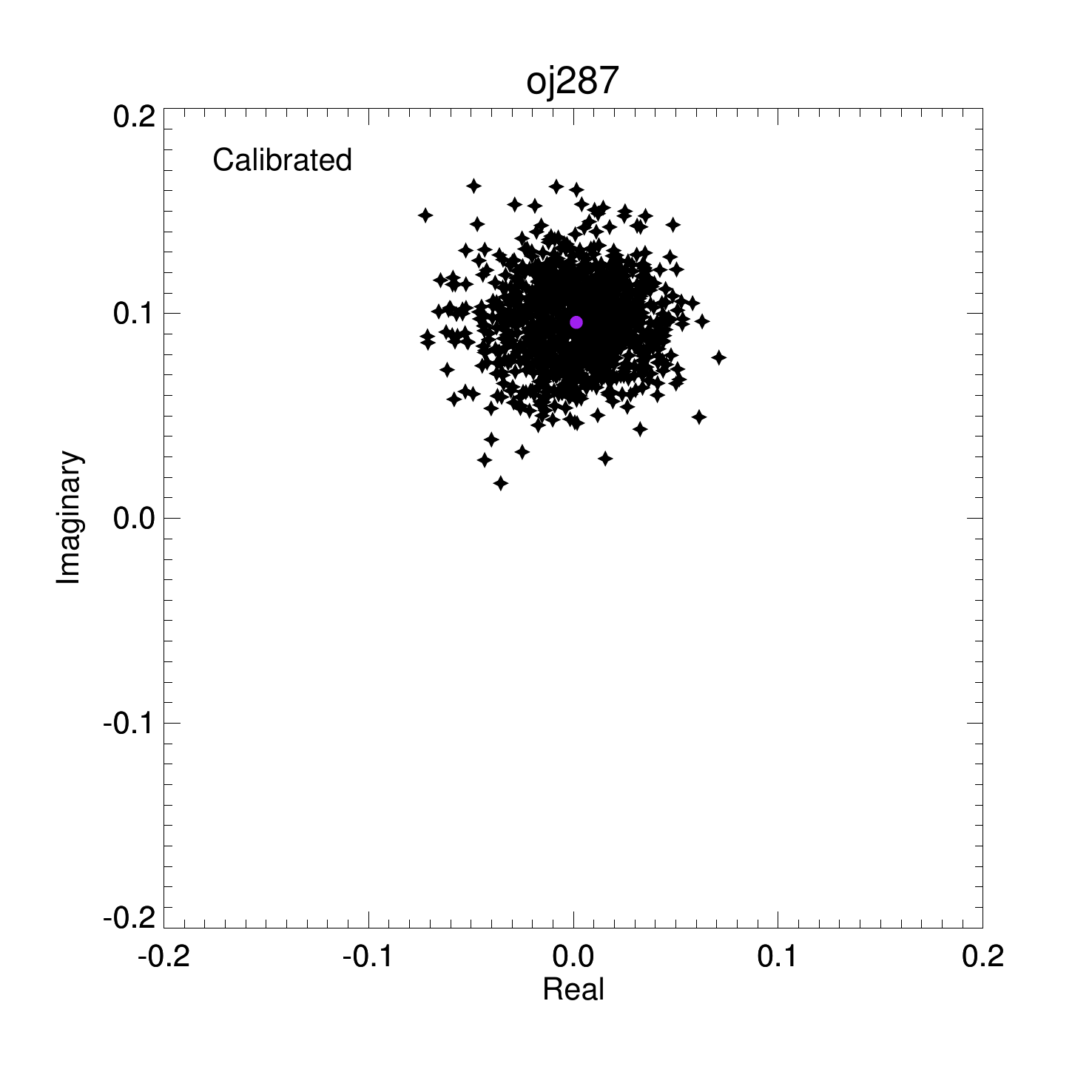}

\caption{\emph{Left:} Visibility ratio LR/LL at 22, 43, and 86~GHz (top to bottom panels) for the baseline Tamna--Yonsei in January 2017 on the complex plane. The plots include: observed ratios (black data points); expected visibility ratios due to instrumental polarization for the Tamna (red) and Yonsei (green) antennas; sum of the ratios from the two antennas (blue). \emph{Center:} Same as the left panels but with data being binned in time. \emph{Right:} Visibility ratios after instrumental polarization calibration. Purple dots mark the centers of distributions. \label{dtermcal}}
\end{figure*}

\begin{figure*}[!t]
\centering 
\includegraphics[trim=4mm 10mm 10mm 7mm, clip, width = 59mm]{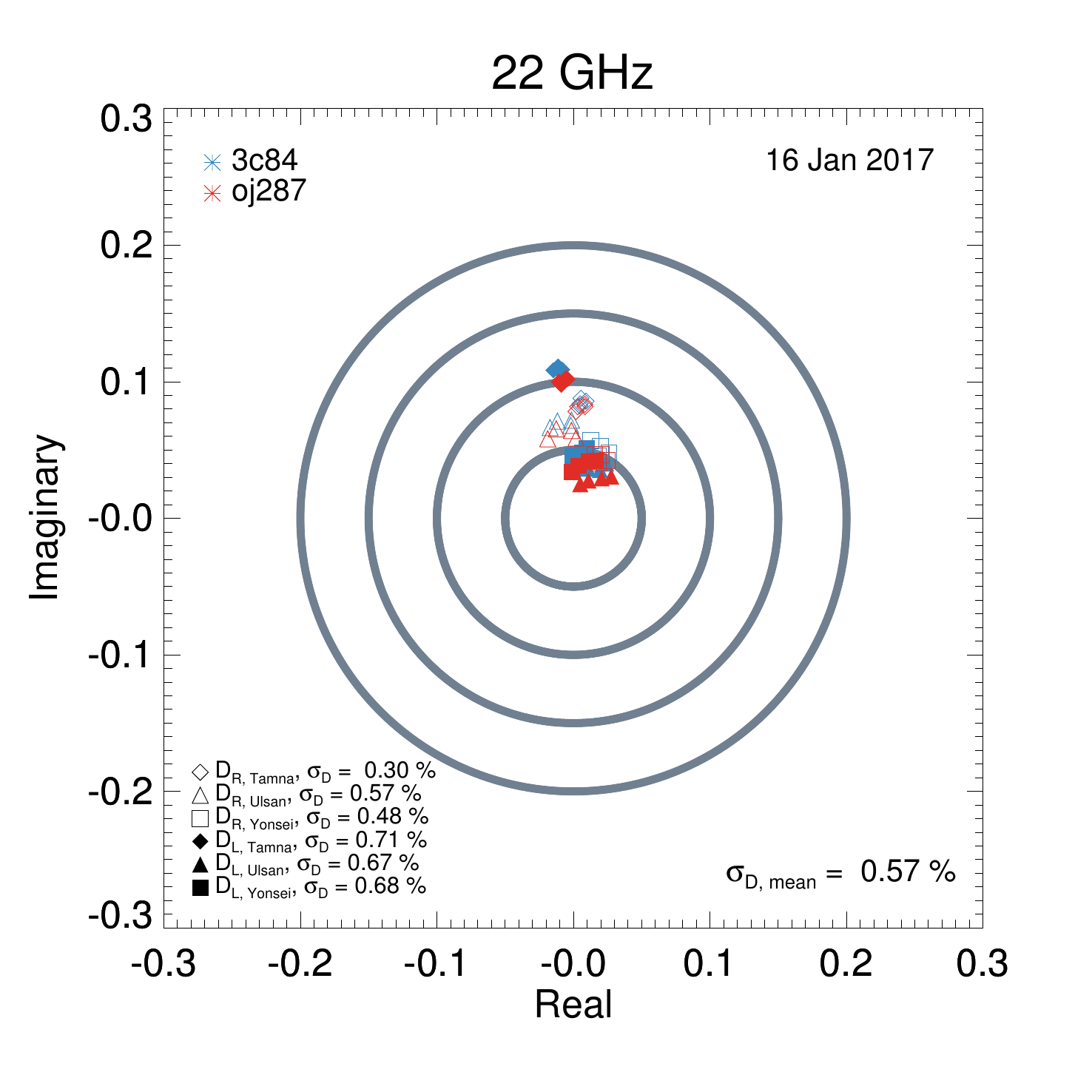}
\includegraphics[trim=4mm 10mm 10mm 7mm, clip, width = 59mm]{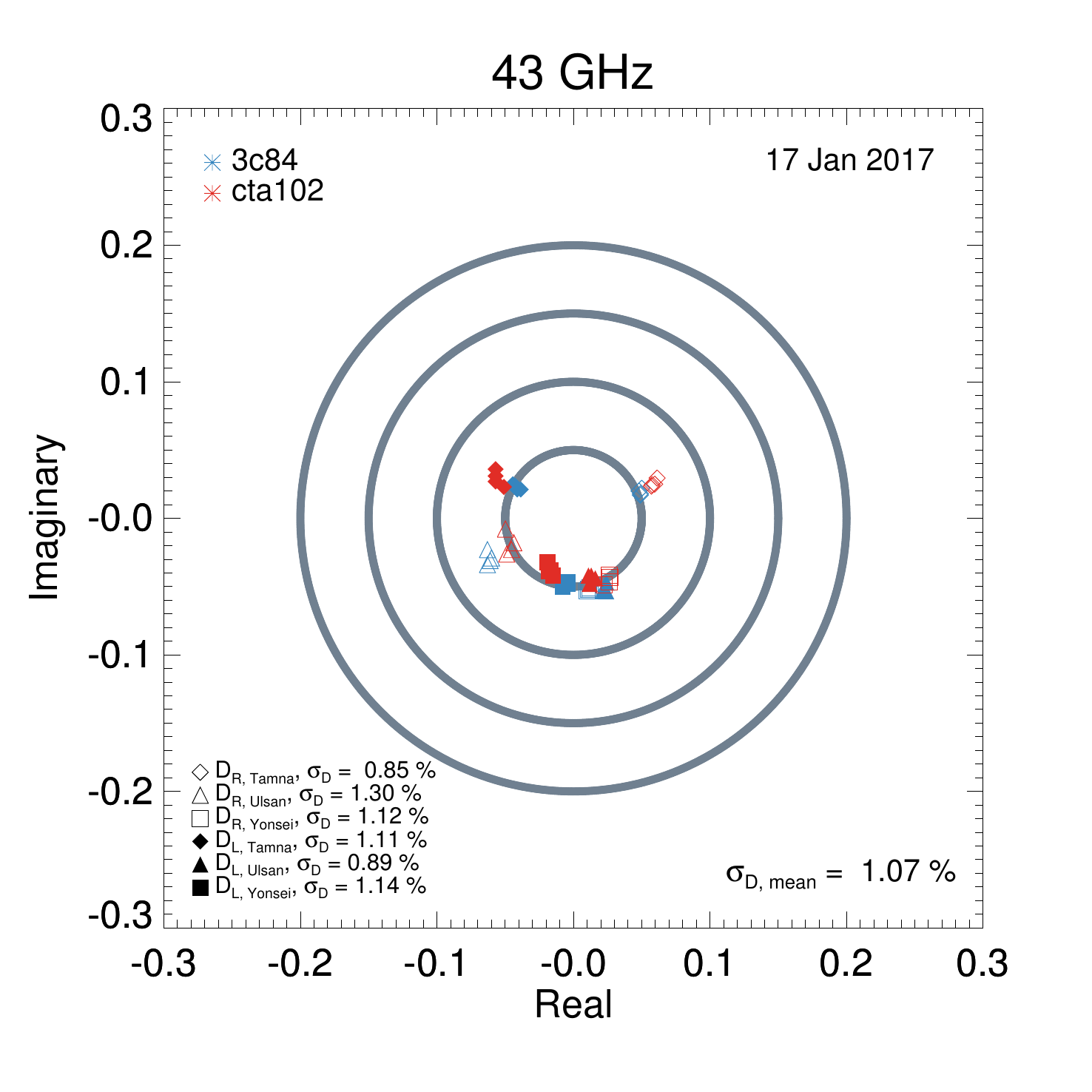}
\includegraphics[trim=4mm 10mm 10mm 7mm, clip, width = 59mm]{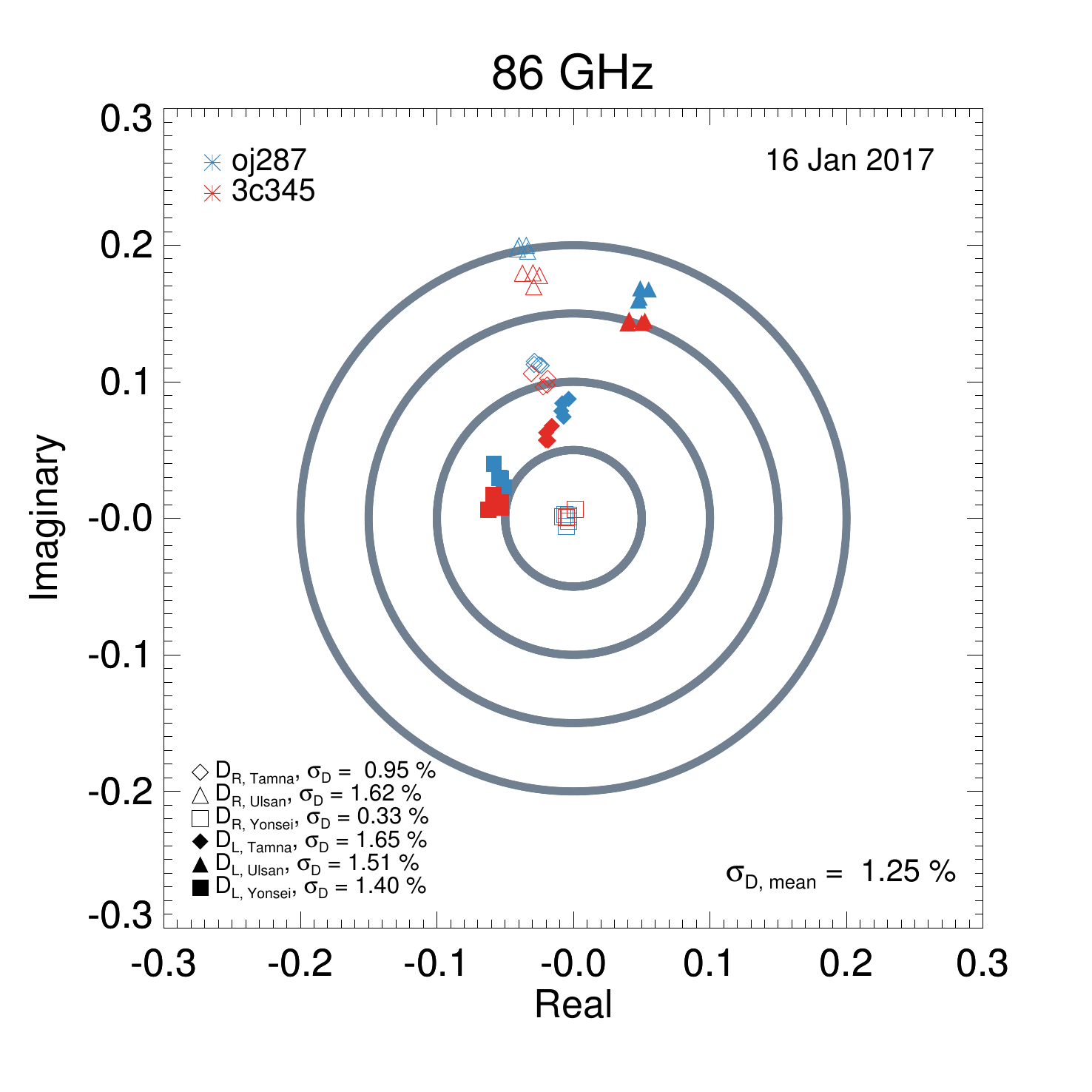}
\caption{D-terms obtained using different calibrator sources (different colors) in the complex plane for the KVN 22, 43, and 86 GHz bands in the left, center, and right panels, respectively. The standard deviations of the D-terms for each antenna and for each polarization are noted on the bottom left of each panel. Their mean values are noted on the bottom right of each panel. \label{dtermcomparison}}
\end{figure*}

\begin{figure*}[!t]
\centering 
\includegraphics[trim=4mm 4mm 10mm 4mm, clip, width = 59mm]{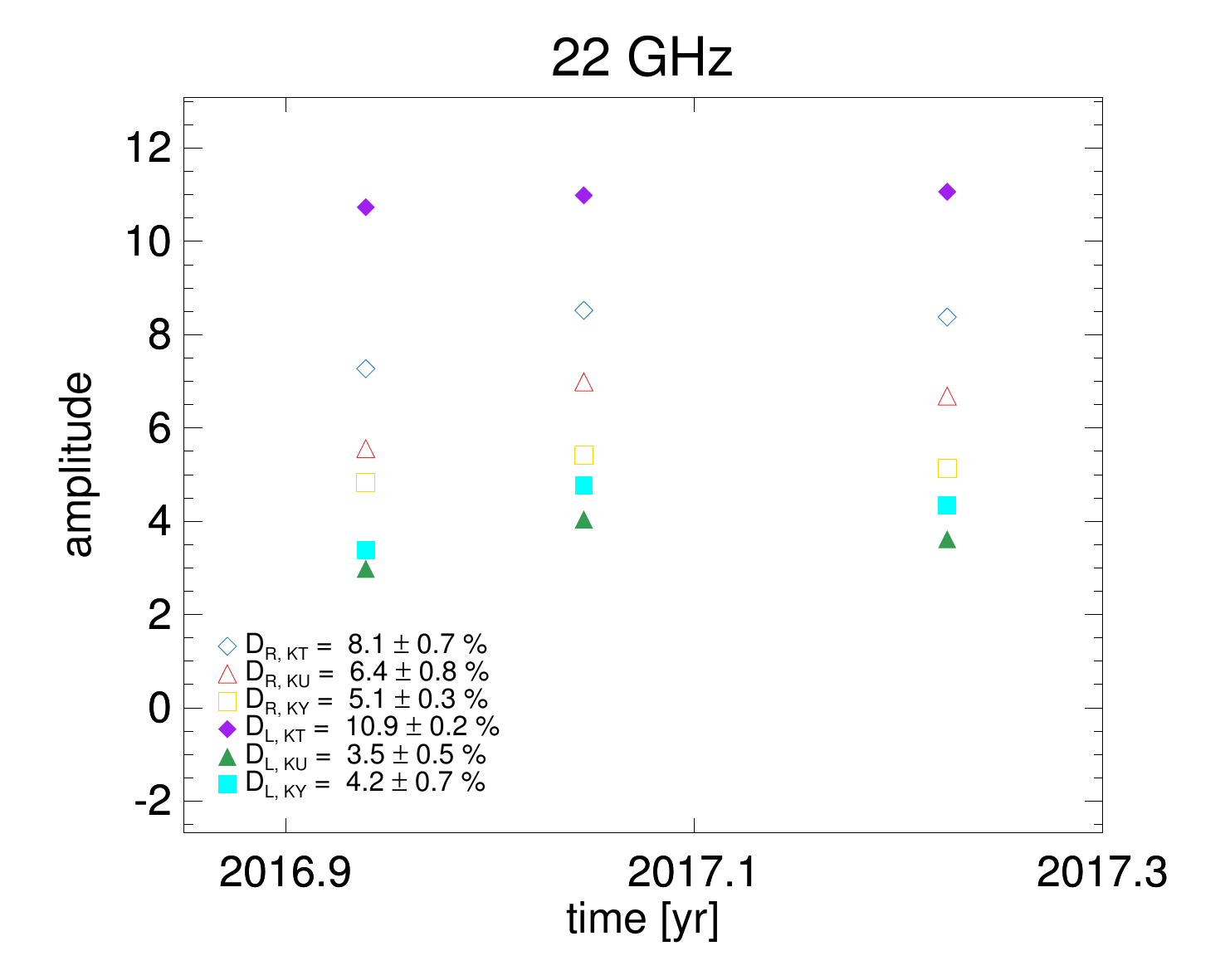}
\includegraphics[trim=4mm 4mm 10mm 4mm, clip, width = 59mm]{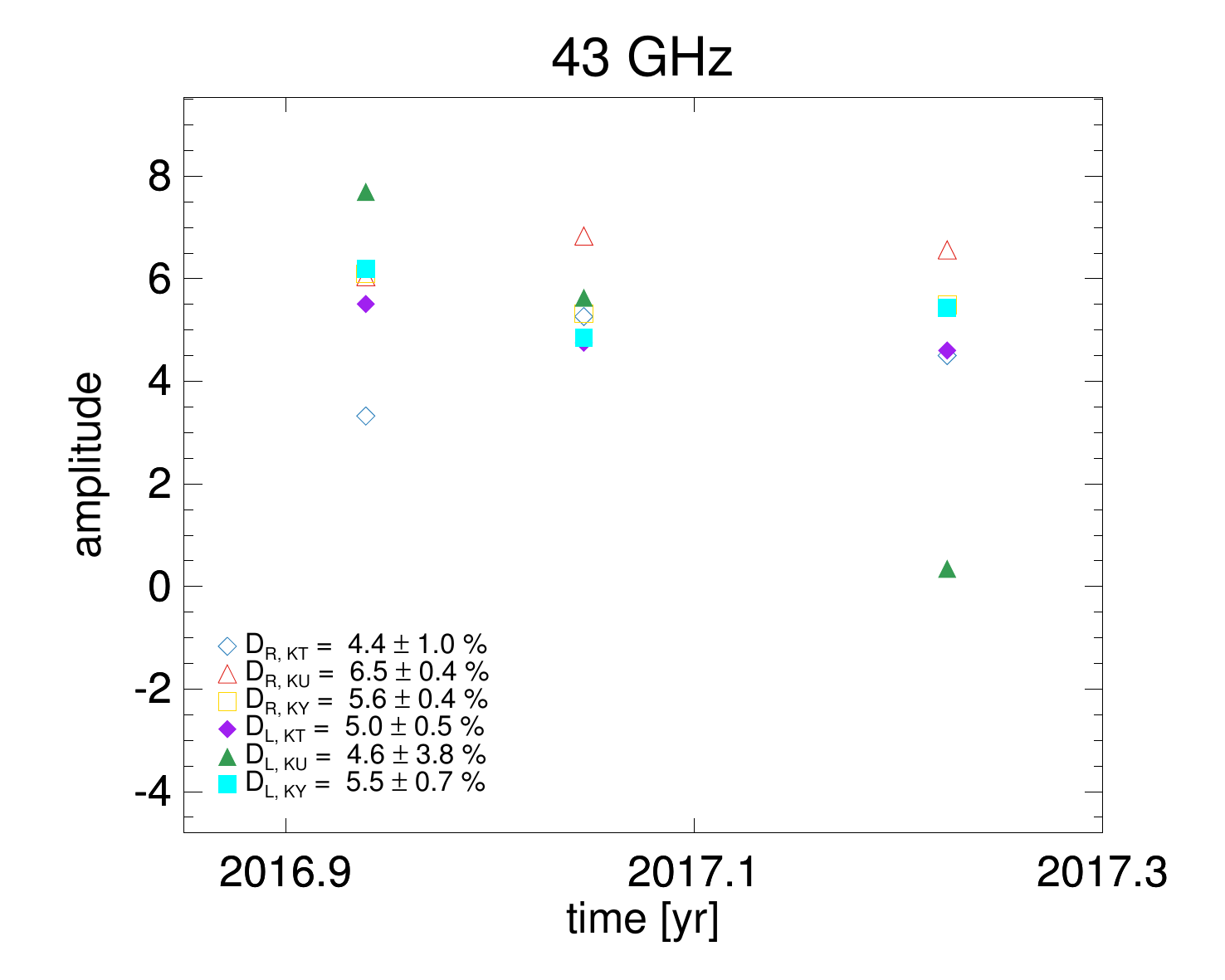}
\includegraphics[trim=4mm 4mm 10mm 4mm, clip, width = 59mm]{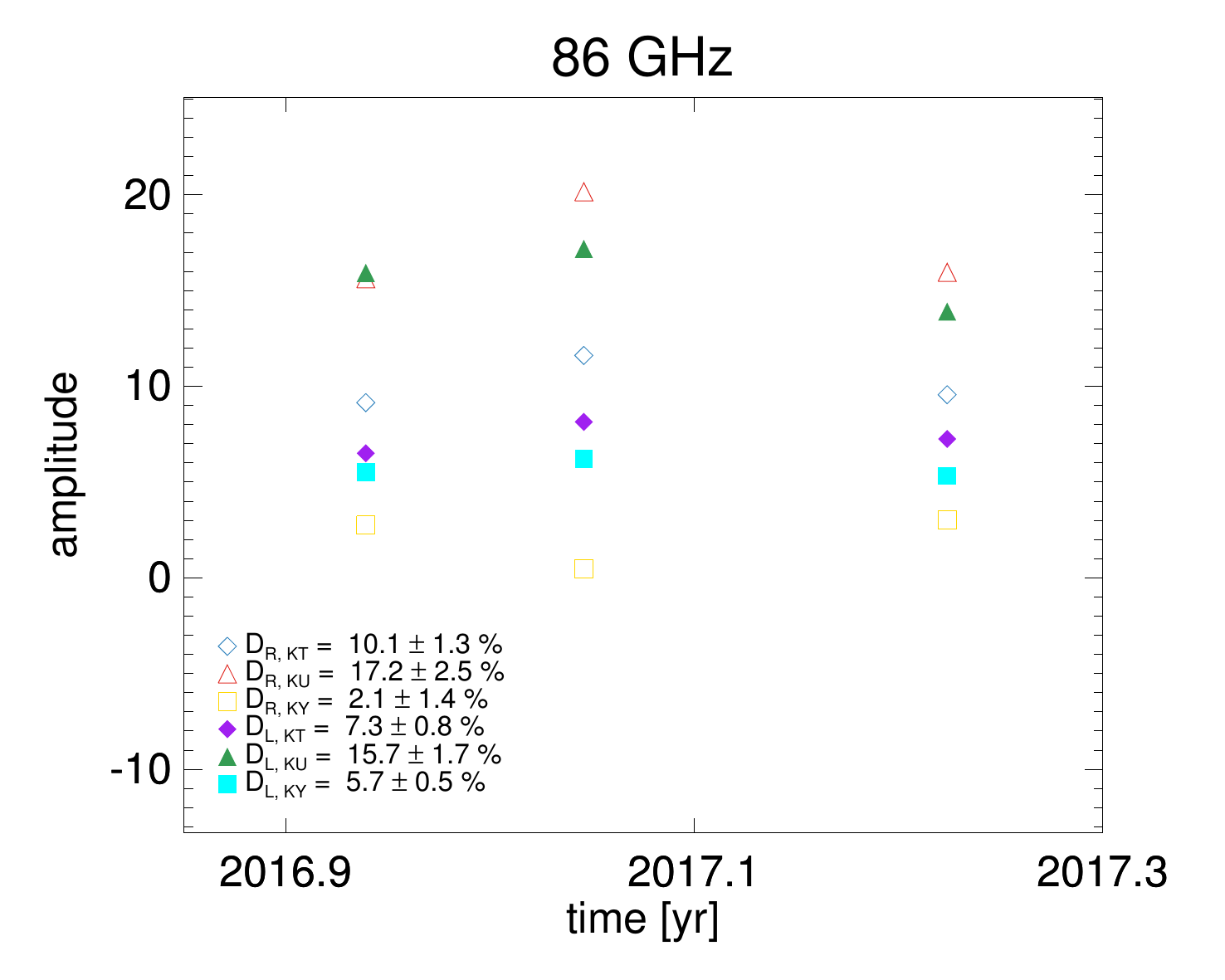}
\caption{D-terms amplitudes as function of time. Means and standard devations are noted on the bottom left of each panel. \label{dtermevolution}}
\end{figure*}

\begin{figure*}[!t]
\centering
\includegraphics[trim=7mm 1mm 6mm 0mm, clip, width = 58mm]{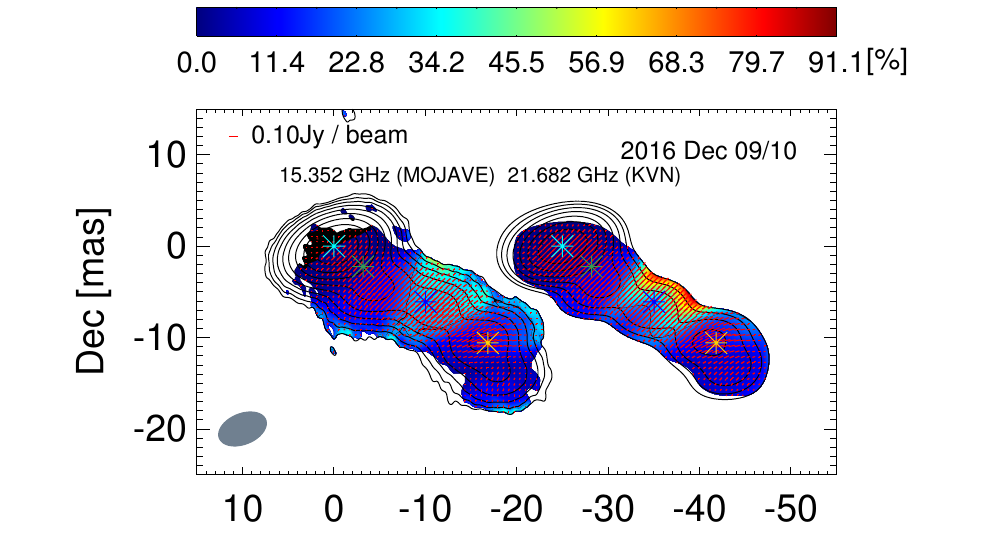}
\includegraphics[trim=7mm 1mm 6mm 0mm, clip, width = 60mm]{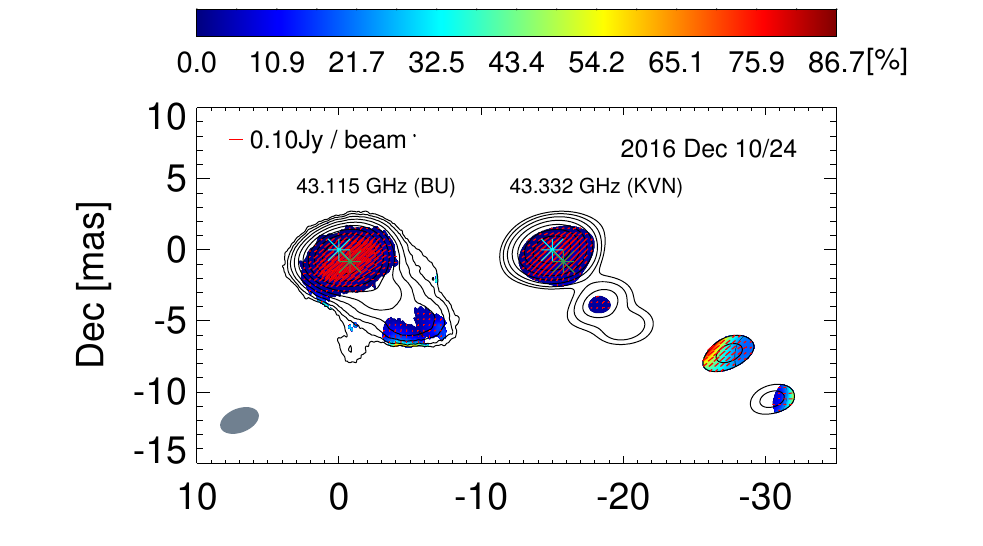}
\includegraphics[trim=10mm 1mm 6mm 0mm, clip, width = 59mm]{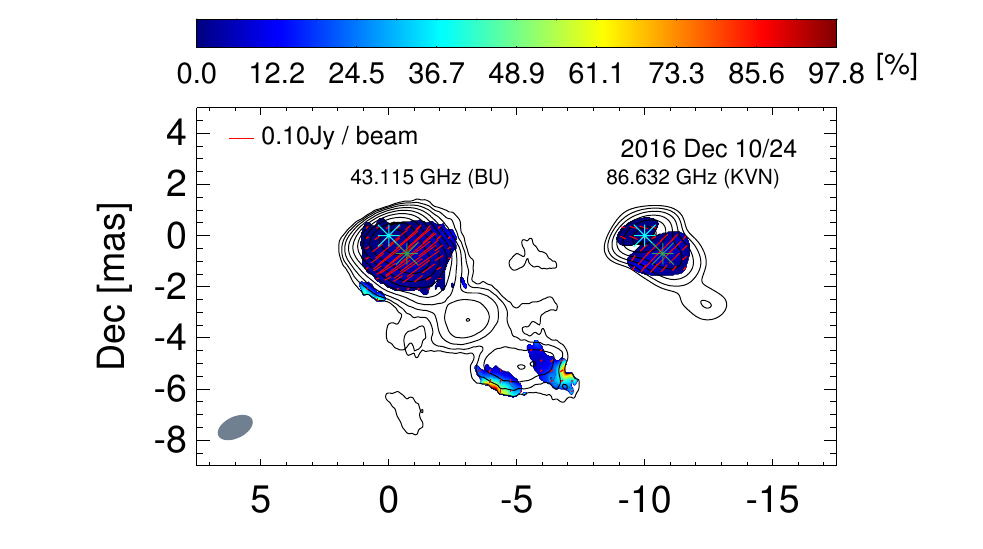}
\includegraphics[trim=5mm 8mm 8mm 2mm, clip, width = 58mm]{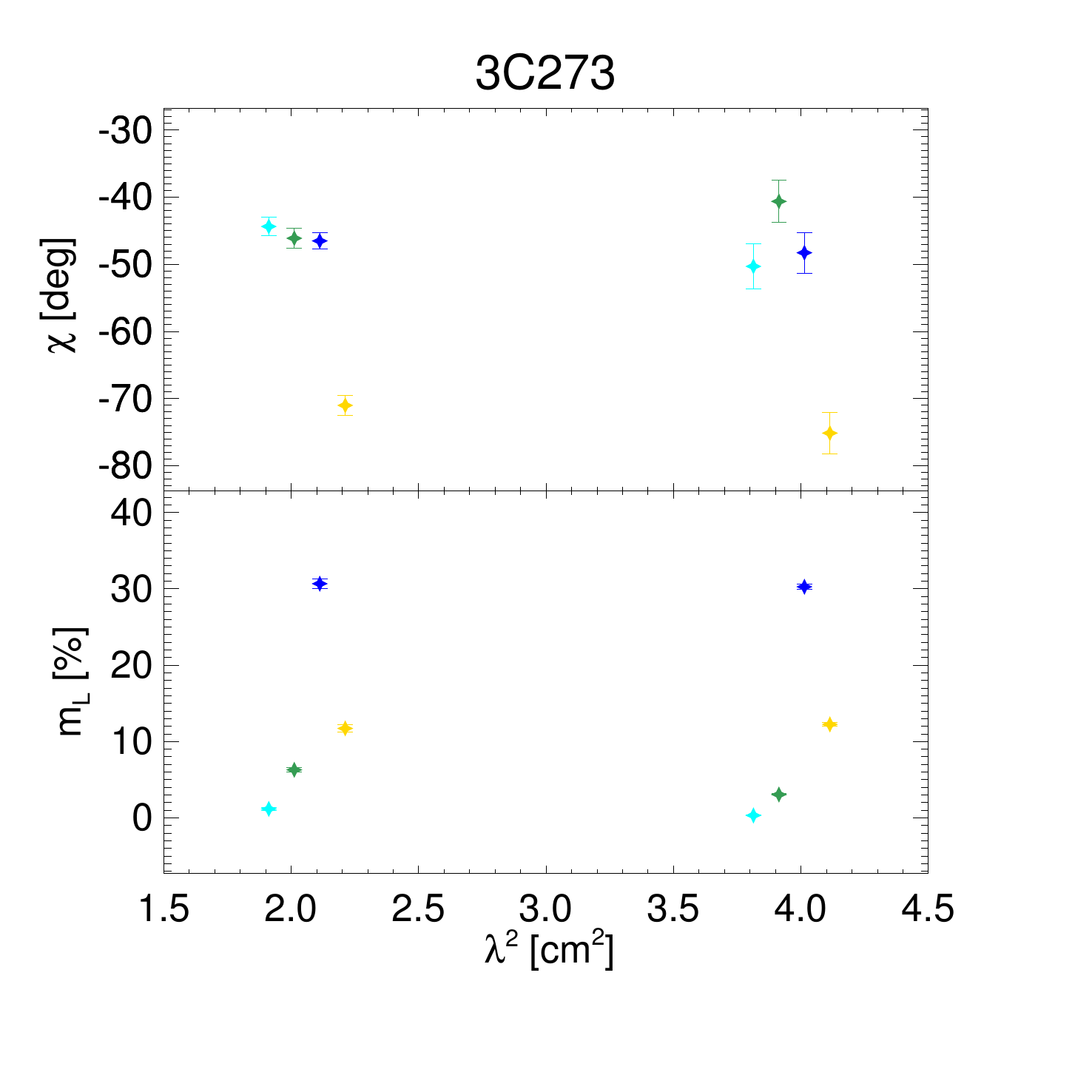}
\includegraphics[trim=5mm 8mm 8mm 2mm, clip, width = 58mm]{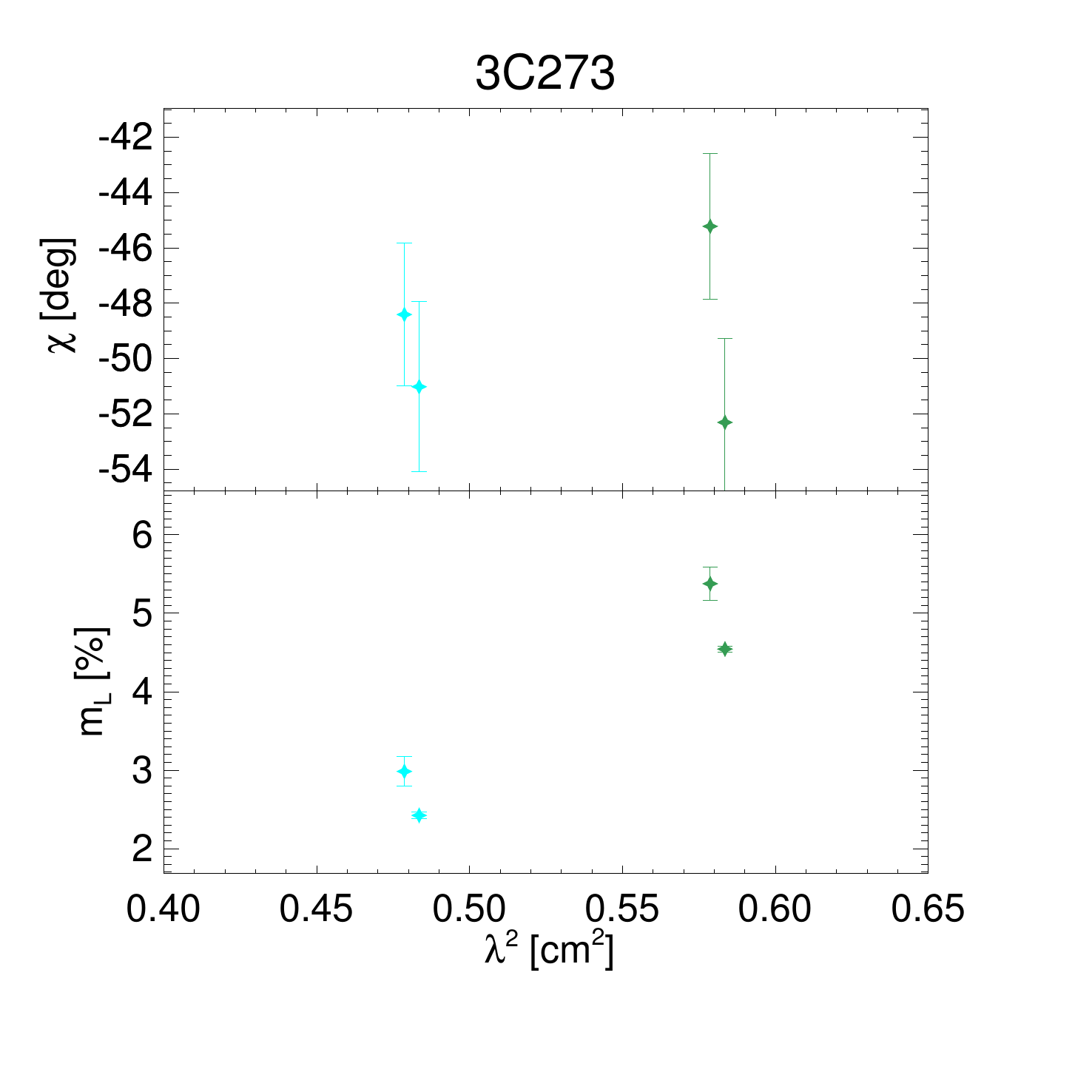}
\includegraphics[trim=5mm 8mm 8mm 2mm, clip, width = 58mm]{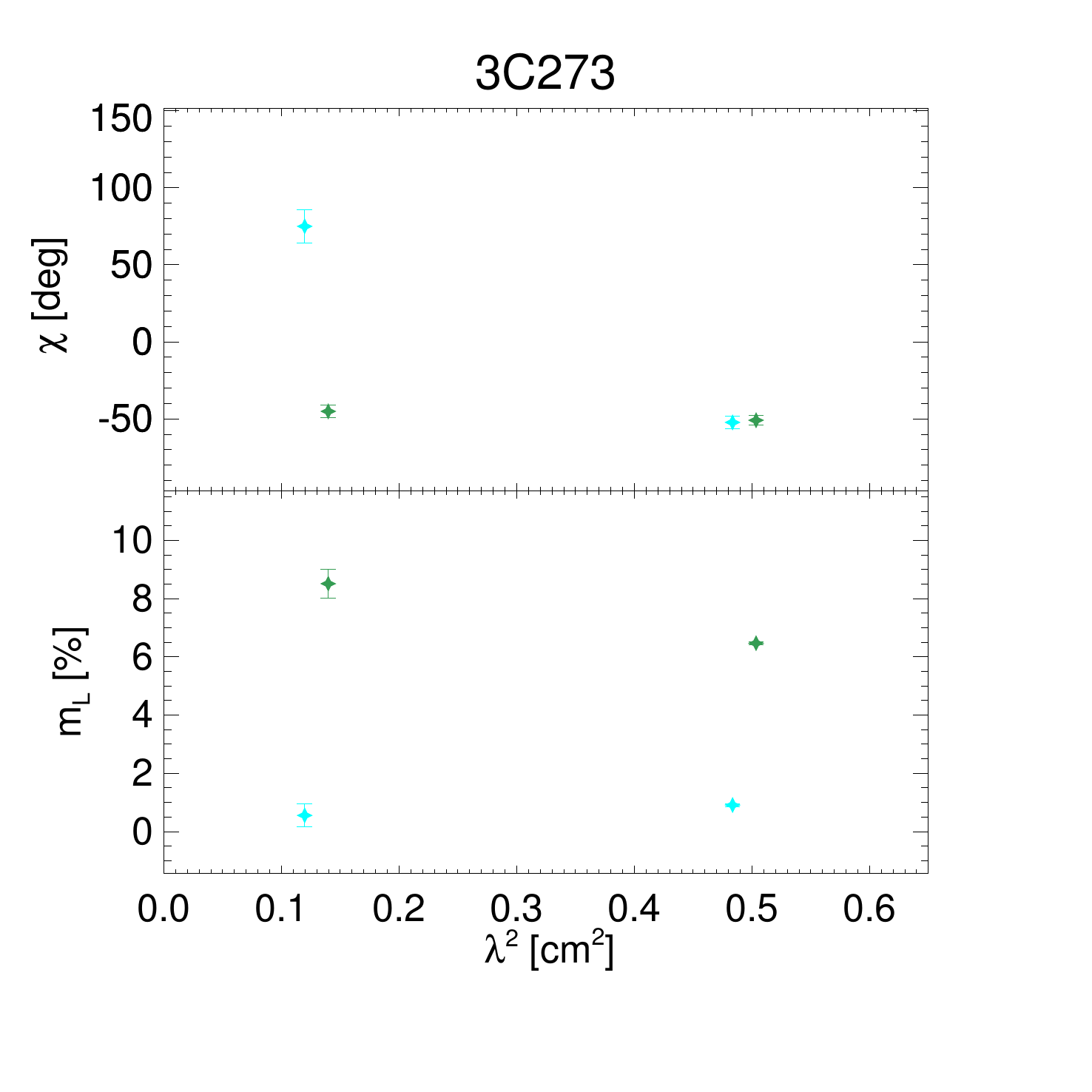}
\caption{\emph{Top panels:} Same as the left panel of Figure~\ref{result} but showing contemporaneous VLBA images along our KVN images of 3C~273. \emph{Left:} MOJAVE 15 GHz vs. KVN 22 GHz, \emph{Center:} BU 43 GHz vs KVN 43 GHz, \emph{Right:} BU 43 GHz vs KVN 86 GHz. The VLBA maps are convolved with the beam of the corresponding KVN maps. \emph{Bottom panels:} EVPA and degree of linear polarization as a function of $\lambda^2$. Colors indicate the polarization quantities at the jet locations marked by asterisks in the upper panel images. For clarity, we shifted values obtained from the same map along the $\lambda^2$ axis. \label{comparison}}
\end{figure*}

We show examples of calibration of instrumental polarization in Figure~\ref{dtermcal}. The calibrated complex visibility ratio, LR/LL with L and R referrring to left and right hand circular polarization, respectively, from the calibrator sources used for D-term calibration of our 2017 January data is shown on the complex plane. KVN antennas are on alt-azimuth mounts and instrumental polarization does not (usually) change over time as antennas change the direction of pointing. Target parallactic angles change over time, causing the polarization signal received by the alt-az antennas to vary. Since we perform parallactic angle correction in the early stage of data pre-processing, the polarization signal intrinsic to the target remains constant in the LR/LL plane while the instrumental polarization signal rotates with parallactic angle. Therefore, the rotation of the visibility ratio on the complex plane in Figure~\ref{dtermcal} is (mostly) due to instrumental polarization. See \cite{roberts1994} and \cite{aaron1997} for details of instrumental polarization calibration.

Amplitudes and phases of D-terms can be derived from the rotating pattern by assuming that the center of rotation does not vary with time. This is true only when (i) the antenna D-terms do not vary during the observation, which is true in most cases, and (ii) the calibration sources are either unpolarized or are polarized but spatially un-resolved. As we have 6 independent visibility ratios from three baselines and 6 unknown D-terms for three antennas and two polarizations, solutions of D-terms can be obtained from fitting the pattern observed in the complex plane.

The green and red lines in the left and the central panels of Figure~\ref{dtermcal} correspond to the expected LR/LL variation caused by the D-terms of two antennas, as obtained from the AIPS task LPCAL. The blue lines are the sum of contributions of the two D-terms, which is in good agreement with the data. The right panels show the visibility ratio after the D-term correction; the data are clustered around fixed points which mark polarization signals intrinsic to the target AGNs. The scatter is mostly due to thermal noise in the data; the noise becomes larger at higher observing frequencies.

Even precise D-term measurements may be affected by substantial systematic errors like non-stationary centers of rotation in the complex plane. Our calibrator sources are not perfectly un-polarized (or are polarized with sub-structure), and the D-terms may not always be constant during an observation run. We can estimate the errors on the D-terms by comparing the values obtained from different instrumental polarization calibrators as we did in Figure~\ref{dtermcomparison}. We use the standard devations of the D-terms obtained using different calibrators as errors. The errors are usually less than 1--2\% but sometimes up to 3\%. These errors will be transferred to the polarization quantities we used in our analysis and we considered these errors as described in Section~\ref{sect2}.

The D-terms of the VLBA antennas are known to vary only on timescales of months or longer (e.g., \citealt{gomez2002}). In Figure~\ref{dtermevolution} we check the stability of the KVN D-terms; their amplitudes seem to be mostly stable over $\approx4$ months but sometimes show non-negligible variability. Their standard deviations are usually less than 1--2\% -- in agreement with the formal errors of the D-terms.

\section{Reliability check of KVN polarimetry \label{appendixb}}
\label{appendixb}

Thanks to the extensive monitoring of blazars, specifically by the MOJAVE program at 15 GHz and the VLBA-BU program\footnote{https://www.bu.edu/blazars/VLBAproject.html} at 43 GHz, we could check if our KVN maps are consistent with contemporaneous VLBA images. We picked 3C 273, which shows complex jet structure in both total intensity and polarization and thus is not used for our analysis of blazar cores, as reference source and compared (a) our KVN 22 GHz data observed on 2016 December 9 with the MOJAVE 15 GHz data observed one day after, (b) our KVN 43 GHz data observed on 2016 December 10 with the BU data observed on 2016 December 24, and (c) our KVN 86 GHz data observed on 2016 December 10 with the BU data in Figure~\ref{comparison}. The top panels show polarization maps generated from the KVN and VLBA data next to each other. The VLBA maps are convolved with the corresponding KVN beam.

The VLBA and KVN distributions of fractional jet polarization at 15 GHz and 22 GHz, respectively, are in good agreement with each other, showing higher degrees of polarization -- up to $\approx70\%$ -- at the northern edge of the jet located $\approx10$ mas from the core. Likewise, the 43-GHz VLBA and KVN data are consistent with each other except that the KVN maps show more polarized emission in the outer jet, i.e., $\approx10$ mas from the core. This might be because there is a time gap of $\approx2$ weeks between the observations and/or the KVN has only short baselines (with the maximum baseline length less than 500 km) and thus is more sensitive to the extended emission. However, the 86 GHz KVN map shows an additional polarization component near the core region, while the 43 GHz VLBA map does not have such a component but has an extended polarization near the core from the inner jet polarization component (at $\approx$ 1 mas from the core) by convolution of a large beam. Interestingly, the core of this source is usually unpolarized (e.g., \citealt{jorstad2005, attridge2005, hada2016}), which has been attributed to strong Faraday depolarization or intrinsically very low polarization at the core. Its inner jet components at $\lesssim1$ mas from the core show $|\rm RM| \approx a\ few \times 10^4\ rad/m^2$ \citep{attridge2005, jorstad2007, hada2016}. Therefore, our result might indicate a detection of core polarization of 3C 273 at 86 GHz presumably because of less depolarization at higher frequency. We will verify this possibility in a forthcoming paper with more data at 86 and 129 GHz (Park et al. in preparation).

The bottom panels of Figure~\ref{comparison} show EVPA and degree of polarization as a function of $\lambda^2$ at a few locations in the jet marked in the top panels. The 15-GHz VLBA and 22-GHz KVN results are, to first order, consistent with each other but show non-negligible differences. This might be due to Faraday rotation with a RM of a few hundred $\rm rad/m^2$ in the jet, as reported by many other studies (e.g., \citealt{hovatta2012}) and possibly different polarization structure in the jet at different frequencies. The VLBA and KVN data at 43 GHz are consistent with each other within errors, especially when considering the time gap of $\approx2$ weeks. Similarly, The VLBA 43 GHz and KVN 86 GHz data are in agreement with each other for the inner jet component at $\approx$ 1 mas from the core, taking into account the time gap and the small rotation measure. Therefore, we conclude that the polarimetry mode of KVN is reliable.


\end{document}